\documentclass[12pt,a4paper]{article}

\usepackage[T1]{fontenc}
\usepackage[utf8]{inputenc}
\usepackage[a4paper,top=2.2cm,bottom=2.2cm,left=2.3cm,right=2.3cm]{geometry}
\usepackage{setspace}
\usepackage{array}
\usepackage{multirow}
\usepackage{tabularx}
\usepackage{booktabs}
\usepackage{lmodern}
\usepackage{amsmath,amssymb,mathtools,bm}
\usepackage{graphicx}
\usepackage{wrapfig}
\usepackage{fvextra}
\usepackage{esint}
\usepackage[normalem]{ulem}
\usepackage{dcolumn}
\newcolumntype{d}[1]{D{.}{.}{#1}}

\newcommand{\dd}{\mathrm d}
\newcommand{\M}{\mathcal M}

\newcommand{\ii}{\mathrm i}

\newcommand{\sigman}{\boldsymbol{\sigma}_n}
\newcommand{\sigmap}{\boldsymbol{\sigma}_p}
\newcommand{\pp}{\boldsymbol{p}}
\newcommand{\Del}{\boldsymbol{\Delta}}
\newcommand{\supercite}{\cite}
\usepackage{orcidlink}
\begin{document}
	\begin{titlepage}
		\pagestyle{empty}
		\author{Jean-Marc Richard\footnote{email: j-m.richard@ip2i.in2p3.fr}\,\orcidlink{0000-0001-6459-765X}\\
			{\small Present address : Universit\'e Lyon 1, Institut de Physique des 2 Infinis de Lyon, CNRS-IN2P3}\\
			{\small 4, rue Enrico Fermi, Villeurbanne, France}}
		\title{\bf Contribution to the theoretical and phenomenological study of nuclear forces}
		\date{\small\today}
		\maketitle
		\begin{abstract}
			This work presents a detailed study of two-pion exchange between two nucleons, including the effects of pion–nucleon rescattering and pion–pion interactions. The peripheral nucleon–nucleon scattering phase shifts are calculated and compared with those extracted from data analyses. The interaction is then supplemented by a phenomenological short-range potential, and the resulting model is used to evaluate nucleon–nucleon scattering observables and the properties of the deuteron. 
			
			This document is the English translation of a thesis defended on June 19, 1975, at the University of Paris, in fulfillment of the requirements for the doctoral degree. It describes some of the basic elements underlying the construction of what is now known as the Paris potential.
		\end{abstract}
	\end{titlepage}
	
	\thispagestyle{empty}
	\begin{center}
		\vspace*{1.15cm}
		
		{\sffamily\Large\MakeUppercase{STATE DOCTORAL THESIS}\par}
		\vspace{0.55cm}
		{\sffamily\Large\MakeUppercase{IN PHYSICAL SCIENCES}\par}
		
		\vspace{1.85cm}
		
		{\ttfamily presented\par}
		\vspace{0.45cm}
		{\ttfamily to the University Pierre et Marie Curie\par}
		
		\vspace{1.75cm}
		
		{\ttfamily by\par}
		\vspace{0.35cm}
		{\ttfamily Mr.~Jean-Marc RICHARD\par}
		
		\vspace{1.75cm}
		
		{\ttfamily in order to obtain\par}
		\vspace{0.35cm}
		{\ttfamily the degree of DOCTOR OF SCIENCE\par}
	\end{center}
	
	\vspace{1.55cm}
	
	{\ttfamily
		Thesis title: CONTRIBUTION TO THE THEORETICAL AND PHENOMENOLOGICAL
		\par
		\phantom{Thesis title: }STUDY OF NUCLEAR FORCES.
		\par}
	
	\vspace{1.55cm}
	
	{\ttfamily
		defended on June 19, 1975
		\par
		\vspace{0.25cm}
		before a jury composed of:
		\par}
	
	\vspace{0.45cm}
	
\begin{center}
	\ttfamily
	\begin{tabular}{lll}
		M.   &R. NATAF  \hspace*{2cm}   & Chairman\\
		Mme. &N. MARTY     & \\
		MM.\  &R. VINH MAU    & \\
		&R. BRYAN   \qquad    & Examiners\\
		&W.N. COTTINGHAM&  \\
		&E. LEADER      & 
	\end{tabular}
\end{center}

\newpage
\thispagestyle{empty}

\newpage
\thispagestyle{empty}
\vspace*{1.35cm}

{
	\setlength{\parindent}{1.2cm}
	\setlength{\parskip}{0.45em}
	\baselineskip=15pt

	The subject of this thesis was suggested to me by Professor
	R.~Vinh Mau, who welcomed me into his research group working
	on strong interactions. His insightful advice and effective help
	in many areas were a decisive source of support throughout my work.
	
	I would also like to thank Mrs.~N.~Marty and Messrs.~R.~Nataf,
	R.~Bryan, W.N.~Cottingham, and E.~Leader for agreeing to serve
	on the thesis jury.
	
	The work presented here was carried out in collaboration with
	R.~Vinh Mau, M.~Lacombe, and B.~Loiseau. This collaboration
	was both stimulating and rewarding. In the first part of the work,
	W.N.~Cottingham generously shared his experience and enthusiasm
	with us. R.~de Tourreil and P.~Pires played an essential role
	in overcoming the difficulties encountered in the second part.
	
	Many discussions with G.E.~Brown and the members of his group,
	together with regular comparisons of our results with theirs,
	helped us clarify a number of subtle points. A.~Gersten's advice
	was also very valuable, particularly regarding Chapter~IV.
	F.~Lehar and J.~Bystricky kindly provided us with preliminary
	results from their phase-shift analyses, while H.~Nielsen and
	G.~Oades shared the results of their studies of pion-nucleon
	interactions.
	
	The research reported here was carried out within the Theoretical
	Physics Division of the I.P.N. The scientific standards and
	working atmosphere of this laboratory, fostered in particular by
	M.~Jean, M.~Vénéroni, and Mrs.~N.~Vinh Mau, provided an
	especially favourable environment for research.
	
	The numerical calculations were performed at the Nuclear Physics
	Computing Centre of the University Pierre et Marie Curie and at
	the Paris-Sud Computing Centre, thanks to the skill and dedication
	of their computing staff and technicians.
	
	The typing, preparation of the figures, and layout of this thesis
	were carried out by J.~Durin, while the printing and binding
	were handled by L.~Duquesne and D.~Bernia.
	
	\vspace{0.55cm}
	\noindent\hspace*{1.2cm}I would like to express my sincere gratitude to all of them.}
\newpage
\vspace*{0.2cm}
\begin{center}
	TABLE OF CONTENTS
\end{center}
\vspace{0.55cm}

\noindent\underline{Introduction}\hfill 1

\vspace{0.55cm}
\noindent Chapter I -- \underline{Relativistic Kinematics of Nucleon-Nucleon Scattering}\hfill\\
\vspace{-0.2cm}\\
\hspace*{1.2cm}1) Invariant amplitudes\dotfill 4\par
\hspace*{1.2cm}2) Direct-channel helicity amplitudes\dotfill 5\par
\hspace*{1.2cm}3) Crossed-channel helicity amplitudes\dotfill 6\par
\hspace*{1.2cm}4) Kinematic singularities\dotfill 9\par

\vspace{0.5cm}
\noindent Chapter II -- \underline{Contribution from Two-Pion Exchange}

\vspace{0.2cm}
\hspace*{1.2cm}1) Analytic properties of the nucleon-nucleon amplitude\par
\hspace*{1.8cm}a-- Mandelstam representation\dotfill 10\par
\hspace*{1.8cm}b-- Asymptotic behaviour in $t$\dotfill 11\par
\hspace*{1.8cm}c-- Asymptotic behaviour in $w$\dotfill 12\par

\vspace{0.2cm}
\hspace*{1.2cm}2) Spectral functions arising from two-pion exchange\par
\hspace*{1.8cm}a-- Expression of the double spectral functions\dotfill 12\par
\hspace*{1.8cm}b-- Subtracted dispersion relations\dotfill 15\par
\hspace*{1.8cm}c-- Consistency\dotfill 16\par
\hspace*{1.8cm}d-- Contribution of the S- and P-waves in $\pi\pi\to N\bar N$\dotfill 17\par
\hspace*{1.8cm}e-- Alternative formulation of the dispersion relations\dotfill 17\par

\vspace{0.45cm}
\noindent Chapter III -- \underline{Evaluation of the Spectral Functions from $\pi N$ and $\pi\pi$ Data}
\\
\vspace{-0.2cm}\\
\hspace*{1.2cm}1) Evaluation of the $\pi N$ absorptive parts in the region $t>4\mu^2$\par
\hspace*{1.8cm}a-- Extrapolation from the physical region\dotfill 21\par
\hspace*{1.8cm}b-- $\pi N$ amplitude in the physical region\dotfill 23\par

\vspace{0.2cm}
\hspace*{1.2cm}2) Determination of the S- and P-waves of $\pi\pi\to N\bar N$ on the two-pion cut\par
\hspace*{1.8cm}a-- Review of the properties of the partial waves $\pi\pi\to N\bar N$\dotfill 24\par
\hspace*{1.8cm}b-- P-wave\dotfill 25\par
\hspace*{1.8cm}c-- S-wave\dotfill 26\par

\newpage
\vspace*{0.75cm}
\noindent Chapter IV -- \underline{Peripheral Nucleon-Nucleon Phase Shifts}

\vspace{0.25cm}
\hspace*{1.2cm}1) Determination of the phase parameters\par
\hspace*{1.8cm}a-- From invariant amplitudes to $\ell-s$ transitions\dotfill 30\par
\hspace*{1.8cm}b-- Froissart-Gribov representation\dotfill 31\par
\hspace*{1.8cm}c-- Definition of the phase shifts and unitarization issues\dotfill 32\par

\vspace{0.25cm}
\hspace*{1.2cm}2) Discussion of the results\par
\hspace*{1.8cm}a-- Summary of the various contributions\dotfill 35\par
\hspace*{1.8cm}b-- Results\dotfill 36\par
\hspace*{1.8cm}c-- Comparison with other works\dotfill 40\par

\vspace{0.45cm}
\noindent Chapter V -- \underline{Nucleon-Nucleon Interaction in the Central Region}

\vspace{0.25cm}
\hspace*{1.2cm}1) Phase-shift analysis\dotfill 42\par

\vspace{0.2cm}
\hspace*{1.2cm}2) Search for a semi-phenomenological potential\par
\hspace*{1.8cm}a-- Description of the interaction in configuration space\dotfill 43\par
\hspace*{1.8cm}b-- Evaluation of the long- and intermediate-range potential\dotfill 44\par
\hspace*{1.8cm}c-- Nonlocal terms\dotfill 44\par
\hspace*{1.8cm}d-- Parametrization of the core\dotfill 46\par
\hspace*{1.8cm}e-- Fitting the core parameters\dotfill 48\par
\hspace*{1.8cm}f-- Results and discussion\dotfill 49\par

\vspace{0.55cm}
\noindent\underline{Conclusion}\hfill 57

\vspace{0.35cm}
\noindent\underline{References}\hfill 58

\newpage\centerline{\hfill\small -- 1 --}\vskip .25cm
\begin{center}
	\underline{INTRODUCTION}
\end{center}
\vspace{0.4cm}

\setlength{\parindent}{1.2cm}
\setlength{\parskip}{0.45em}
The large number of publications devoted to the study of nuclear forces
reflects the interest in this problem, which is certainly one of the most
important in contemporary physics. The various contributions are, of course,
of unequal significance, but it must be acknowledged that, taken as a whole,
considerable progress has been made over the past few decades in
understanding the nucleon-nucleon interaction at low energies. A complete
account of all the work carried out in this field would require an entire
book. Therefore, before presenting the approach adopted here, we shall
restrict ourselves to recalling the main lines of research pursued so far,
while accepting in advance the risk of giving an overly brief critical
assessment.

The exchange of a $\pi$ meson (O.P.E.), proposed by Yukawa, provides
a good description of long-range effects. It was therefore natural to
attempt to account for the less peripheral forces through the exchange
of heavier mesons. It is easy to understand why many advances in the
theory of nuclear forces have been closely tied to developments in
meson physics.

One such approach consists in including the contributions of mesons
such as the $\rho$ or $\omega$. This has led, in particular, to the
construction of one-boson-exchange models\supercite{ref1} (O.B.E.),
whose agreement with experiment is sought by varying the coupling
constants. This is therefore a semi-phenomenological approach which,
owing to its simplicity, would be attractive if satisfactory results
could be obtained without resorting to the exchange of one or two
unrealistic scalar isoscalar mesons (whose masses are sometimes
different for proton-proton and neutron-proton interactions, and are
in any case too low compared with the possible experimental values).

It is also known, however, from the work of Lévy\supercite{ref2}
and Klein\supercite{ref3}, which has been repeatedly revisited and
improved\supercite{ref4}, that two-pion exchange (T.P.E.) plays
an important role and provides, at least qualitatively, a good
description of the intermediate-range properties of nuclear forces,
starting from a very simple Lagrangian. However, because of the
very strong attraction arising from fourth-order diagrams, it becomes
necessary either to introduce corrections to the perturbative
calculation (such as pair suppression) or to include additional
contributions.

This is the approach adopted in some mixed models, in which
boson exchange is supplemented by two-pion exchange. Although
this type of approach generally provides fairly good agreement
with experiment, it nevertheless remains somewhat empirical and
lacks a fully satisfactory theoretical justification. It loses the
simplicity of a Lagrangian theory involving only a small number
of fundamental fields, while retaining the ambiguities associated
with the exchange of scalar ``mesons'' such as $\epsilon$ or $\sigma$.

More generally, a satisfactory theory must describe not only
the nucleon-nucleon interaction, but also meson-meson
collisions and meson scattering on matter. Unfortunately,
no theory has yet succeeded in achieving this.

\newpage
\centerline{\hfill\small -- 2 --}
\vspace{0.45cm}
Dispersion relations offer a different, broader, although perhaps
less ambitious, way of approaching this problem. In this framework,
intermediate-range forces are no longer obtained by summing the
perturbation series to sufficiently high order (while also testing
the Lagrangian), but rather by evaluating the discontinuities
associated with the nearest singularities of the $S$-matrix.
These discontinuities are themselves related, through unitarity,
to amplitudes for reactions such as
$\pi\pi\longrightarrow N\bar N$,
$\pi\pi\longrightarrow\rho\ldots$.
In the absence of a comprehensive theory of strong interactions
satisfying these unitarity conditions and the other general
properties of the $S$-matrix, that is, the Bootstrap equations,
one may simply attempt to derive nuclear forces from the well-established
properties of pion-pion and pion-nucleon scattering and compare
the resulting predictions with experiment.

The groundwork for such an approach was laid in 1960 by the
studies of Goldberger, Grisaru, Mac Dowell, and Wong\supercite{ref5},
and by those of Amati, Leader, and Vitale\supercite{ref6},
who investigated the analytic properties of the nucleon-nucleon
amplitude and proposed a Mandelstam representation to describe
the full contribution from two-pion exchange. Subsequent developments
by Cottingham and Vinh Mau\supercite{ref7} led to the construction
of a theoretical nuclear potential which reproduced, at least
qualitatively, the essential features of the long- and intermediate-range
nuclear forces. At that time, the only well-established properties
of mesons were the formation of the $(3/2,3/2)$ resonance in
$\pi N$ scattering and that of the $\rho$ in the P-wave of
$\pi\pi$ scattering. Our knowledge of these reactions is much
more extensive today. Furthermore, Partovi and Lomon\supercite{ref8}
showed that part of the excessively strong attraction found by
Cottingham and Vinh Mau arose solely from the conversion of the
scattering amplitude into a potential. The method used, following
Charap and Tausner\supercite{ref9}, involved overly drastic
approximations.

This approach was taken up again some years ago by several groups
bringing together elementary-particle physicists and nuclear physicists,
with various motivations for using a Mandelstam representation to study
the nucleon-nucleon interaction. Our contribution is presented in
the first part of this thesis.

We first critically examined the foundations of the earlier work,
particularly with regard to the convergence of the various dispersion
relations employed. The subtractions we found necessary allowed us
to isolate the significant contribution of the S- and P-waves in
$\pi\pi\longrightarrow N\bar N$, which we studied in detail,
taking particular account of recent estimates of the pion-pion
phase shift $\delta^0_0$. We also sought to make the fullest
possible use of our knowledge of the $\pi N$ interaction,
reconstructing the amplitude from the data and extrapolating it
to the region corresponding to virtual pions exchanged between
the nucleons.

\newpage
\centerline{\hfill\small -- 3 --}
\vspace{0.35cm}

The results obtained can be expressed either in terms of an equivalent
potential or by calculating the peripheral partial waves directly.
As a first step, we chose the latter approach, since it allows
our predictions to be compared with phenomenological phase-shift
analyses.

\vspace{0.25cm}
However, scattering experiments are far from being the only means
of investigating nuclear forces. The study of nuclear properties
provides another powerful tool, which led us also to consider
the viewpoint of low-energy nuclear physicists. In the absence of
a satisfactory theory of the nucleon-nucleon interaction at all
distances, they use phenomenological potentials in nuclear-structure
calculations, at least in those cases where effective interactions
need not be introduced. These potentials are constructed numerically
to reproduce the nucleon-nucleon scattering data and the proton-neutron
bound state. Some achieve this successfully, but they differ
considerably from one another. To reduce this arbitrariness,
one may fix the long-range part of the interaction (one-pion exchange)
and restrict the remaining terms to physically reasonable
parametrizations, for example, a superposition of Yukawa terms.
Despite these constraints, significant differences between the
potentials appear already at distances of 1.5 or 2 fm, these
differences being compensated for by cores of differing hardness.

\vspace{0.25cm}
We have also attempted to construct a nuclear potential, but
retaining as intermediate-range forces those derived in our
previous work from the $\pi N$ and $\pi\pi$ interactions and
from $\omega$ exchange. The free parameters therefore affect
only the core region. Their relatively small number reduces
the ambiguities in their determination through numerical
procedures. Moreover, this approach provides an indirect test
of the theoretical part of the potential.

From another point of view, if one has confidence in the results
obtained from dispersion relations, one may hope to construct,
by adding a core, a genuinely realistic interaction based on
sound theoretical foundations and therefore suitable for use
in nuclear calculations. The methods employed in this
phenomenological study, and especially the results obtained,
are presented in the second part of this thesis. We also discuss
the possible application of a better theoretical understanding
of intermediate-range forces to nucleon-nucleon phase-shift
analyses.
\newpage
\centerline{\hfill\small -- 4 --}
\vspace{0.35cm}
\noindent\underline{Chapter I} -- RELATIVISTIC KINEMATICS OF NUCLEON-NUCLEON SCATTERING

\vspace{0.35cm}
The relativistic kinematics of collisions between two spin-$1/2$
particles has been discussed many times, but it remains rather
complicated, and incorrect formulae are often encountered in the
literature. In this chapter, we shall try to specify as clearly
as possible the notation used throughout the following sections,
particularly with regard to the decomposition of the amplitudes
into partial waves in the direct and crossed channels.
We shall then briefly discuss the problem of kinematic singularities.

\vspace{0.3cm}
\noindent 1) \underline{Invariant amplitudes}

In terms of the momenta defined in Fig.~1, the Mandelstam
variables are given by:\\
\centerline{
	\begin{minipage}{.49\textwidth}	
		\begin{equation*}
			\begin{aligned}
				w&=-(p_1+n_1)^2,\\
				t&=-(p_2-p_1)^2,\\
				\bar t&=4m^2-w-t,
			\end{aligned}
			\qquad \text{(1.1)}
		\end{equation*}
	\end{minipage}
	\hfill
	\begin{minipage}{.45\textwidth}
		\vskip .4cm
		\includegraphics[width=0.7\linewidth]{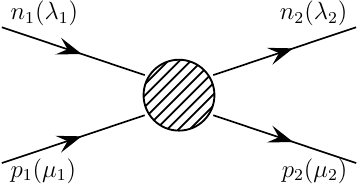}\\[3pt]
		\small Fig.~1. Nucleon-nucleon collision
	\end{minipage}\hfill}
\vskip .2cm
\noindent
or, in the direct-channel center-of-mass frame, using obvious notation,
\begin{equation*}
	\begin{aligned}
		w&=4E^2=4(k^2+m^2),\\
		t&=-2k^2(1-\cos\theta),\\
		\bar t&=-2k^2(1+\cos\theta),
	\end{aligned}
	\qquad \text{(1.2)}
\end{equation*}

The $S$-matrix element is written as
\begin{equation*}
	S_{21}=\delta_{21}+i\delta^4(m_1+p_2-n_1-p_1)
	\left(\frac{m}{2\pi E}\right)^2
	\left[\M_{21}\right]_{\text{antisymmetrized}} .
	\tag{1.3}
\end{equation*}

\begin{equation}
	\M_{21}=\bar u(n_2)\,\bar u(p_2)\,
	\chi_{n2}^\dagger \chi_{p2}^\dagger\,\M\, 
	u(n_1)\,u(p_1) \chi_{n1} \chi_{p1},
	\tag{1.4}
\end{equation}
In isospin space, we have the decomposition
\begin{equation}
	\M=3\M^+ +2\M^-\,\vec\tau_n\cdot\vec\tau_p .
	\tag{1.5}
\end{equation}
Thus, the amplitudes with definite isospin $I$ in the direct
channel are
\begin{equation}
	\begin{aligned}
		\M^{I=1}_0&=3\M^+ +2\M^-,\\
		\M^{I=0}_0&=3\M^+ -6\M^-.
	\end{aligned}
	\tag{1.6}
\end{equation}
\newpage
\centerline{\hfill\small -- 5 --}\vskip .3cm
As far as the spins are concerned, one of the commonly used bases
is that formed by the so-called ``perturbative'' invariants,
namely the operators\supercite{ref6}
\begin{equation}
	\begin{aligned}
		P_1&=1^n\cdot 1^p,\\
		P_2&=(i\gamma^n\cdot P)\,1^p
		+ 1^n\,(i\gamma^p\cdot N),\\	
			P_3&=(i\gamma^n\cdot P)\,(i\gamma^p\cdot N),\\
		P_4&=\gamma^n\cdot\gamma^p,\\	
			P_5&=\gamma_5^n\cdot\gamma_5^p,
	\end{aligned}
	\tag{1.7}
\end{equation}
with
\begin{equation}
	N=\frac12(n_1+n_2),\qquad
	P=\frac12(p_1+p_2).
	\tag{1.8}
\end{equation}

We can therefore write
\begin{equation}
	\M^\pm=\sum_{i=1}^{5}p_i^\pm(w,t)P_i,
	\tag{1.9}
\end{equation}
thereby defining the invariant amplitudes $p_i^\pm(w,t)$.

The particles $(n)$ and $(p)$ are now treated as distinct,
since the Pauli principle has been explicitly taken into account
in the definition (1.3) of the transition matrix\supercite{ref7}.

%
\noindent 2) \underline{Direct-channel helicity amplitudes}

\vspace{0.2cm}
To study the asymptotic behaviour in $t$, or to calculate the
phase parameters, we shall need the direct-channel helicity
amplitudes $\langle\lambda_2\mu_2|\phi|\lambda_1\mu_1\rangle$
(using the notation of Fig.~1). Only five of these are independent,
and we may choose\supercite{ref5}
\begin{equation}
	\begin{aligned}
		\varphi_1&=\langle++|\phi|++\rangle
		=\sum_J(2J+1)T^J_1d^J_{00}(z),\\
		\varphi_2&=\langle++|\phi|--\rangle
		=\sum_J(2J+1)T^J_2d^J_{00}(z),\\
		\varphi_3&=\langle+-|\phi|+-\rangle
		=\sum_J(2J+1)T^J_3d^J_{11}(z),\\
		\varphi_4&=\langle+-|\phi|-+\rangle
		=\sum_J(2J+1)T^J_4d^J_{1,-1}(z),\\
		\varphi_5&=\langle++|\phi|+-\rangle
		=\sum_J(2J+1)T^J_5d^J_{10}(z).
	\end{aligned}
	\tag{1.10}
\end{equation}
where $z=\cos\theta$.

The transformation from the $p_i(w,t)$ to the $\varphi_j$
can readily be obtained\supercite{ref5} by saturating the
perturbative invariants with helicity spinors
\begin{equation}
	\langle\lambda_2\mu_2|\phi|\lambda_1\mu_1\rangle
	=\sum_{i=1}^{5}p_i(w,t)\,
	\bar u_{\lambda_2}(p_2)\bar u_{\mu_2}(n_2)
	P_i\, u_{\lambda_1}(p_1)u_{\mu_1}(n_1).
	\tag{1.11}
\end{equation}
\newpage\centerline{\hfill \small -- 6 --}\vskip .4cm
It follows that
\begin{equation}
	\begin{aligned}
		\varphi_1={}&\frac{1+z}{2}
		(p_1-2m D p_2+m^2D^2p_3)
		+(D-\frac{1-z}{2})p_4,\\
		\varphi_2={}&-\frac{1-z}{2}
		\left(\frac{E^2}{m^2}p_2-\frac{2E^2}{m^2}p_3
		+p_3+p_4+\frac{k^2}{m^2}p_5\right),\\
		\varphi_3={}&\frac{1+z}{2}
		(p_1-2m D p_2+m^2D^2p_3)+D p_4,\\
		\varphi_4={}&\frac{1-z}{2}
		\left(\frac{E^2}{m^2}p_1-\frac{2E^2}{m^2}p_2
		+E^2p_3+p_4-\frac{k^2}{m^2}p_5\right),\\
		\varphi_5={}&\frac{E\sqrt{1-z^2}}{2m}
		\left(-p_1+\frac{2E^2}{m^2}p_2-m^2 D p_3-p_4\right),
	\end{aligned}
	\tag{1.12}
\end{equation}

	\begin{equation*}
		D=\frac{E^2+k^2}{m^2}.
	\end{equation*}
	
	The inverse relations are given by
	\begin{equation}
		\begin{aligned}
			p_5&=-\frac{m^2}{k^2(1-z)}
			(\varphi_2+\varphi_4),\\
			p_4&=-\frac{2}{1-z}(\varphi_2-\varphi_3),\\
			p_3&=\frac{1}{2E^2(1-D^2)}
			\left[\frac{2}{1-z}(\varphi_4-\varphi_2)
			-\frac{4E^2}{m^2}\frac{\varphi_3}{1+z}
			-2\left(1-\frac{E^2}{m^2}D\right)p_4\right],\\
			p_2&=\frac{m}{2(E^2-m^2D)}
			\left[\frac{2\varphi_2}{1+z}
			+\frac{2m\varphi_5}{E\sqrt{1-z^2}}
			-m^2D(D-1)p_3-(D-1)p_4\right],\\
			p_1&=\frac{2\varphi_3}{1+z}+2m D p_2-m^2D^2p_3-D p_4.
		\end{aligned}
		\tag{1.13}
	\end{equation}
	%
	\vspace{0.35cm}
	\noindent 3) \underline{Crossed-channel helicity amplitudes}
	
	According to the general principles of $S$-matrix theory,
	$\M_{t}$ is the analytic continuation of the elastic nucleon--antinucleon
	scattering amplitude (Fig.~2).
	
	In the center-of-mass frame of the $t$-channel, the Mandelstam
	variables are given by
	
	\vskip .15cm
	\centerline{\begin{minipage}{.49\textwidth}
			\vskip .9cm
			\begin{equation}
				\left\{
				\begin{aligned}
					t&=4W^2=4(\vec p^{\,2}+m^2),\\
					w&=-2p^2(1+x),\\
					\bar t&=-2p^2(1-x),
				\end{aligned}
				\right.
				\tag{1.14}
			\end{equation}
			\vskip.1cm
			where $x=\cos\psi$.
			\vskip .1cm
			The $S$-matrix element for the reaction
			\[
			m_1+(-m_2)\longrightarrow p_2+(-p_1)
			\]
			can be written as
		\end{minipage}\hfill
		\begin{minipage}{.49\textwidth}
					\centering	\includegraphics[width=0.6\linewidth]{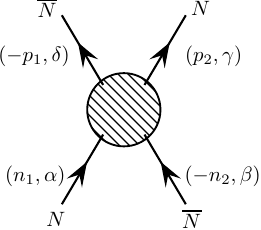}\\
					\vskip .2cm
					{\small Fig.~2. Elastic nucleon--antinucleon \\ scattering}
			\end{minipage}}
			
			\newpage\centerline{\hfill\small -- 7 --}\vskip .35cm
			\begin{equation}
				f_{fi}=f^{\,0}_{fi}
				+i\delta^4(n_1-n_2-p_2+p_1)
				\left(\frac{m}{2\pi W}\right)\M_{fi}.
				\tag{1.15}
			\end{equation}
			\begin{equation}
				\M_{fi}=\bar v(-n_2)\,u(p_2)\,\M\,
				\bar u(n_1)\,v(-p_1).
				\tag{1.16}
			\end{equation}
			
			The transitions with definite isospin $T$ are, according to (1.5),
			\begin{equation}
				\M^{T=1}_t=4\M^-,
				\qquad
				\M^{T=0}_t=6\M^+.
				\tag{1.17}
			\end{equation}
			
			The helicity amplitudes, analogous to those in the direct channel,
			can be expressed in terms of the invariant amplitudes $p_i(w,t)$\supercite{ref10}
			\begin{equation}
				\langle\gamma,\delta|\widetilde\phi|\alpha,\beta\rangle
				=\sum_{i=1}^{5}P_i(w,t)\,
				\bar v_\beta(-m_2)u_\gamma(p_2)
				p_i u_\alpha(m_1)v_\delta(-p_1).
				\tag{1.18}
			\end{equation}
			which gives
			\begin{equation}
				\begin{aligned}
					\widetilde\varphi_1
					&=\langle++|\widetilde\phi|++\rangle
					=\frac{p^2}{m^2}p_1-\frac{2p^2}{m}p_2
					+p^2x^2p_3-xp_4-\frac{W^2}{m^2}p_5,\\
					\widetilde\varphi_2
					&=\langle++|\widetilde\phi|--\rangle
					=\frac{p^2}{m^2}p_1-\frac{2p^2}{m}p_2
					+p^2x^2p_3-xp_4+\frac{W^2}{m^2}p_5,\\
					\widetilde\varphi_3
					&=\langle+-|\widetilde\phi|+-\rangle
					=-\frac{W^2}{m^2}(1+x)
					[p^2(1-x)p_3+p_4],\\
					\widetilde\varphi_4
					&=\langle+-|\widetilde\phi|-+\rangle
					=\frac{W^2}{m^2}(1-x)
					[p^2(1+x)p_3-p_4],\\
					\widetilde\varphi_5
					&=\langle++|\widetilde\phi|+-\rangle
					=\frac{W}{m}\sin\psi
					\left[\frac{p^2}{m^2}p_2-p^2 x p_3+p_4\right].
				\end{aligned}
				\tag{1.19}
			\end{equation}
and the inverse relations
\begin{equation}
	\begin{aligned}
		p_1={}&\frac{m^2}{2p^2}
		(\widetilde\varphi_1+\widetilde\varphi_2)
		+\frac{m^4x}{2W^2p^2}
		\left(\frac{\widetilde\varphi_3}{1+x}
		+\frac{\widetilde\varphi_4}{1-x}\right)
		+\frac{m^3x}{p^2W\sin\psi}\widetilde\varphi_5,\\
		p_2={}&\frac{m^3}{2W^2p^2}
		\left(\frac{\widetilde\varphi_3}{1+x}
		+\frac{\widetilde\varphi_4}{1-x}\right)
		+\frac{m^2}{p^2W\sin\psi}\widetilde\varphi_5,\\
		p_3={}&-\frac{m^2}{2p^2W^2}
		\left(\frac{\widetilde\varphi_3}{1+x}
		-\frac{\widetilde\varphi_4}{1-x}\right),\\
		p_4={}&-\frac{m^2}{2W^2}
		(\widetilde\varphi_3+\widetilde\varphi_4),\\
		p_5={}&-\frac{m^2}{2W^2}
		(\widetilde\varphi_1-\widetilde\varphi_2).
	\end{aligned}
	\tag{1.20}
\end{equation}

The partial waves are defined by the expansion
\begin{equation}
	\widetilde\varphi_j
	=\langle\gamma,\delta|\widetilde\phi|\alpha,\beta\rangle
	=\sum_J(2J+1)M^J_i\,d^J_{\alpha-\beta,\gamma-\delta}(x),
	\tag{1.21}
\end{equation}
but it is more convenient to construct nucleon--antinucleon states
with natural parity $(-)^J$
\begin{equation}
	\begin{aligned}
		&|1\rangle&=\frac{1}{\sqrt2}\,\{|J,++\rangle+|J,--\rangle\},\\
		&|2\rangle&=\frac{1}{\sqrt2}\,\{|J,+-\rangle+|J,-+\rangle\},
	\end{aligned}
	\tag{1.22}
\end{equation}
\newpage
\centerline{\hfill\small-- 8 --}\vskip .3cm
or unnatural parity
\begin{equation}
	\begin{aligned}
		|3\rangle&=\frac{1}{\sqrt2}\,
		\{|J,++\rangle-|J,--\rangle\},
		\\
		|4\rangle&=\frac{1}{\sqrt2}\,
		\{|J,+-\rangle-|J,-+\rangle\}.
	\end{aligned}
	\tag{1.23}
\end{equation}

and hence transitions with definite parity
\begin{equation}
	\begin{aligned}
		\langle1|\M^J|1\rangle&=M^J_1+M^J_2,\\
		\langle2|\M^J|2\rangle&=M^J_3+M^J_4,\\
		\langle1|\M^J|2\rangle&=2M^J_5,\\
		\langle3|\M^J|3\rangle&=M^J_1-M^J_2,\\
		\langle3|\M^J|4\rangle&=M^J_3-M^J_4.
	\end{aligned}
	\tag{1.24}
\end{equation}

Retaining, from now on, only transitions with natural parity
in the $t$-channel (in particular, two-pion exchange), one obtains,
from (1.21) and (1.22), the partial-wave expansions of the
invariant amplitudes
\begin{align}
	p_1={}&\frac{m^2}{2p^2}\sum_J(2J+1)
	\langle1|\M^J|1\rangle P_J(x)
	+\frac{m^4}{2p^2W^2}\sum_J\frac{2J+1}{J(J+1)}
	\langle 2|\M^J|2\rangle
	\left[P'_J(x)+xP''_J(x)\right]\notag\\
	&-\frac{m^3}{p^2W}\sum_J(2J+1)
	\frac{\langle2|\M^J|1\rangle}{\sqrt{J(J+1)}}P'_J(x),\notag\\
	p_2={}&-\frac{m^3}{2p^2w^2}\sum_J\frac{(2J+1)^2}{J(J+1)}
	\left\{2\langle 1|M^J|1\rangle P_J'(x)+xP_J''(x)\right\} \notag\\
	={}&-\frac{m^2}{2p^2w^2}\sum_J(2J+1)
	\frac{\langle 1|M^J|1\rangle}{\sqrt{J(J+1)}}P_J'(x),
	\tag{1.25}\\[1ex]
	p_3={}&-\frac{m^2}{2p^2w^2}\sum_J\frac{(2J+1)}{J(J+1)}
	\langle 2|M^J|2\rangle P_J''(x),\notag\\
	p_4={}&-\frac{m^2}{2w^2}\sum_J\frac{(2J+1)}{J(J+1)}
	\langle 2|M^J|2\rangle P_J'(x).\notag
\end{align}

In particular, one finds that
\begin{equation}
	p_5=0,\qquad p_3=-\frac{1}{p^2}\frac{\dd p_4}{\dd x}.
	\tag{1.26}
\end{equation}
The preceding relations make it possible to separate the contributions
of the $S$ and $P$ waves in the amplitudes.

Writing
\begin{align*}
	p_j^+(w,k)&=(p_j^+)_S+(J\geq 2),\\
	p_j^-(w,k)&=(p_j^-)_P+(J\geq 3),
\end{align*}
\newpage\centerline{\hfill\small -- 9 --}\vskip .5cm
\noindent one obtains the projections
\begin{align}
	(p_1^+)_S={}&\frac12\int_{-1}^{+1}
	\left(p_1^+-2m x\,p_2^++m^2x^2p_3^+
	-\frac{m^2}{p^2}x p_4^+\right)\dd x,\notag\\
	(p_1^-)_P={}&\frac32\int_{-1}^{+1}\Bigl[p_1^-x^2
	+m x p_2^-(1-3x^2)
	+m^2 p_3^-(2x^3-x)\notag\\
	&\hspace{38mm}
	+\frac{m^2}{2p^2}p_4^-(1-3x^2)\Bigr]\dd x,
	\tag{1.27}\\
	(p_2^-)_P={}&\frac34\int_{-1}^{+1}(1-x^2)
	(p_2^- -m x p_3^- )\dd x,\notag\\
	(p_4^-)_P={}&\frac34\int_{-1}^{+1}(1-x^2)p_4^-\dd x.\notag
\end{align}
\noindent 4) \underline{Kinematic singularities}

Before writing any dispersion relation, it is essential to ensure that the amplitudes involved are free of kinematic singularities. In the nucleon--nucleon case, it has been shown\supercite{ref5} that the amplitudes associated with the Fermi invariants satisfy this requirement.

They are defined by
\begin{equation}
	\mathcal M=F_1\,1\otimes1+F_2\,\gamma^\mu\otimes\gamma_\mu
	+F_3\,(\sigma^{\mu\nu}\otimes\sigma_{\mu\nu})
	+F_4\,(\gamma^\mu\gamma_5\otimes\gamma_\mu\gamma_5)
	+F_5\,(\gamma_5\otimes\gamma_5),
	\tag{1.28}
\end{equation}
and are expressed in terms of the amplitudes $P_i(w,t)$ through the relations
\begin{align}
	F_1&=p_1+\frac{\bar t-w}{4m}p_2,\qquad 
	&F_2&=-m p_2-\frac{\bar t-w}{4} p_4,\notag\\
	F_3&=\frac{\bar t}{4m}p_2,\tag{1.29}\qquad 
	&F_4&=-\frac{\bar t}{4m}p_3,\notag\\
	F_5&=\frac{\bar t-w}{4m}p_2+m^2 p_3+p_5.\notag
\end{align}

In general, the $p_i(w,t)$ are all expected to be singular at $t=0$, although their residues $r_i(x)$ are subject to the following constraints\supercite{ref11}:
\begin{align}
	r_1-mxr_2&=0,\notag\\
	-m r_2+m^2xr_3+r_4&=0,\tag{1.30}\notag\\
	-m x r_2+m^2r_3+r_5&=0.\notag
\end{align}
However, it can be shown that these singularities disappear when only natural-parity exchanges are considered. For instance, equations (1.26) imply the following relations among the residues:
\begin{equation}
	\left\{\begin{aligned}
		r_5(x)&=0,\\
		m^2r_3(x)&=\frac{\dd r_4}{\dd x}.
	\end{aligned}\right.
	\tag{1.31}
\end{equation}
Together with the preceding constraints, these conditions imply that
\begin{equation*}r_i(x)=0,\qquad i=1,2,\ldots,5.\end{equation*}
This therefore justifies the choice of perturbative invariants for studying the exchange of two $\pi$ mesons.

\newpage\centerline{\hfill\small -- 10 --}\vspace{0.35cm}
\centerline{\underline{Chapter II}: \hfill CONTRIBUTION OF TWO-PION EXCHANGE TO THE NUCLEAR FORCE}
\vskip .2cm
As recalled in the introduction, a few years ago a Mandelstam representation was proposed\supercite{ref5,ref6} to describe the contribution of two-pion exchange to the nucleon--nucleon amplitude. This provided a new impetus for theoretical studies of nuclear forces\supercite{ref7,ref12,ref13}. However, in the works cited above, as well as in more recent ones, insufficient attention has been paid, in our view, to the asymptotic behavior of the amplitudes required for the convergence of the dispersion relations being considered. In this chapter, we discuss the subtractions that must be introduced assuming Regge behavior, and show that, despite this restriction, the expressions derived by Amati, Leader, and Vitale\supercite{ref6}, relating the double spectral functions arising from two-pion exchange to pion--nucleon amplitudes, remain valid. Finally, the subtraction functions will be expressed in terms of the lowest partial waves of $\pi\pi\to N\bar N$.

\noindent 1) \underline{Analytic properties of the nucleon--nucleon amplitude}

\quad a -- \dashuline{Mandelstam representation}

We make the usual assumption, suggested by perturbation theory, that the only singularities of the amplitudes $p_i^\pm(w,t)$ are particle-exchange poles and the cuts generated from these poles by unitarity, with the discontinuities themselves having analogous analytic properties. In other words, we assume the existence of a Mandelstam representation:
\begin{equation}
	p_i^\pm(w,t)=\begin{pmatrix}0\\
		-\dfrac12\dfrac{g^2}{\mu^2-t}\,\delta_{i5}\end{pmatrix}
	+\frac{1}{\pi}\int\limits_{4\mu^2}^{\infty}\frac{\dd t'}{t'-t}\,\rho_i^\pm(w,t')
	+\frac{1}{\pi^2}\int\limits^{\infty}\frac{\dd t'}{t'-t} 
	\rho_i^\pm(\bar t,t')\quad\tag{2.1}
\end{equation}
with, for example,
\begin{equation}
	\rho_i^\pm(w,t)=\frac{1}{\pi}\int_{\mu^2}^{\infty}
	\frac{y_i^\pm(w',t)}{w'-w}\,\dd w'.
	\tag{2.2}
\end{equation}
\begin{figure}[h]
	\centering
	\includegraphics[width=.5\linewidth]{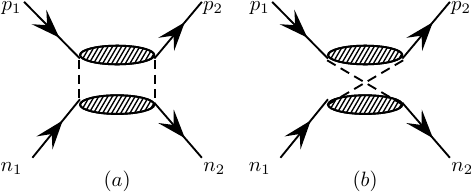}\\[3pt]
	\small Fig. 3: Contribution from two-pion exchange
\end{figure}
The contribution from the crossed diagrams (Fig. 3b) can be obtained from that of the ladder diagrams (a) by the substitution $\eta_1\leftrightarrow-\eta_2$, which transforms $w$ into $\bar t$ and changes the sign of the invariants $P_1$, $P_3$, and $P_5$, while leaving $P_2$ and $P_4$ unchanged. The isospin operators $1$ and $\bm\tau_n\cdot\bm\tau_p$ also behave differently under this exchange. Consequently, for two-pion exchanges, which will be the only contributions considered from now on,
\begin{equation}
	\widetilde \rho_i^\pm(t,t')=\pm(-1)^i \rho_i^\pm(\bar t,t').
	\tag{2.3}
\end{equation}
\newpage
\centerline{\hfill\small -- 11 --}\vskip .2cm 
By defining
\begin{equation}
	a_i^\pm(w',t)=\frac{1}{\pi}\int_{4\mu^2}^{\infty}
	\frac{\gamma_i^\pm(w',t')}{t'-t}\,\dd t',
	\tag{2.4}
\end{equation}
we can also write a dispersion relation at fixed $w$:
\begin{equation}
	\begin{aligned}
		p_i^\pm(w,t)={}\begin{pmatrix} 0 \\ -\dfrac12\dfrac{g^2}{\mu^2-t}\delta_{i5}\end{pmatrix}
		&{}+\frac{1}{\pi}\int\limits_{4\mu^2}^{\infty}\frac{\dd t'}{t'-t}
		\left[\rho_i^\pm(w,t')+(-1)^i \rho_i^\pm(4\mu^2-w-t',t')\right]\notag\\
		&{}+(-1)^i\,\frac{1}{\pi}\int\limits_{4\mu^2}^{\infty}
		\frac{\dd w'}{w'-\bar t}\,a_i^\pm(w',4m^2-w-w').
	\end{aligned}
	\tag{2.5}
\end{equation}

In this equation, the last term is not fundamentally different from the preceding one, since, as we recall, the amplitude will have to be antisymmetrized under the interchange $t\leftrightarrow u$. It nevertheless corresponds to very heavy exchanges ($t>4\mu^2$), and may therefore be neglected since we are only concerned here with the peripheral part of the nucleon--nucleon interaction. In practice, this approximation amounts to setting $\bar t=4\mu^2-w-t'$ in (2.1).

\quad b -- \dashuline{Asymptotic behavior in $t$}

As far as dispersion relations in $t$ are concerned, the issue of possible subtractions is somewhat academic. Indeed, we will never evaluate the integrals appearing in (2.1) or (2.5) themselves, but rather their partial-wave projections or the corresponding potentials.

The Froissart--Gribov projection
\begin{equation}
	f_\ell(w)\propto\int_{4\mu^2}^{\infty}\dd t'\,
	\rho(w,t')\,Q_\ell\!\left(1+\frac{t'}{2k^2}\right)
	\tag{2.6}
\end{equation}
is not affected by peripheral partial waves if one or two subtractions are performed in $t$.

The potential, on the other hand, is given by a relation of the form
\begin{equation}
	V(r)\propto\int_{4\mu^2}^{\infty}\dd t'\,\rho(w,t')\,
	\frac{e^{-r\sqrt{t'}}}{r},
	\tag{2.7}
\end{equation}
and the subtractions would only affect the short-range part, where we do not aim to predict $V(r)$.

In the Regge model, the behavior of $\rho(w,t)$ as $t\to\infty$ is governed by the trajectories in the $w$ channel. The deuteron may belong to such a trajectory $\alpha(w)$, but no resonance with baryon number 2 and spin greater than 1 is known. Moreover, the choice of perturbative invariants is particularly convenient (though legitimate only for natural-parity exchanges), since a pole of spin $\alpha(w)$ implies $P_i(w,t)\sim t^{\alpha(w)-1}$ for all $i$. Thus, for $\alpha(w)<2$, at most one subtraction is required.

\newpage
\centerline{\hfill\small -- 12 --}\vskip .2cm

\quad c --  \dashuline{Asymptotic behavior in $w$.}

In the previous chapter, we derived the decomposition of the $P_i(w,t)$
in terms of angular-momentum states in the $t$ channel. (This can be
obtained by replacing $p_i$ with $P_i$ and $M^{J}$ with
$\operatorname{Im}M^{J}$ in equations (1.25).) The functions $M^J$
are the partial-wave amplitudes for $NN\to NN$ and contain contributions
from mesonic resonances. If we associate these resonances with Regge
poles and assume that they dominate at high energies (including for
$t>4\mu^2$, which takes us beyond the usual range of Regge-pole
phenomenology), then
\begin{equation}
	\begin{aligned}
		w\to\infty:\quad &p_1^\pm(w,t)\sim w^{\alpha(t)},\\
		&p_2^\pm(w,t)\sim w^{\alpha(t)-1},\\
		&p_3^\pm(w,t)\sim w^{\alpha(t)-2}.
	\end{aligned}
	\tag{2.8}
\end{equation}
For the $\oplus$ amplitudes, corresponding to isospin $T=1$ in the
$t$ channel, the leading trajectory contains the $\rho$ and $g$ mesons:
\begin{equation}
	\operatorname{Re}\alpha_\rho(t)\simeq 0.5+0.02\,(t/\mu^2).
	\tag{2.9}
\end{equation}
What matters here is not the linearity of the trajectory, but rather
the fact that $t<m_\rho^2\simeq120\mu^2$ implies
$\operatorname{Re}\alpha_\rho(t)<3$. Thus, if we restrict the
integrals (2.6) and (2.7) to these values of $t$, a subtraction
function must be introduced in $P_1^\pm$, $P_2^\pm$, and $P_4^\pm$,
taking into account the symmetry $w\leftrightarrow\bar t$.

For the $\ominus$ amplitudes (isospin $T=0$ in the crossed channel),
the dominant contribution is initially associated with an approximately
flat Pomeron, and then, as $t$ increases, with the $f$ trajectory,
degenerate with the $\rho$ trajectory. Restricting ourselves to
$t<m_\rho^2\simeq80\mu^2$, we have $\operatorname{Re}\alpha(t)<2$,
so that only one subtraction function is required, in $p_1^-$.
\vskip .1cm 

\noindent 2) \underline{Double spectral functions arising from two-pion exchange}

We first recall the method developed by Amati, Leader, and Vitale
\supercite{ref6}, and then discuss the modifications required to account
for the asymptotic behavior of the amplitudes.

\quad a -- \dashuline{Expression for the double spectral functions}

The generalized unitarity relation, restricting the intermediate states
in the $t$ channel to two pions, reads
\begin{equation}
	-\ii\langle p_2\bar p_1|T-T^\dagger|m_1\bar m_2\rangle
	=\sum_{(2\pi)}\langle p_2\bar p_1|T^\dagger|q_1q_2\rangle
	\langle q_1q_2|T|m_1\bar m_2\rangle.
	\tag{2.10}
\end{equation}
or, equivalently,
\begin{equation}
	\sum_j\left[p_i^+(w,t)+(-1)^i p_i^-(w,t)\right]P_j
	=N\int\dd\Omega_Q\,[\bar A_p^*+\ii\bm\sigma\cdot\bm Q\,B_p^*]
	[ A_m+\ii\bm\sigma\cdot\bm Q\,B_m].
	\tag{2.11}
\end{equation}
where
\begin{equation*}
	N=\frac{1}{128\pi^2}\sqrt{\frac{t-4\mu^2}{t}},
\end{equation*}
and the momenta of the exchanged pions (Fig. 4), in their center-of-mass
frame, are given by
\begin{equation}
	q_1=\left(\frac12\sqrt t,-\bm Q\right),\qquad
	q_2=\left(\frac12\sqrt t,\bm Q\right).
	\tag{2.12}
\end{equation}
\newpage\centerline{\hfill\small -- 13 --}\vskip -.1cm
\noindent
Here, $A$ and $B$ are the usual amplitudes for the reactions
$NN\to2\pi$ and $\pi N\to\pi N$, respectively. The associated
Mandelstam variables, in addition to $t$, are
\begin{equation}
	\text{at vertex}\ (n)\quad \left\{
	\begin{aligned}
		s_n&=-(n_1-q_1)^2=m^2+\mu^2-\frac t2-2\bm N\cdot\bm Q,\notag\\
		\bar s_n&=-(n_1-q_2)^2=m^2+\mu^2-\frac t2+2\bm N\cdot\bm Q,
	\end{aligned}\right.
	\tag{2.13}\end{equation}
\begin{equation}
	\text{at vertex}\ (p)\quad \left\{
	\begin{aligned}	s_p&=-(p_1+q_1)^2=m^2+\mu^2-\frac t2-2\bm P\cdot\bm Q,\notag\\
		\bar s_p&=2m^2+2\mu^2-t.
	\end{aligned}\right.
	\tag{2.14}\end{equation}

\begin{figure}[ht]
	\centering	
	\includegraphics[width=0.4\linewidth]{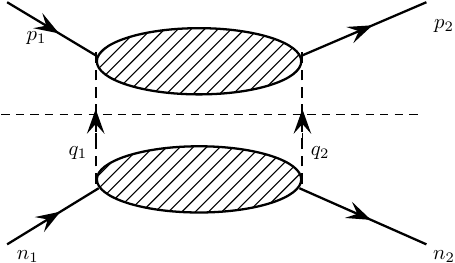}
	%
	\vskip .2cm	
	\centerline{\small Fig. 4: Kinematics of two-pion exchange}
\end{figure}

Equation (2.11) can be decomposed in terms of the invariants as
\begin{equation}
	\int\dd\Omega_Q\,[A_p^*+\ii\bm\gamma\cdot\bm Q B_p^*]
	[A_m+\ii\bm\gamma\cdot\bm Q B_n]
	=(a+2mc+m^2e)P_1-(b+m d)P_2+e P_3+f P_4,
	\tag{2.15}
\end{equation}
where
\begin{align}
	a&=\int A_p^*A_m\,\dd\Omega_Q,\qquad 
	b=\int B_p^*A_m\left[\frac{\bm Q\cdot\bm M}{M^2}+\frac{\bm Q\cdot\bm K}{K^2}\right]\dd\Omega_Q,\quad\notag\\ 
	c&=\int B_p^*A_m\left[\frac{\bm Q\cdot\bm M}{M^2}-\frac{\bm Q\cdot\bm K}{K^2}\right]\dd\Omega_Q,\tag{2.16}\\
	d &=
	\frac{1}{M^{2}K^{2}}
	\int \beta_{p}^{*} B_{m}
	\left[
	\frac{(\mathbf{Q}\cdot\mathbf{M})^{2}}{M^{2}}
	\left(2K^{2}+M^{2}\right)
	+\frac{(\mathbf{Q}\cdot\mathbf{K})^{2}}{K^{2}}
	\left(2M^{2}+K^{2}\right)
	-Q^{2}\left(K^{2}+M^{2}\right)
	\right]\,\mathrm{d}\Omega_{Q}, \notag
	\\[1ex]
	e &=
	\frac{1}{M^{2}K^{2}}
	\int \beta_{p}^{*} B_{m}
	\left[
	\frac{(\mathbf{Q}\cdot\mathbf{M})^{2}}{M^{2}}
	\left(2K^{2}-M^{2}\right)
	+\frac{(\mathbf{Q}\cdot\mathbf{K})^{2}}{K^{2}}
	\left(K^{2}-2M^{2}\right)
	-Q^{2}\left(K^{2}-M^{2}\right)
	\right]\,\mathrm{d}\Omega_{Q},\notag
	\\[1ex]
	f &=
	\int \beta_{p}^{*} B_{m}
	\left[
	Q^{2}
	-\frac{(\mathbf{Q}\cdot\mathbf{M})^{2}}{M^{2}}
	-\frac{(\mathbf{Q}\cdot\mathbf{K})^{2}}{K^{2}}
	\right]\,\mathrm{d}\Omega_{Q}.\notag
\end{align}

\begin{equation}
	\bm M=\frac12(\bm n_1+\bm n_2+\bm p_1+\bm p_2),\qquad
	\bm K=\frac12(\bm n_1+\bm n_2-\bm p_1-\bm p_2).
	\tag{2.17}
\end{equation}

Writing the dispersion relations at fixed $t$, without subtractions
for the time being (we shall return to this issue in the following
sections),
\begin{align}
	A^\pm(\lambda,t)&=\frac{1}{\pi}\int_{(m+\mu)^2}^{\infty}\dd s'
	\,G_A^\pm(\lambda',t)\left[\frac{1}{s'-s}\pm\frac{1}{s'-\bar s}\right],
	\tag{2.18}\\
	B^\pm(s,t)&=\frac{1}{\pi}\int_{(m+\mu)^2}^{\infty}\dd s'
	\,G_B^\pm(s',t)\left[\frac{1}{s'-s}\mp\frac{1}{s'-\bar s}\right],
	\tag{2.19}
\end{align}
\vskip .5cm
\newpage\centerline{\hfill\small -- 14 --}\vskip .1cm
\noindent
including, if desired, the nucleon pole,
$g^2\bigl(1/(m^2-s)\pm1/(m^2-\bar s)\bigr)$, the integrals appearing
in (2.16) take the form
\begin{equation*}
	I=\int\dd\Omega_Q\left(\frac{1}{s'_p-s_p}\pm\frac{1}{s'_m-s_m}\right),
\end{equation*}
sometimes with spin-dependent terms in the numerator. They can be
expressed in the form of a spectral representation in $t$ or $w$,
either using the method proposed by Mandelstam\supercite{ref16} or,
more directly, by applying the Cutkosky rules\supercite{ref17}.
This makes it possible to evaluate all integrals of this type in a
single calculation.

For example,
\begin{equation}
	I=\frac{8\pi}{\sqrt{t-4\mu^2}}\int_{w_0}^{\infty}
	\frac{\dd w'}{w'-t}\,K(s',s'',w',t).
	\tag{2.20}
\end{equation}
Introducing
\begin{equation}
	\left(a^\pm,b^\pm,\ldots f^\pm\right)
	=\iint^{\infty}\dd s'\dd s''
	\,\sigma_{A\atop B}^\pm(s')\sigma_{A\atop B}^\pm(s'')
	\left[\frac{1}{w'-\bar t}\pm\frac{1}{w'-w}\right]\left[\tilde a(w),\tilde b(w),\ldots, \tilde f(w)\right],
	\tag{2.21}
\end{equation}
we obtain
\begin{equation}
\begin{aligned}
	\widetilde{a}(\omega)
	&=
	\frac{16\pi^{3}}{\sqrt{t-4\mu^{2}}}\,
	K(s',s'',w,t),
	\nonumber\\
	\widetilde{b}(w)
	&=
	\frac{8\pi^{3}}{\sqrt{t-4\mu^{2}}}\,
	K\,(Z-Y),
	\nonumber\\
	\widetilde{c}(w)
	&=
	\frac{8\pi^{3}}{\sqrt{t-4\mu^{2}}}\,
	K\,(Z+Y),
	\nonumber\\
	\widetilde{d}(w)
	&=
	\frac{4\pi^{3}}{\sqrt{t-4\mu^{2}}}\,
	K\left[
	Z^{2}+Y^{2}
	-\frac{w+\bar{t}}{(w\bar{t}\,\kappa)^{2}}
	\right],
	\nonumber\\
	\widetilde{e}(w)
	&=
	\frac{4\pi^{3}}{\sqrt{t-4\mu^{2}}}\,
	\mathcal{K}\left[
	Z^{2}-Y^{2}
	-\frac{w-\bar{t}}{(w\bar{t}\,\kappa)^2}
	\right],
	\nonumber\\
	\widetilde{f}(w)
	&=
	-\frac{4\pi^{3}}{\sqrt{t-4\mu^{2}}}\,
	\frac{1}{\omega\bar{t}\,\kappa}.
\end{aligned}	\tag{2.22}\end{equation}
where
\begin{equation}
	\begin{aligned}
		&Y = \frac{s'-s''}{w},\qquad 	Z = \frac{s'+s''+t-2(\mu^{2}+w^{2})}{\bar t}\\[2pt]
		&	K(s',s'',w,t)=
		\left[
		-w(\bar{t}Z)^{2}
		-\bar{t}\,w\left(w Y^{2}+t-4\mu^{2}\right)
		\right]^{-1/2},\\[2pt]
		&	\bar{t} = 4m^{2}-w-t.
	\end{aligned}\tag{2.23}
\end{equation}	
	The integrals in (2.21) are restricted to the region where
	$K^{2}$ is positive.
		
\newpage\centerline{\hfill\small -- 15 --}\vskip .2cm
		Using (2.21), (2.15), and (2.11), we obtain the double spectral
functions defined in~(2.2).	
\begin{align}
	&y_1^\pm(w,t)
	= \pm\frac{1}{8\pi^3\sqrt{t-4\mu^2}}
	\iint K\Bigg\{
	\Sigma_{AA}^\pm
	+m_c(z+y)\Sigma_{BA}^\pm
	+\frac{m_c^2}{4}
	\left[
	\frac{\bar t-w}{(w\bar t\,K)^2}
	+z^2-y^2
	\right]\Sigma_{BB}^\pm
	\Bigg\}\,\mathrm{d}s'\,\mathrm{d}s'',
	\notag\\[1ex]
	&y_2^\pm(w,t)
	= \pm\frac{1}{16\pi^3\sqrt{t-4\mu^2}}
	\iint K\left\{
	\Sigma_{BA}^\pm(z-y)
	+\frac{m_c}{2}
	\left[
	z^2+y^2
	-\frac{w+\bar t}{(w\bar t\,\kappa)^2}
	\right]\Sigma_{BB}^\pm
	\right\}\,\mathrm{d}s'\,\mathrm{d}s'',
	\notag\\[1ex]
	&y_3^\pm(w,t)
	= \pm\frac{1}{32\pi^3\sqrt{t-4\mu^2}}
	\iint K
	\left[
	\frac{\bar t-w}{(w\bar t\,\kappa)^2}
	+z^2-y^2
	\right]\Sigma_{BB}^\pm\,
	\mathrm{d}s'\,\mathrm{d}s'',\tag{2.24}
	\\[1ex]
	&y_4^\pm(w,t)
	= \pm\frac{1}{32\pi^3\sqrt{t-4\mu^2}}
	\iint K
	\left(-\frac{1}{w\bar t\,K^2}\right)
	\Sigma_{BB}^\pm\,
	\mathrm{d}s'\,\mathrm{d}s''.\notag
\end{align}
where
\begin{align}
	\Sigma_{AA}^\pm
	&= \sigma_A^{\pm *}(s',t)\,\sigma_A^\pm(s'',t),
	\nonumber\\
	\Sigma_{BA}^\pm
	&= \operatorname{Re}\!\left[
	\sigma_B^{\pm *}(s',t)\,\sigma_A^\pm(s'',t)
	\right],
	\tag{2.25}\\
	\Sigma_{BB}^\pm
	&= \sigma_B^{\pm *}(s',t)\,\sigma_B^\pm(s'',t).\nonumber
\end{align}
One can readily verify that these satisfy the relations derived
from the kinematic analysis in Chapter I (Eq.~(1.26)):
\[
\left\{
\begin{aligned}
	y_5^\pm(w,t) &= 0,\\
	y_3^\pm(w,t) &= 2\,\frac{\mathrm{d}y_4^\pm}{\mathrm{d}w}.
\end{aligned}
\right.
\]

\quad b -- \dashuline{Dispersion relations with subtractions}

One way to make equations (2.2) convergent, while taking into
account the asymptotic behavior discussed in Section 1, is to subtract
a suitable number of partial waves in the $t$ channel. Using the
projection formulae (1.27), we obtain
\begin{align}
	\rho_1^+(w,t)+\rho_1^+(\bar t,t)
	&=(\rho_1^+)_S
	+\frac{1}{\pi}\int
	y_1^+(w',t)
	\left[
	\frac{1}{w'-w}+\frac{1}{w'-\bar t}
	+\frac{1}{p^2}Q_0(x')
	\right]\mathrm{d}w'
	\nonumber\\
	&\quad+\frac{1}{\pi}\int
	\frac{Q_1(x')}{p^2}
	\left\{
	-2m\,y_2^+ +m^2x'y_3^+
	-\frac{m^2}{p^2}y_4^+
	\right\}\mathrm{d}w',
	\nonumber\\[1.2ex]
	\rho_{2,4}^+(w,t)-\rho_{2,4}^+(\bar t,t)
	&=\frac{1}{\pi}\int\mathrm{d}w'\,
	y_{2,4}^+(w',t)
	\left[
	\frac{1}{w'-w}-\frac{1}{w'-\bar t}
	\right],
	\tag{2.26}\\[1.2ex]
	\rho_3^+(w,t)+\rho_3^+(\bar t,t)
	&=\frac{1}{\pi}\int\mathrm{d}w'\,
	y_3^+(w',t)
	\left[
	\frac{1}{w'-w}+\frac{1}{w'-\bar t}
	\right].
	\notag
\end{align}

\newpage\centerline{\hfill\small -- 16 --}\vskip .1cm
\begin{align}
	\rho_1^-(w,t)-\rho_1^-(\bar t,t)
	&=(\rho_1^-)_P
	+\frac{1}{\pi}\int y_1^-(w',t)
	\Bigg\{
	\frac{1}{w'-w}-\frac{1}{w'-\bar t}
	\nonumber\\
	&\qquad
	+\frac{3x'}{5p^2}
	\Bigg[
	Q_3(x')
	\left(
	-6m\,y_2^-+4m^2x'y_3^-
	-\frac{m^2}{p^2}y_4^-
	\right)
	\nonumber\\
	&\qquad\qquad
	+Q_1(x')
	\left(
	5y_1^- -4m\,y_2^-+m^2x'y_3^-
	-\frac{2m^2}{p^2}y_2^-
	\right)
	\Bigg]\Bigg\}\,\mathrm{d}w',
	\nonumber\\[1.2ex]
	\rho_2^-(w,t)+\rho_2^-(\bar t,t)
	&=(\rho_2^-)_P
	+\frac{1}{\pi}\int\mathrm{d}w'\,
	\Bigg\{
	y_2^-(w',t)
	\left[
	\frac{1}{w'-w}+\frac{1}{w'-\bar t}
	+\frac{3}{2p^2}\left(Q_0-x'Q_1(x')\right)
	\right]
	\nonumber\\
	&\qquad
	-y_3^-(w',t)\frac{3m}{5p^2}
	\left(Q_1(x')-Q_3(x')\right)
	\Bigg\},
	\nonumber\\[1.2ex]
	\rho_3^-(w,t)-\rho_3^-(\bar t,t)
	&=\frac{1}{\pi}\int\mathrm{d}w'\,
	y_3^-(w',t)
	\left[
	\frac{1}{w'-w}-\frac{1}{w'-\bar t}
	\right],
	\tag{2.27}\\[1.2ex]
	\rho_4^-(w,t)+\rho_4^-(\bar t,t)
	&=(\rho_4^-)_P
	+\frac{1}{\pi}\int\mathrm{d}w'\,
	y_4^-(w',t)
	\left[
	\frac{1}{w'-w}+\frac{1}{w'-\bar t}
	+\frac{3}{2p^2}\left(Q_0-x'Q_1(x')\right)
	\right].
	\nonumber
\end{align}
where
\[
\left(w \atop \bar t\right)=-2p^2(1\pm x)
\qquad\text{and, similarly,}\qquad
w'=-2p^2(1+x').
\tag{2.28}
\]

\quad c -- \dashuline{Consistency}

We have thus taken into account the asymptotic behavior of the
nucleon--nucleon amplitudes. However, in deriving the double spectral
functions $\psi_i^\pm$, we used in equations (2.18) and (2.19)
dispersion relations for the $\pi N$ system that are not consistent
with this behavior. To ensure consistency, we must introduce
subtractions, precisely those required by the Regge poles discussed
in Section I, and therefore write, for $t<80\mu^2$:
\begin{equation}
	\left\{
	\begin{aligned}
		A^+(s,t)
		={}&-\frac{4\pi}{p^2}f_0^+(t)
		+\frac{1}{\pi}\int \mathrm{d}s'\,
		\Bigg\{
		\sigma_A^+(s',t)
		\left[
		\frac{1}{s'-s}
		+\frac{1}{s'-\bar{s}}
		-\frac{1}{pq}Q_0(u')
		\right]
		\\[-1mm]
		&\hspace{42mm}
		+\sigma_B^+(s',t)\frac{m}{p^2}Q_1(u')
		\Bigg\},
		\\[2ex]
		B^+(s,t)
		={}&\frac{1}{\pi}\int \mathrm{d}s'\,
		\sigma_B^+(s',t)
		\left[
		\frac{1}{s'-s}-\frac{1}{s'-\bar{s}}
		\right].
	\end{aligned}
	\right.
	\tag{2.29}
\end{equation}

\newpage\centerline{\hfill\small -- 17 --}\vskip .2cm

\noindent
and for $t\lesssim120\mu^2$:

\begin{equation*}
	\left\{
	\begin{aligned}
		\widetilde A^-(s,t)
		={}&
		12\pi\frac{q}{p}
		\left[
		\frac{m}{\sqrt{2}}p_0^1(t)-p_+^1(t)
		\right]\cos\psi
		\\
		&+\frac{1}{\pi}\int\mathrm{d}s'\,
		\sigma_A^-(s',t)
		\left[
		\frac{1}{s'-s}
		-\frac{1}{s'-\bar{s}}
		-\frac{3\cos\psi}{pq}Q_1(u')
		\right]
		\\
		&+\frac{1}{\pi}\int\mathrm{d}s'\,
		\frac{m}{p^2}\sigma_B^-(s',t)
		(3\cos\psi)Q_2(u'),\hspace*{5cm} (2.30)
		\\[2ex]
		\widetilde B^-(s,t)
		={}&
		\frac{12\pi}{\sqrt{2}}p_-^1(t)
		+\frac{1}{\pi}\int\mathrm{d}s'\,
		\sigma_B^-(s',t)
		\left[
		\frac{1}{s'-s}
		+\frac{1}{s'-\bar{s}
		}-\frac{1}{pq}\bigl(Q_0(u')-Q_2(u')\bigr)
		\right].
	\end{aligned}
	\right.
\end{equation*}

\medskip
\noindent
where we have used the standard kinematic notation for the $\pi N$
amplitudes in the $t$ channel\supercite{ref18,ref19} and defined
\[
u'=\frac{s'+p^2+q^2}{2pq}.
\]

Substituting these expressions into the angular integrals (2.11)
or (2.16), we must bear in mind that, owing to angular-momentum
conservation, many terms do not interfere. More importantly, it turns
out that the convergence of all the integrals requires the subtraction
of a certain number of partial waves, and that it is equivalent to
perform these subtractions either at the level of the $\pi N$ amplitudes
or at the final stage, on the $N N$ amplitudes. The expressions
derived in the preceding sections are therefore correct.

\quad d-- \dashuline{Contribution of the S and P waves of
	${\pi\pi\longrightarrow N\overline N}$}

By writing the unitarity relation in the $t$ channel for each partial
wave, it is straightforward to express the subtraction functions in terms
of the S and P waves of $\pi\pi\longrightarrow N\overline N$,
$f_{\pm}^{J}(t)$.

One finds\supercite{ref20}
\begin{equation}\begin{aligned}
		(\rho_1^+)_{S}
		&= \frac{\pi q}{\sqrt{t}\,p^4}
		\left|f_+^0\right|^2,
		\\[1ex]
		(\rho_1^-)_{P}
		&= -\frac{3}{2}\frac{\pi q^3}{\sqrt{t}\,p^4}
		\left|f_+^1-\frac{m_N}{\sqrt{2}}f_-^1\right|^2,
		\\[1ex]
		(\rho_2^-)_{P}
		&= \frac{3\pi q^3}{\sqrt{2t}\,p^2}
		\operatorname{Re}\left\{
		f_-^{1*}\left(f_+^1-\frac{m_N}{\sqrt{2}}f_-^1\right)
		\right\},
		\\[1ex]
		(\rho_4^-)_{P}
		&= -\frac{3\pi q^3}{2\sqrt{t}}
		\left|f_-^1\right|^2 .
	\end{aligned}
	\tag{2.31}\end{equation}

\quad e-- \dashuline{An alternative form of the dispersion relations}

Alternatively, one may write convergent dispersion relations by
subtracting the first terms of the Taylor expansion about a point
$w_0$. Here, the choice
\[
w=\bar{t}=w_0=-2p^2
\]
is the most convenient, given the symmetries.

\newpage\centerline{\hfill\small -- 18 --}\vskip .2cm

On the other hand, the fourth-order diagrams (i.e., the terms in
which only the nucleon pole is retained in $\pi N$) satisfy an
unsubtracted Mandelstam representation. They can therefore be
separated from the other contributions, which also makes the
calculation of the nucleon--nucleon potential easier\supercite{ref14,ref21,ref22}.

Let us write the absorptive part in $t$ of the ladder diagram:
\begin{equation}
	\left(3-2\,\vec{\tau}_m\cdot\vec{\tau}_p\right)
	\sum_{i=1}^{4}\rho_i^B( w,t)\,P_i .
	\tag{2.32}
\end{equation}
where
\begin{equation}
	\rho_i^B( w,t)
	= \frac{1}{\pi}\int
	\frac{y_i^B( w, w')}{ w'- w}\, \dd w' .
	\tag{2.33}
\end{equation}
(This integral can be evaluated analytically.)

The corresponding expression for the crossed diagram is
\begin{equation}
	\left(3+2\,\vec{\tau}_m\cdot\vec{\tau}_p\right)
	\sum_{i=1}^{5}(-1)^{i+1}\rho_i^B(\bar t,t)\,P_i .
	\tag{2.34}
\end{equation}
By defining
\begin{equation}
	y_i^B( w,t)=y_i( w,t)-y_i^0( w,t)
	\tag{2.35}
\end{equation}
we can write\supercite{ref20}
\begin{equation}
	\begin{aligned}
		\rho_1^+( w,t)+\rho_1^+(\bar t,t)
		={}& S_1^+(t)
		+\frac{( w- w_0)^2}{\pi}
		\int
		\frac{y_1'( w',t)}{( w'- w_0)^2}
		\left[
		\frac{1}{ w'- w}
		+\frac{1}{ w'-\bar t}
		\right] \dd  w'
		\nonumber\\
		&+\rho_1^B( w,t)+\rho_1^B(\bar t,t),
		\\[1.2ex]
		\rho_{2,4}^+( w,t)-\rho_{2,4}^+(\bar t,t)
		={}&\frac{ w- w_0}{\pi}
		\int
		\frac{y'_{2,4}( w',t)}
		{ w'- w_0}
		\left[
		\frac{1}{ w'- w}
		+\frac{1}{ w'+\bar t}
		\right] \dd  w'
		\nonumber\\
		&+\rho_{2,4}^B( w,t)-\rho_{2,4}^B(\bar t,t),
		\\[1.2ex]
		\rho_3^+( w,t)+\rho_3^+(\bar t,t)
		={}&\frac{1}{\pi}\int
		y'_3( w',t)
		\left[
		\frac{1}{ w'- w}
		+\frac{1}{ w'-\bar t}
		\right] \dd  w'
		\nonumber\\
		&+\rho_3^B( w,t)+\rho_3^B(\bar t,t).
	\end{aligned}
	\tag{2.36}
\end{equation}
\begin{align}
		\rho_1^-( w,t)-\rho_1^-(\bar t,t)
		={}&S_1^-(t)( w- w_0)
		+\frac{( w- w_0)^3}{\pi}
		\int \dd w'\,
		\frac{y'_1( w',t)}{( w'- w_0)^3}
		\left[
		\frac{1}{ w'- w}
		+\frac{1}{ w'-\bar t}
		\right]
		\notag\\
		&+\rho_1^B(\bar t,t)-\rho_1^B( w,t),
		\notag\\[1ex]
		\rho_{2,4}^-( w,t)+\rho_{2,4}^-(\bar t,t)
		={}&S_{2,4}^-(t)
		+\frac{( w- w_0)^2}{\pi}
		\int  \dd  w'\,
		\frac{y_{2,4}^-( w',t)}{( w'- w_0)^2}
		\left[
		\frac{1}{ w'- w}
		+\frac{1}{ w'-\bar t}
		\right]
		\notag\\
		&+\rho_{2,4}^B(\bar t,t)-\rho_{2,4}^B( w,t),
	\notag	\\[1ex]
		\rho_3^-( w,t)-\rho_3^-(\bar t,t)
		={}&\frac{ w- w_0}{\pi}
		\int_{4m_\pi^2}^{\infty}\dd w'\,
		\frac{y'_3( w',t)}{ w'- w_0}
		\left[
		\frac{1}{ w'- w}
		+\frac{1}{ w'-\bar t}
		\right]
			\tag{2.37}\\
		&+\rho_3^B(w,t)-\rho_3^B(\bar t,t).\notag
	\end{align}

\newpage\centerline{\hfill\small -- 19 --}\vskip .2cm

The new subtraction functions $S_1^+(t)$ and $S_1^-(t)$
are readily obtained by comparing equations (2.26), (2.27), and
(2.31), with $w=\bar t=w_0$.
One finds$^{21}$
\begin{align}
	S_1^+(t)
	={}&
	\frac{4\pi q}{\sqrt{t}\,w_0^2}
	\left(\left|f_+^0\right|^2-\left|f_{B+}^0\right|^2\right)\notag\\
	&\quad{}-\frac{2}{\pi w_0}
	\int dw'\,Q_1(w')
	\left[
	\frac{y_1^{1+}(w',t)}{w'}
	-2m_Ny_2^{1-}(w',t)
	+m_N^2y_3^{1+}
	+\frac{2m_N^2}{w_0}y_4
	\right],
	\notag	\\[1ex]
	S_1^-(t)
	={}&
	-\frac{6\pi q^3}{\sqrt{t}\,w_0^2}
	\left[
	\left|f_+^1-\frac{m_N}{\sqrt{2}}f_-^1\right|^2
	-\left|f_{B+}^1-\frac{m_N}{\sqrt{2}}f_{B-}^1\right|^2
	\right] \tag{2.38}
	\\
	&+\frac{6}{5\pi w_0}
	\int_{4m_\pi^2}^{\infty}dw'\,
	\Bigl\{
	2Q_3(w')
	\left[
	-\frac{y_4^{1-}(w',t)}{w'^2}
	+\frac{3m_Ny_2^{1-}}{w'^2}
	\right]
	\notag\\
	&\hspace{3.4cm}
	+Q_1(w')
	\left[
	\frac{3y_4^{1-}}{w'^2}
	+\frac{4m_Ny_2^{1-}}{w'^2}
	-m_Ny_3^{1-}
	-\frac{4m_N^2y_4^{1-}}{w_0w'^2}
	\right]
	\Bigr\},
	\notag\\[1ex]
	S_2^-(t)
	={}&
	\frac{3\sqrt{2}\,\pi q^3}{\sqrt{t}\,w_0}
	\operatorname{Re}\left[
	f_+^{1*}\left(f_+^1+\frac{m_N}{\sqrt{2}}f_-^1\right)
	-f_{B+}^{1*}\left(f_{B+}^1+\frac{m_N}{\sqrt{2}}f_{B-}^1\right)
	\right]
	\notag\\
	&+\frac{6}{5\pi w_0}
	\int dw'\,
	\left[Q_3(w')-Q_1(w')\right]
	\left[
	\frac{y_2^{1-}(w',t)}{w'}
	-m_Ny_3^{1-}(w',t)
	\right],
	\notag\\[1ex]
	S_4^-(t)
	={}&
	-\frac{3\pi q^3}{2\sqrt{t}}
	\left[\left|f_-^1\right|^2-\left|f_{B-}^1\right|^2\right]
	+\frac{6}{5\pi w_0}
	\int dw'\,
	\left[Q_3(w')-Q_1(w')\right]
	\frac{y_4^{1-}(w',t)}{w'} .\notag
\end{align}
where $f_{B\pm}^{J}(t)$ represents the contribution of nucleon exchange
to the S- and P-wave $\pi\pi\longrightarrow N\overline{N}$ amplitudes.

The preceding equations establish the connection between the medium-range
nuclear forces and the $\pi N$ and $\pi\pi\longrightarrow N\overline{N}$
interactions.

The amplitudes $\sigma_{A,B}^{+}(s')$, $\sigma_{A,B}^{+}(s'')$
act as weight functions for diagrams in which isobars of masses
$\sqrt{s'}$ and $\sqrt{s''}$ are produced in intermediate states
(Fig.~5(a)). In the crossed channel, this role is played by the partial
waves
$
M^J(t)\propto |f_{\pm}^{J}(t)|^2
\ (J=0,1),
$
which describe a continuous exchange of meson systems with total mass
$\sqrt{t}$ and angular momentum $J$ (Fig.~5(b)).

\begin{center}
	\includegraphics[width=0.5\linewidth]{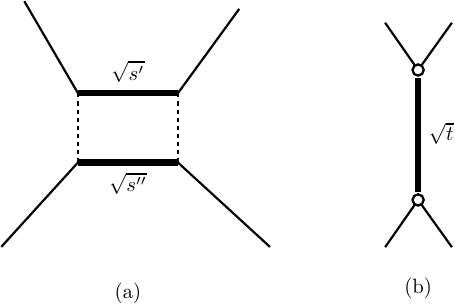}
	\\[4pt]
	\begin{small}	\underline{Fig. 5} Two aspects of two-pion exchange\\
		(a) Nucleon isobars in intermediate states\\
		(b) Pion correlations in the crossed channel
	\end{small}
\end{center}
\newpage\centerline{\hfill\small -- 20 --} \vskip .25cm

The difference becomes clear when compared with O.B.E. or
O.B.E. + T.P.E. models, which treat these quantities as delta functions
centered at the resonance masses. This approximation is not justified
when the resonances are broad, particularly in the case of the
S-wave pion-pion interaction. Moreover, rather than treating the
coupling constants as free parameters, we have attempted to determine
the functions $\sigma_{A,B}^{\pm}$ and $f_{\pm}^{J}(t)$ from
experimental results on meson scattering. This is a rather delicate task,
since these amplitudes must be evaluated outside their physical region.
This part of the work is presented in the following chapter.
\newpage\centerline{\hfill \small -- 21 --} \vskip .2cm
\noindent\underline{\textbf{Chapter III}} -- EVALUATION OF THE SPECTRAL FUNCTIONS
FROM \\\hspace*{2.9cm} EXPERIMENTAL $\pi N$ AND $\pi\pi$ DATA

This chapter is devoted to determining the absorptive parts of the
pion-nucleon amplitudes and the S- and P-wave
$\pi\pi\longrightarrow N\overline{N}$ amplitudes along the cut
$t>4\mu_\pi^2$, using the direct or indirect experimental information
available on these reactions. This will allow us, as discussed above,
to evaluate the spectral functions $P_i^\pm(w,t)$ of the
nucleon-nucleon amplitude.

The problem is not straightforward: these amplitudes are only
imperfectly known in their physical region, and their rigorous
analytic properties are not sufficient to define an unambiguous
extrapolation procedure. We will therefore have to rely on physical
arguments to select the most plausible solutions.

\noindent1)\  \underline{Evaluation of the $\pi N$ absorptive parts in the region
	$t>4\mu^2$}

\quad a-- \dashuline{Extrapolation from the physical region}

The calculation of the double spectral contribution (the terms
involving $y_i'(w,t)$ in Eqs.~(2.36) and (2.37)) requires knowledge
of the absorptive parts $\sigma_{A,B}^{\pm}(s',t)$ of the
pion-nucleon amplitudes $A^\pm$ and $B^\pm$ over the domain covered
by the integrals in Eq.~(2.24), namely, the region where the
Mandelstam function $K^{-2}(s',s'',w',t)$ is positive.

The first variable always lies in the physical region
$(s',s''>(m+\mu)^2)$ and, for small values of $w'$, the integration
over $s'$ or $s''$ is cut off before reaching the upper energy limit
of the $\pi N$ phase-shift analyses. For large $w'$, the
$\sigma_{A,B}^{\pm}(s',t)$ are, in principle, also required at high
energies. However, the contributions from the resonance region remain
dominant. Furthermore, precisely because of the analytic subtractions
performed in Chapter~2, large values of $w'$ contribute little to the
low-energy nucleon-nucleon spectral functions $P_i^\pm(w,t$).
The integrals are of the form
\begin{equation*}
	\rho_i^\pm(w,t)
	= \frac{1}{\pi}\int
	y_i'(w',t)\,dw'\,
	\left[
	\frac{1}{w'-w}
	-\frac{1}{w'-w_0}
	\right].
\end{equation*}

For these two reasons, the energy range covered by the $\pi N$
phase-shift analyses is sufficient for our purposes, and there is no
need to introduce a model for the $\sigma_{A,B}^{\pm}(s,t)$ in the
Regge region and beyond.

Conversely, in our calculation, the momentum transfer takes
nonphysical values ($t>\mu_\pi^2$). The analyticity domain of the
functions $\sigma_{A,B}^{\pm}(s,t)$ is determined by the well-known
Mandelstam boundary, which is reproduced in Fig.~6.
\begin{center}
	\includegraphics[width=0.55\linewidth]{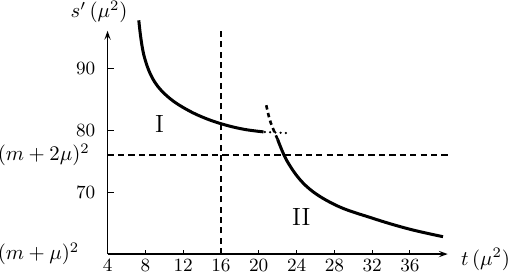}
	\vskip .25cm
	{\small	\underline{Fig. 6}: Analyticity domain of the functions. See Ref.~18.}
\end{center}
\newpage\centerline{\hfill\small -- 22 --} \vskip .25cm

This domain is quite large for small values of $s'$, which, as we
have seen, play a dominant role. Moreover, several very plausible
physical arguments suggest that the effective analyticity domain is
much wider. Indeed, if pion production below the threshold for
$\pi N\longrightarrow \pi\pi N$ is neglected, corresponding to
$s\simeq 150\mu_\pi^2$, the support of the double spectral function
associated with two-pion exchange (curve I) is pushed far away.
In this case, only four-pion exchange determines the analyticity
domain (curve II).


This boundary occurs at $t=16\mu^2$, but, owing to the four-pion
phase-space factors, the effective boundary can be expected to set in
only at larger values of $t$.

The phase-shift analyses provide the functions
$\sigma_{A,B}^{\pm}$ in the physical region as polynomials of degree
4 in $t$. Although more sophisticated and, in principle, more accurate
methods could be considered~\supercite{ref23}, our extrapolation
consists in using these polynomials for $t>4\mu^2$. Of course, it is
impossible to rigorously justify the validity of this procedure.
Nevertheless, its reasonableness can be assessed by examining the
polynomial coefficients. Tables I and II give the values of the
functions $C_n(s)$ in the expansion
\begin{equation*}
	\sigma_{A,B}^{\pm}
	= \sum_{n=0}^{4} C_n(s)
	\left(\frac{t}{s_0}\right)^n .
\end{equation*}

For two energy values, $s=132\mu^2$ and
$s=194\mu^2$, we find that, for $t<80\mu^2$,
the higher powers of $t$ contribute relatively little. This provides
an \textit{a posteriori} argument in support of this extrapolation
procedure.

{\small\begin{center}
			
			\textsc{Table I} -- Coefficients of the expansion
			$\displaystyle{
				\sigma_{A,B}^{\pm}
				= \sum_{n=0}^{4} C_n(s)
				\left(\frac{t}{s_0}\right)^n
			}$
			for $s=132\mu^2$.\\
			\vskip .2cm
			\renewcommand{\arraystretch}{1.12}
			\begin{tabular}{c l r r r r r}
				\toprule
				& & $C_0$ & $C_1$ & $C_2$ & $C_3$ & $C_4$ \\
				\midrule
				$\sigma_A^+$ & CERN    & $-12.63$ & $-32.53$ & $-19.67$ & $2.251$  & $0.2117$ \\
				& Glasgow & $-8.620$ & $-19.566$ & $-12.62$ & $0.3817$ & $0.3961$ \\
				\addlinespace[2pt]
				$\sigma_B^+$ & CERN    & $3.379$ & $9.329$ & $7.106$ & $0.1647$ & $0.0116$ \\
				& Glasgow & $2.782$ & $7.716$ & $7.712$ & $1.781$  & $0.0217$ \\
				\addlinespace[2pt]
				$\sigma_A^-$ & CERN    & $-8.572$ & $-27.74$ & $-25.50$ & $-1.254$ & $0.0249$ \\
				& Glasgow & $-7.756$ & $-29.39$ & $-37.10$ & $-9.966$ & $0.0322$ \\
				\addlinespace[2pt]
				$\sigma_B^-$ & CERN    & $1.828$ & $6.865$ & $6.539$ & $0.0202$ & $0.0014$ \\
				& Glasgow & $1.718$ & $7.050$ & $8.225$ & $1.306$  & $0.0018$ \\
				\bottomrule
			\end{tabular}
	\end{center}}

	\newpage\centerline{\hfill\small -- 23 --} \vskip .25cm
	
	{\small
		\begin{center}
			\textsc{Table II} -- Coefficients of the expansion
			$\displaystyle{
				\sigma_{A,B}^{\pm}
				= \sum_{n=0}^{4} C_n(s)\left(\frac{t}{s_0}\right)^n,
				\ \text{for } s=194\mu^2.}$\\
			
			\vskip .2cm
			\renewcommand{\arraystretch}{1.12}
			\begin{tabular}{c l r r r r r}
				\toprule
				& & $C_0$ & 
				$C_1$ & $C_2$ & $C_3$ & $C_4$ \\
				\midrule
				$\sigma_A^+$ & CERN     & $17.86$  & $54.29$  & $40.82$  & $9.711$  & $0.402$ \\
				& Glasgow  & $21.10$  & $55.82$  & $40.61$  & $8.494$  & $0.058$ \\
				\addlinespace[2pt]
				$\sigma_B^+$ & CERN     & $0.2176$ & $1.1387$ & $2.841$  & $1.2194$ & $0.0194$ \\
				& Glasgow  & $-0.0112$& $1.4486$ & $3.340$  & $1.432$  & $0.0029$ \\
				\addlinespace[2pt]
				$\sigma_A^-$ & CERN     & $-8.937$ & $-31.06$ & $-26.81$ & $-5.954$ & $0.325$ \\
				& Glasgow  & $-23.57$ & $-69.65$ & $-53.51$ & $-11.73$ & $0.0576$ \\
				\addlinespace[2pt]
				$\sigma_B^-$ & CERN     & $0.6961$ & $2.059$  & $1.436$  & $0.3067$ & $0.0158$ \\
				& Glasgow  & $2.023$  & $5.206$  & $3.133$  & $0.4534$ & $0.0028$ \\
				\bottomrule
			\end{tabular}
	\end{center}}

	\quad b-- \dashuline{The $\pi N$ amplitudes in the physical region}
	
	Several phase-shift analyses of pion-nucleon scattering are available
	for determining the amplitudes $\sigma_{A,B}^{\pm}$ in the physical
	region. They are qualitatively similar, but sometimes exhibit
	appreciable quantitative differences. These differences can be
	illustrated by comparing the CERN analysis~\supercite{ref24}
	(denoted by C) with Glasgow A~\supercite{ref25} (denoted by G).
	We have therefore chosen to carry out our calculations using both
	solutions. Indeed, Tables I and II show that the differences between
	them are quite noticeable, both for $s=132\mu^2$ and for
	$s=194\mu^2$.
	
	Using these two phase-shift analyses, we calculated the double
	spectral functions $y_i(w',t)$ in the region
	$(w'<1000\mu^2,\ t<80\mu^2)$. For moderate values of $w'$ and
	$t$, the results are very similar, but the differences become
	significant as $w'$ and $t$ increase. To illustrate these ambiguities,
	we have plotted the double spectral functions obtained from the
	solutions (C) and (G), retaining, for each isospin state, the
	dominant helicity amplitude.
	
	For isospin $\oplus$, the helicity-non-flip amplitude is important.
	Figure~7 shows, at $t=30\mu^2$, the combination
	\[
	y_{-+}^{+}
	= y_1^{+} - 2m y_2^{+}
	+ m^2 y_3^{+} + D y_4^{+},
	\qquad
	\left(D=\frac{w-2m^2}{2m^2}\right).
	\]
	%
	\noindent For isospin $\ominus$, we have chosen the double
	helicity-flip amplitude, associated with the spectral function
	\[
	y_{+---+}^{-}
	= y_1^{-}-2m y_2^{-}+m^2y_3^{-}
	+\frac{4m^2}{w}\,y_4^{-},
	\]
	which is shown in Fig.~8 for $t=30\mu^2$.

	\newpage\centerline{\hfill\small -- 24 --} \vskip .25cm

	The resulting functions $y_i'(w',t)$ can be used, through
	Eqs.~(2.36) and (2.37), to calculate the contribution of the
	$\pi N$ data to the single spectral functions $\rho_i^\pm(w,t)$,
	which directly determine the nucleon-nucleon phase shifts or
	potential. For $t<50\mu^2$, the two analyses (C) and (G) yield
	essentially the same results, and the differences increase rapidly
	only beyond this range. We can therefore expect that the ambiguity
	in the nuclear forces arising from the interpretation of the
	pion-nucleon data will become significant only at short distances.
	
	
	\begin{center}
		\begin{minipage}[t]{0.46\textwidth}
			\centering
			\includegraphics[width=0.85\linewidth]{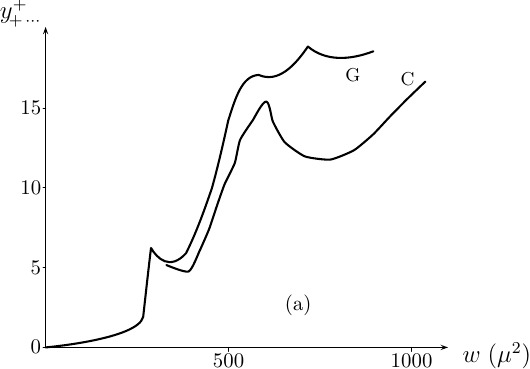}\\ [3pt]
			{\small	\underline{Fig.\ 7}\quad 
				$y_{-+-+}^{+}(w,t)$ for $t=30\mu^2$}
		\end{minipage}
		\hfill
		\begin{minipage}[t]{0.46\textwidth}
			\centering
			\includegraphics[width=0.85\linewidth]{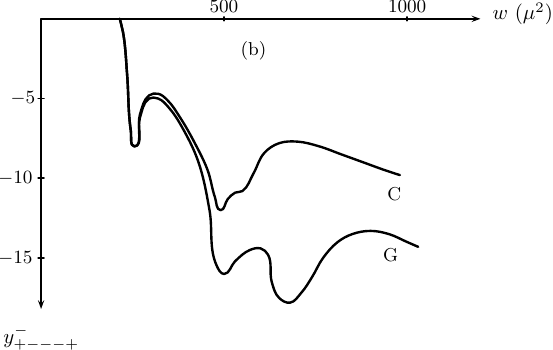}\\	[3pt]
			{\small \underline{Fig.\ 8}\quad 
				$y_{+---+}^{-}(w,t)$ for $t=30\mu^2$}
		\end{minipage}
	\end{center}
	
	\noindent
	2) \underline{Determination of the S- and P-wave
		$\pi\pi\longrightarrow N\bar N$ amplitudes along the
		two-pion cut}
	
	\vskip .2cm
	
	The second contribution of two-pion exchange to the nuclear forces
	is contained in the subtraction functions $S_1^+(t)$ and
	$S_{1,2,4}^-(t)$. Their expressions are given by Eqs.~(2.38),
	in which the integral term is very small and varies very little
	from one model to another (CERN or Glasgow A). The functions
	$S_i^\pm(t)$ therefore depend essentially on the lowest partial
	waves of $\pi\pi\longrightarrow N\bar N$. The amplitudes
	$f_+^0(t)$ and $f_+^1(t)$ thus play an important role in the
	theory of nuclear forces, but they also arise in other problems,
	such as peripheral pion-nucleon scattering~\supercite{ref26} and
	the study of the nucleon form factor~\supercite{ref27}. This
	explains why many physicists have worked on determining them along
	the cut $t\geq 4\mu^2$. We shall review their methods and results.
	
	\quad a-- \dashuline{Some properties of the partial waves
		$\pi\pi\longrightarrow N\bar N$}
	
	The analytic structure of the functions $f_\pm^J(t)$ is
	summarized in Fig.~9. They have a right-hand cut starting at
	$t=4\mu^2$, and a left-hand cut for $t<0$, which extends,
	for the Born term only, up to
	$a=4\mu^2\left(1-{\mu^2}/{(4m^2)}\right)$.
	
	\begin{center}
		\includegraphics[width=.7\linewidth]{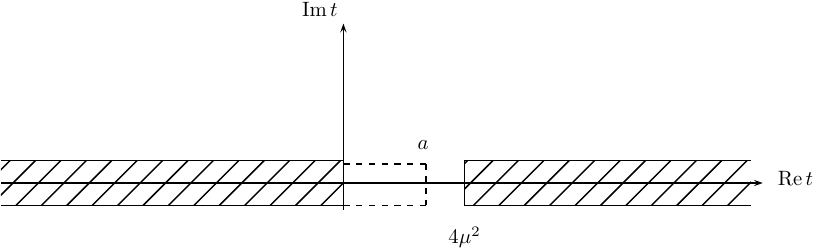}\\ [3pt]
		{\small\underline{Fig. 9}: Cuts of the functions $f_\pm^J(t)$}
	\end{center}
	
	\newpage\centerline{\hfill\small -- 25 --} \vskip .25cm
The amplitude $\operatorname{Im} f_+^J(t)$ can be calculated
\supercite{ref18,ref28} down to $t\simeq -26\mu^2$ along the
left-hand cut, using pion-nucleon phase shifts. This limit,
$-26\mu^2$, corresponds to the endpoint of the Lehmann ellipse
in the partial-wave expansion of the $s$-channel, assuming a
Mandelstam representation. In the low-energy region
$(0,270\ \mathrm{MeV})$, where accurate data are lacking, the
pion-nucleon phase shifts must be reconstructed analytically
\supercite{ref29}, or an appropriate model must be used
(e.g. the C.G.L.N. model\supercite{ref30}).

The real part $\operatorname{Re} f_+^J(t)$ can then be evaluated
in the domain $-26\mu^2<t<4\mu^2$ by writing fixed-$t$
dispersion relations for $A^\pm$ and $B^\pm(s,t)$, which also
include high-energy contributions that must be taken into account.

At the beginning of the right-hand cut, generalized unitarity
requires the phase of $f_\pm^J(t)$ to coincide, modulo $\pi$,
with the pion-pion phase shift in the corresponding partial wave.
This is, in principle, valid from $t=4\mu^2$ to $t=16\mu^2$,
and probably beyond, since inelasticity in pion-pion scattering
does not really set in until the $K\bar K$ channel opens.

Some authors have evaluated the functions $f_\pm^J(t)$
($J=0$ or $1$) along the two-pion cut ($t>4\mu^2$), using
all or part of the above properties, i.e. by analytic continuation
from the region $t<0$, with or without imposing a constraint on
the phase along the right-hand cut. We shall now examine the
results they obtained.

\quad b-- \dashuline{P wave}

As far as isospin $\circleddash$ is concerned, the P wave,
dominated by the $\rho$ resonance, appears to be well determined.
For the calculation of the peripheral phase shifts, we used two
models\supercite{ref31,ref32} which differ only slightly and
which, as can be seen in Fig.~10, lead to very similar results,
comparable to those obtained by treating $\rho$ exchange as a pole.
In this case, the coupling constants are well known. For the
potential calculation (Chapter V), we adopted a more recent
solution by Nielsen and Oades\supercite{ref33}.

\vskip .25cm
\centerline{\includegraphics[width=0.45\linewidth,height=.4\linewidth]{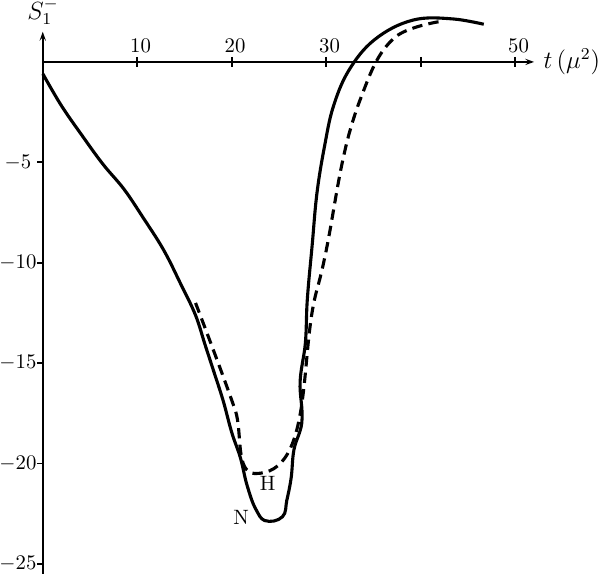}}
\vskip .2cm
\begin{center}
	{\small	\underline{Fig. 10}: The subtraction function $S_1^-(t)$ obtained
		using the amplitudes $f_+^1(t)$ \\ from Ref.~\cite{ref31} (curve H)
		or Ref.~\cite{ref32} (curve N).}
\end{center}
\newpage\centerline{\hfill\small -- 26 --} \vskip .20cm

\quad c-- \dashuline{S wave}\\
\vspace{0.3cm}
\indent\hspace*{2cm}$\bullet$ \emph{General remarks}

It has long been known that the fourth-order diagrams
(Fig.~11) in perturbation theory contribute too much attraction
to the nucleon-nucleon amplitude. A substantial part of this effect
comes from the S wave in the crossed channel, which is the square
of the partial amplitude
$f_{B+}^0(t)$ for $\pi\pi\longrightarrow N\bar N$,
with only nucleon exchange retained (Fig.~12).
\vskip .2cm

\begin{minipage}{.46\linewidth}
	\begin{center}
		\scalebox{1}{\includegraphics[width=0.8\textwidth]{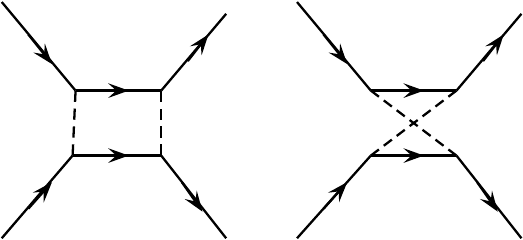}}\\ [4pt]
		{\small \underline{Fig. 11}: Fourth-order diagrams in nucleon-nucleon scattering}		
	\end{center}
\end{minipage}
\begin{minipage}{.46\linewidth}
	\begin{center}
		\scalebox{1}{		\includegraphics[width=0.8\textwidth]{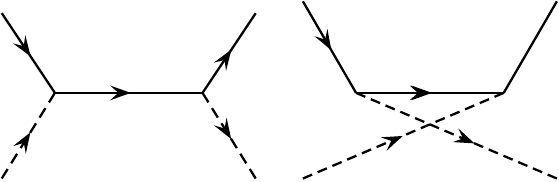}}\\ [8pt]
		{\small \underline{Fig. 12}: Nucleon pole in $\pi\pi\to N\overline N$}		
	\end{center}	
\end{minipage}
\vskip .2cm

Some authors have attempted to reduce the attraction arising from
the fourth-order terms through the mechanism known as
``pair suppression''. This consists in using, in the evaluation
of the diagrams above, an effective propagator\supercite{ref9}
\begin{equation*}
	\frac{1}{2w_p}
	\left[
	\frac{\gamma_4 w_p-i\vec{\gamma}\cdot\vec{p}+m}
	{w_p-p^0}
	+
	\lambda
	\frac{-\gamma_4 w_p-i\vec{\gamma}\cdot\vec{p}+m}
	{w_p+p^0}
	\right]
	\qquad
	\left(w_p=\sqrt{\vec{p}^{\,2}+m^2}\right)
\end{equation*}
where $\lambda\in[0,1]$ is the ``suppression factor''.

Although this procedure is partly motivated by estimates of certain
radiative corrections, it is far from satisfactory. Nevertheless,
it clearly illustrates the difficulties encountered in the study
of nuclear forces, and in particular the need for a more realistic
description of the S-wave $\pi\pi\longrightarrow N\bar N$
amplitude than the simple nucleon pole.

Let us recall that this amplitude $\phi_+^0(k)$ must satisfy
three constraints:

\begin{enumerate}
	\item[(i)] Its phase must be compatible with what is known about
	the pion-pion phase shift $\delta_0^0(k)$.
	\item[(ii)] Through its left-hand cut, $f_+^0(t)$ depends on the
	$\pi N$ amplitude and is particularly sensitive to the values
	of the scattering lengths\supercite{ref19}.
	
	\item[(iii)] The subtraction function $S_+^t(t)$
	\[
	S_+^t(t)\propto
	\left(|f_{++}^0(t)|^2-|f_{+-}^0(t)|^2\right)
	\]
	plays an essential role in the medium-range nuclear forces.
	Their study, as we shall see in Chapter IV, appears to require
	\[
	S_+^t(t)<0
	\qquad
	(\text{or } |f_{++}^0|<|f_{+-}^0|)
	\]
	for $t\simeq 4-30\mu^2$. This is the expression of
	``pair suppression'', recast in terms of dispersion relations.
\end{enumerate}
\newpage\centerline{\hfill\small -- 27 --} \vskip .20cm

We shall now review the main models proposed for $f_+^0(t)$
in light of these three criteria.

%
\noindent
\hspace*{2cm}\textbullet\ \emph{Review of the main models}

The amplitude $f_+^0(t)$ proposed by Nielsen, Petersen and
Pietarinen\supercite{ref35} has a small modulus along the
right-hand cut, corresponding to strong damping, treated in a
relativistic framework. In this model, $f_+^0(t)$ is indeed
calculated using analyticity, crossing symmetry, and pion-nucleon
scattering data, which make it possible to reconstruct the beginning
of the left-hand cut. This cut is then explicitly removed from the
amplitude, and the extrapolation is of type I $\longrightarrow$ B
(interior $\longrightarrow$ boundary; see Fig.~13).

\begin{center}
	\includegraphics[width=.7\linewidth]{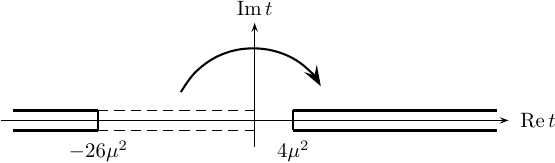}
	\vskip .2cm
	{\small \underline{Fig. 13}: Extrapolation $(t<0) \to (t>4\mu^2)$ of type I $\to$ B}	
\end{center}

The method allows both $\operatorname{Re} f_+^0(t)$ and
$\operatorname{Im} f_+^0(t)$ to be calculated. Unfortunately,
the phase obtained by these authors is very small and incompatible
with the results of the most recent pion-pion phase-shift analyses.
In the following, we shall refer to this model as N-R (non-resonant).

\noindent -- In the model of Nielsen and Oades\supercite{ref32},
which we shall henceforth denote by R (for resonant),
$f_+^0(t)$ is constructed to have the BDI (between-down)
pion-pion phase of Morgan and Shaw\supercite{ref36}. Indeed,
the right-hand cut is explicitly removed by factoring the Omnès
function out of $f_+^0(t)$:

\[
\exp\left(
-\frac{t}{\pi}
\int\limits_{4\mu^2}^{50\mu^2}
\frac{\delta_0^0(t')\,dt'}{t'(t'-t)}
\right)
\]

The left-hand cut is also known down to $-26\mu^2$, and the
analytic continuation of the remaining amplitude is of type
I $\longrightarrow$ I (Fig.~14), which is therefore, in principle,
more ``reliable''.

\begin{center}
	\includegraphics[width=.7\linewidth]{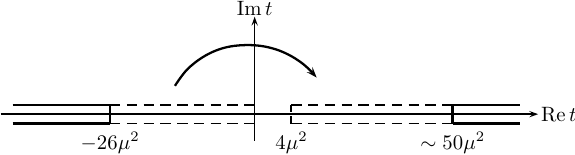}
	\vskip .2cm
	{\small \underline{Fig. 14}: Extrapolation $(t<0) \to (t>4\mu^2)$ of type I $\to$ I}	
\end{center}

\newpage\centerline{\hfill\small -- 28 --} \vskip .20cm

The partial wave $f_+^0(t)$ constructed in this way, consistent
with the $\pi N$ and $\pi\pi$ data, has a large modulus,
greater than that of the Born term, contrary to pair suppression.
This result appears to be confirmed by a study by Epstein and
Mac Kellar\supercite{ref37}, who use a method similar to that of
Nielsen and Oades and the same phase shift $\delta_0^0(t)$.
The large modulus of their S-wave amplitude $f_+^0(t)$ is
responsible for the excessive attraction in their potential
\supercite{ref15}. In a similar study, Bohannon and Signell
\supercite{ref38} obtain values of $|f_+^0(t)|$ that are only
slightly smaller.

\noindent
-- Furuichi, Kanada and Watanabe\supercite{ref39} also reconstructed
$f_+^0(t)$ from the left-hand cut, restricted to the nucleon and
$N^*_{1/2,3/2}$ contributions, and used, on the right-hand cut,
either a UD (Up Down) or a strongly resonant DU pion-pion phase-shift
solution. In the latter case, they obtained values of
$|f_+^0(t)|$ that were much too large.

\medskip

\noindent
-- Brown and Durso\supercite{ref34} proposed a very simple model
in which the amplitude $B$ is reduced to the Born term, while
$A$ is taken at its value at the Adler point. It is already known,
and was one of the successes of the ``soft-pion'' approach, that
these two terms partially cancel in the calculation of the
$\pi N$ scattering lengths. The same favorable cancellation
is found in the amplitude $\operatorname{Re} f_+^0(t)$,
which these authors obtain to be very small. The imaginary part
is then inferred from a non-resonant pion-pion phase shift
(of the UD type), and the subtraction function $S_A^+(t)$
constructed from this model is smaller than that obtained with
(N.R.). Verwest and Riska\supercite{ref40} modified
$\operatorname{Im} f_+^0(t)$ to account for the more recent
pion-pion phase-shift analyses, but this affects
$|f_+^0(t)|$ only in the region $t>20\mu^2$, which is less
important than the beginning of the cut.

\hspace*{2cm}\textbullet \emph{Current situation}

At the time of the latest series of publications
\supercite{ref43} on the application of dispersion relations to
peripheral nucleon-nucleon scattering, physicists faced the following
dilemma concerning the S-wave
$\pi\pi\longrightarrow N\bar{N}$:

(i) adopt the rather questionable (N.R.) model\supercite{ref33},
although it is instructive in terms of the methods employed.
Similarly, although the Brown and Durso model, justified for
$4\mu^2<t\lesssim 10\mu^2$, appears to remain reasonable beyond
this range, this is, in the authors' own words, a case of
``Mehr Glück als Verstand'' (more luck than understanding).

(ii) accept the (R) model, which appears theoretically more sound,
but leads to nuclear forces that are too attractive.

The problem is actually to reconcile, within the framework of S-matrix
theory, some experimental information on low-energy non-strange
hadronic systems, namely $\pi N$, $\pi\pi$, and $NN$. However, an
examination of the models discussed above shows that increasing
$\delta_0^0(k)$ to a value consistent with the latest phenomenological
estimates automatically increases $|f_+^0(k)|$, thereby leading to
too much attraction in the nucleon-nucleon system.

Yet Nielsen and Oades have recently obtained\supercite{ref33}
two solutions (R$_2$ and R$_3$) that have both a small modulus,
comparable to that of (N.R.), and a phase consistent with the most
recent determinations of the pion-pion phase shift.

\newpage\centerline{\hfill\small -- 29 --} \vskip .20cm

This result is not, in fact, particularly surprising. To understand
it, however, it is preferable to abandon the viewpoint adopted in
Refs.~\cite{ref34} and \cite{ref40}, where $\operatorname{Re} f_+^0(t)$
is calculated first and where
$
|f_+^0(t)| =
{\operatorname{Re} f_+^0(t}/
{\cos\delta_0^0(t)}$ 
appears, at first sight, to be an increasing function of the
phase shift $\delta_0^0$. In reality, $\operatorname{Re} f_+^0(t)$
also depends on $\delta_0^0$. This is also true of the model
mentioned in Ref.~\supercite{ref34}, in which the real part is
calculated from soft-pion properties, which, as shown by
Weinberg\supercite{ref44}, are not independent of
$\delta_0^0(t)$.

The connection between the modulus and the phase of $f_+^0(t)$
becomes clearer when a phase relation is written for this amplitude.
Neglecting possible zeros, we have
\begin{equation*}
	\log\left|
	\frac{f_+^0(t)}{f_+^0(t_0)}
	\right|^2
	=
	\frac{2(t-t_0)}{\pi}
	\int_{-\infty}^{a}
	\frac{\delta_L(t')\,\dd t'}
	{(t'-t)(t'-t_0)}
	+
	\frac{2(t-t_0)}{\pi}
	\fint_{4\mu^2}^{\infty}
	\frac{\delta_R(t')\,\dd t'}
	{(t'-t)(t'-t_0)}
\end{equation*}
where
\[
\delta_R(t')=\delta_0^0(t')
\qquad
\text{for}\qquad
4\mu^2\leq t\lesssim 45\mu^2.
\]

In this same range, we see that $|f_+^0(t)|^2$, through the
principal-value integral, is less sensitive to the value of the phase
shift than to how rapidly it varies. With all other factors along
the left-hand cut held fixed, a DU (Down-Up) solution involving a
rather narrow $\epsilon$ resonance is associated with a very large
modulus $|f_+^0(t)|$ near $t\simeq m_\epsilon^2$, whereas a
slowly varying resonant solution may lead to values of
$|f_+^0(t)|$ similar to those obtained with a non-resonant UD
phase shift.

We shall return in the next chapter to the importance of the
$f_+^0(t)$ partial wave for the study of medium-range nuclear
forces. For the time being, we can illustrate the differences
between the models by compiling in tables the predicted or assumed
values of $\operatorname{Arg}(f_+^0)$ and the corresponding values
of $|f_+^0|$.

\centerline{\begin{minipage}{.49\linewidth}
		\centering
		{\small \underline{Table III}: Phase of $f_+^0(t)$ (in degrees)}\\[3pt]
		\begin{tabular}{|c|c|c|c|c|c|}
			\hline\\[-12pt]
			$t$ & N.R. & R$_0$ & D.B. & R$_2$ & R$_3$ \\
			\hline\\[-12pt]
			5  &  9.9 &  5.5 &  7.7 &  6.2 & 16.9 \\
			10  & 44.2 & 25.7 & 25.9 & 26.6 & 42.1 \\
			20  & 43.1 & 52.0 & 42.3 & 59.2 & 64.1 \\
			30  & 29.7 & 69.1 & 45.0 & 79.1 & 76.0 \\
			40  & 20.6 & 86.9 & 43.1 & 96   & 88.5\\
			\hline
		\end{tabular}
	\end{minipage}
	\begin{minipage}{.49\linewidth}
		\centering 
		{\small \underline{Table IV}: Modulus of $f_+^0(t)$}
		\vskip .2cm
		\begin{tabular}{|c|c|c|c|c|c|c|}
			\hline\\[-12pt]
			$t$ & N.R. & R & D.B. & R$_2$ & R$_3$ & Born\\
			\hline\\[-12pt]
			5  & 46.7 & 52.1 & 27.2 & 43.6 & 50.7 & 70.6 \\
			10  & 48.8 & 73.9 & 30.0 & 57.1 & 51.4 & 68.0 \\
			20  & 42.4 & 86.3 & 43.3 & 56.3 & 43.5 & 57.7 \\
			30  & 40.3 & 81.3 & 48.6 & 39.2 & 37.1 & 50.1 \\
			40  & 42.7 & 64.1 & 48.3 & 18.3 & 16.6 & 44.0\\
			\hline
		\end{tabular}
	\end{minipage}\hfill}
\vskip .1cm
\noindent
-- N.R. : Nielsen, Petersen and Pietarinen\supercite{ref35}\quad 
-- R. : Nielsen and Oades\supercite{ref32}\\
\noindent
-- D.B. : Brown and Durso\supercite{ref34}\quad 
-- R$_2$, R$_3$ : Nielsen and Oades\supercite{ref33}\\
\noindent
-- Born : nucleon pole in
$\pi\pi\longrightarrow N\bar{N}$

\vskip .1cm
\centerline{\begin{minipage}{.51\linewidth}
		The large difference between the (N.R.) and (R) models, which we
		have used successively in our study of nucleon-nucleon scattering,
		can also be seen in Fig.~15 in terms of the subtraction function
		$S_\lambda^+(t)$.
		\\[3pt]
		{\small \underline{Fig. 15}: The subtraction function $S_\lambda^+(t)$
			calculated from the $\pi\pi\rightarrow N\bar{N}$ partial wave\\
			of Ref.~\cite{ref35} (N.R.) or Ref.~\cite{ref32} (R.)}
	\end{minipage}\hfill
	\begin{minipage}{.44\linewidth}
		\centering
		\includegraphics[width=1.05\linewidth]{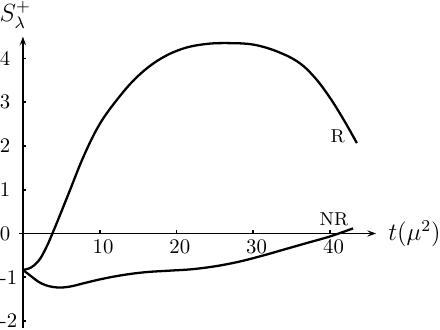}
		\vskip .2cm
		
	\end{minipage}\hfill}
\newpage\centerline{\hfill\small -- 30 --} \vskip .20cm
\centerline{\scalebox{1.1}{\underline{ Chapter IV } PERIPHERAL NUCLEON-NUCLEON PHASE SHIFTS}\hfill}

In the preceding chapters, we have presented the ingredients needed
to calculate the contribution of two-pion exchange to the
nucleon-nucleon scattering amplitude. It is straightforward to add
single-pion exchange and $\omega$-meson exchange, which accounts
for part of the three-pion effects. We have thus, in principle,
evaluated the longest-range forces and can calculate the peripheral
partial waves ($J>2$), which, at least at low energies, are
practically insensitive to the other contributions.

At the beginning of this chapter, we give the kinematic formulae
needed to extract the phase-shift parameters from the dispersion
relations written for the invariant amplitudes. After summarizing
the various ingredients of the calculation, we shall discuss the
results obtained by comparing them with experimental phase-shift
analyses, and examine the importance of the uncertainties
associated with the evaluation of the $\pi N$ and $\pi\pi$
interactions outside their physical regions.

\vskip .1cm
\noindent
{1) \underline{Determination of the phase-shift parameters}}

\quad{a--\dashuline{From invariant amplitudes to ``$\ell-s$'' transitions}}

In Chapter~1, we defined five independent helicity amplitudes
in the direct channel. The corresponding partial waves are given
by the projections

\begin{equation}
	\begin{aligned}
		T_1^J(E)
		&= \frac{1}{2}\int_{-1}^{+1} dz\,
		\varphi_1(E,z)\,d_{00}^J(\theta)
		\\[2mm]
		T_2^J(E)
		&= \frac{1}{2}\int_{-1}^{+1} dz\,
		\varphi_2(E,z)\,d_{00}^J(\theta)
		\\[2mm]
		T_3^J(E)
		&= \frac{1}{2}\int_{-1}^{+1} dz\,
		\varphi_3(E,z)\,d_{11}^J(\theta)
		\\[2mm]
		T_4^J(E)
		&= \frac{1}{2}\int_{-1}^{+1} dz\,
		\varphi_4(E,z)\,d_{-1,1}^J(\theta)
		\\[2mm]
		T_5^J(E)
		&= \frac{1}{2}\int_{-1}^{+1} dz\,
		\varphi_5(E,z)\,d_{1,0}^J(\theta)
	\end{aligned}
	\tag{4.1}
\end{equation}

The matrix elements of the unitary transformation from helicity
states to ``$\ell-s$'' states are\supercite{ref45}

\begin{equation}
	\begin{aligned}
		\langle J,\ell,s|J,\lambda_1,\lambda_2\rangle
		={}&
		\left(\frac{2\ell+1}{2J+1}\right)^{1/2}
		C\left(\frac12,\frac12,s;\lambda_1,-\lambda_2\right)
	\end{aligned}
	\tag{4.2}
\end{equation}
where the $C$ are the usual Clebsch-Gordan coefficients.
The transitions can therefore be evaluated as
\begin{equation*}
	f_J =
	2k\frac{m^2}{4\pi E}
	\langle J,\ell',s|M_{21}|J,\ell,s\rangle
\end{equation*}
\newpage\centerline{\hfill\small -- 31 --} \vskip .20cm
Taking the Pauli exclusion principle into account as well,
we obtain:\\
\noindent
the singlet amplitude

\begin{equation}
	f_{J,1}^{(I)}
	=
	\frac{1-(-)^{I+J}}{2}
	\cdot
	\frac{km^2}{2\pi E}
	\left(T_1^J-T_2^J\right)
	\tag{4.3}
\end{equation}
\noindent
the uncoupled triplet amplitude
\begin{equation}
	f_{J,2}^{(I)}
	=
	\frac{1+(-)^{I+J}}{2}
	\frac{km^2}{2\pi E}
	\left(T_3^J-T_4^J\right)
	\tag{4.4}
\end{equation}
\noindent and the transitions between coupled triplet states\\[3pt]
\noindent $\ell=J-1\longrightarrow\ell=J-1$
\begin{equation}
	\begin{aligned}
		f_{J,3}^{(I)}
		={}&
		\frac{1-(-)^{I+J}}{2}
		\cdot
		\frac{km^2}{2\pi E(2J+1)}
		\times
		\left[
		J(T_1^J+T_2^J)
		+(J+1)(T_3^J+T_4^J)
		+4\sqrt{J(J+1)}\,T_5^J
		\right]
	\end{aligned}
	\tag{4.5}
\end{equation}
\noindent
$\ell=J+1\longrightarrow\ell=J+1$
\begin{equation}
	\begin{aligned}
		f_{J,4}^{(I)}
		={}&
		\frac{1-(-)^{I+J}}{2}
		\cdot
		\frac{km^2}{2\pi E(2J+1)}
		\times
		\left[
		(J+1)(T_1^J+T_2^J)
		+J(T_3^J+T_4^J)
		-4\sqrt{J(J+1)}\,T_5^J
		\right]
	\end{aligned}
	\tag{4.6}
\end{equation}

\noindent
$\ell=J-1\longleftrightarrow\ell=J+1$
\begin{equation}
	\begin{aligned}
		f_{J,5}^{(I)}
		={}&
		\frac{1-(-)^{I+J}}{2}
		\cdot
		\frac{km^2}{2\pi E(2J+1)}
		\times
		\left[
		\sqrt{J(J+1)}
		\left(-T_1^J-T_2^J+T_3^J+T_4^J\right)
		-2T_5^J
		\right]
	\end{aligned}
	\tag{4.7}
\end{equation}

The preceding equations, together with relations (1.19),
ultimately allow us to express these $f_{J,\alpha}^{(I)}$
in terms of the invariant amplitudes $p_j(w,t)$ as

\begin{equation}
	\begin{aligned}
		f_{J,\alpha}^{(I)}
		={}&
		\frac{1-(-)^{I+\ell+s}}{2}
		\sum_{j=1}^{5}
		\frac{k m^2}{4\pi E}
		\int_{-1}^{+1} \dd z\,
		\beta_{J,\alpha}^{j}(k,z)\,
		p_j^{(I)}(w,t)
	\end{aligned}
	\tag{4.8}
\end{equation}
where the $\beta_{J,\alpha}^{j}(k,z)$ are combinations
of Legendre polynomials. Their expressions were given by
Furuichi\supercite{ref13}. For example,

\begin{equation}
	\begin{aligned}
		\beta_{J,1}^{4}(k,z)
		={}&
		\left(1+\frac{k^2}{2m_\pi^2}\right)P_J(z)
		-\frac{k^2}{2m_\pi^2}
		\left[
		\frac{J+1}{2J+1}P_{J+1}(z)
		+
		\frac{J}{2J+1}P_{J-1}(z)
		\right].
	\end{aligned}
	\tag{4.9}
\end{equation}

\quad b--\dashuline{Froissart-Gribov representation}

By substituting the fixed-energy dispersion relations (2.5)
into (4.8), the angular integration can be carried out
explicitly, yielding the representation

\begin{equation}
	\begin{aligned}
		f_{J,\alpha}^{(I)}
		={}&
		\frac{1-(-)^{I+\ell+s}}{2}
		\sum_{j=1}^{5}
		\frac{m_\pi^2}{4\pi^2 E k}
		\int_{4\mu^2}^{\infty}
		d t'\,
		\rho_j^{(I)}(w,t')\,
		\gamma_{J,\alpha}^{j}(k,a)
	\end{aligned}
	\tag{4.10}
\end{equation}
retaining only the dominant contribution from the right-hand cut.

\newpage\centerline{\hfill\small -- 32 --} \vskip .20cm
We have defined
\begin{equation}
	\gamma_{J,\alpha}^{j}(a)
	=
	\frac{1}{2}
	\int_{-1}^{+1} dz\,
	\frac{\beta_{J,\alpha}^{j}(k,z)}
	{a-z}
	\tag{4.11}
\end{equation}
and
\begin{equation}
	a=1+\frac{t'}{2k^2}
	\tag{4.12}
\end{equation}
For example,
\begin{equation}
	\gamma_{J,1}^{1}(k,a)
	=
	\left(1+\frac{k^2}{2m_\pi^2}\right)Q_J(a)-
	\frac{k^2}{2m_\pi^2}
	\left[aQ_J(a)-\delta_{J,0}\right]
	\tag{4.13}
\end{equation}
and, in general, the $\gamma_{J,\alpha}^j(k,b)$ are obtained
from the $\beta_{J,\alpha}^j(k,z)$ by replacing
$P_\ell(z)\longrightarrow Q_\ell(b)$.
As mentioned in Chapter II, a subtraction in $t$ in the
dispersion relations would affect only the S-wave, which we
shall not calculate.

We have also seen that, in principle, there is a left-hand cut
in $t$ arising from crossed-pion exchanges, once their double
spectral representation in $t$ and $\bar t$ has been written
as a dispersion relation at fixed $w$. These short-range
contributions ($t>4m^2$) are usually negligible, but their
contribution to the partial waves can readily be evaluated
if desired. It is given by

\begin{equation}
	\begin{aligned}
		f_{J,\alpha}^{(I)}
		={}&
		\frac{1-(-)^{I+\ell+s}}{2}
		\sum_{j=1}^{5}
		\frac{m_\pi^2}{4\pi^2 E k}
		(-)^j
		\int_{4m_\pi^2}^{\infty} dw'\,
		\gamma_{J,\alpha}^{j}(k,b)\,
		a_j^{I}(4m_\pi^2-w-w',w')
	\end{aligned}
	\tag{4.14}
\end{equation}
where $b=-1-{w'}/{2k^2}$ is the argument of the Legendre
functions of the second kind.

\vskip .1cm
\quad c-- \dashuline{Definition of the phase shifts and unitarization issues}

We adopt here the Stapp parametrization\supercite{ref46}
for the partial waves, which is the most commonly used:

\begin{equation}
	\begin{aligned}
		\text{singlet}\qquad
		&1+i f_{J,1}=e^{2i\delta_J}
		\\[3mm]
		\text{triplet}\quad \ell=J\qquad
		&1+i f_{J,2}=e^{2i\delta_{JJ}}
		\\[3mm]
		\text{coupled triplet}\qquad
		&1+i f_{J,3}
		=\cos 2\epsilon_J\,e^{2i\delta_{J-}}
		\\[2mm]
		&1+i f_{J,4}
		=\cos 2\epsilon_J\,e^{2i\delta_{J+}}
		\\[2mm]
		&f_{J,5}
		=\sin 2\epsilon_J\,
		e^{i(\delta_{J+}+\delta_{J-})}
	\end{aligned}
	\tag{4.15}
\end{equation}

A rigorous identification of the phase shifts is possible only
if $S_J=1+if_J$ lies on the unit circle, which is not
necessarily the case for the amplitudes obtained from our
calculations. Unitarity corrections are therefore required.
To this end, we have tried several different methods.

\newpage\centerline{\hfill\small -- 33 --}\vskip .2cm
\hspace*{2cm}\textbullet\ \emph{Geometrical unitarization}

This method consists in identifying, whenever possible,
the imaginary parts of (4.15), namely, when $f_J$ is real,

\begin{equation}
	\begin{aligned}
		\sin 2\delta_J &= f_{J,1} \\[2mm]
		\sin 2\delta_J &= f_{J,2} \\[2mm]
		\cos 2\epsilon_J\,\sin 2\delta_{J-}
		&= f_{J,3} \\[2mm]
		\cos 2\epsilon_J\,\sin 2\delta_{J+}
		&= f_{J,4} \\[2mm]
		\sin 2\epsilon_J\,\cos(\delta_{J-}+\delta_{J+})
		&= f_{J,5}
	\end{aligned}
	\tag{4.16}
\end{equation}

\hspace*{2cm}\textbullet\ \emph{$K$ matrix}

In this method, the amplitude
$S_J=1+if_J$ ($f_J$ real) obtained from the calculation
is replaced by
$\displaystyle
{(2+i f_J)}/{(2-i f_J)}$,
which is unitary.

For uncoupled states, we obtain

\begin{equation}
	2\tan\delta_J = f_{J,1}\qquad 
	2\tan\delta_{JJ} = f_{J,2}
	\tag{4.17}
\end{equation}
and, in the coupled triplet channel,

\begin{equation}
	\begin{aligned}
		2\delta_{J\pm}
		={}&
		\arctan\left(
		\frac{2(f_{J,3}+f_{J,4})}{4-d(f)}
		\right)\pm
		\arctan\left(
		\frac{2(f_{J,3}-f_{J,4})}{4+d(f)}
		\right)
		\\[4mm]
		\sin 2\epsilon_J
		={}&
		\frac{4f_{J,5}}
		{\sqrt{
				(4-d(f))^2+
				4(f_{J,3}+f_{J,4})^2
		}} \quad \text{where} \quad d(f)=f_{J,3}f_{J,4}-f_{J,5}^2~.
	\end{aligned}
	\tag{4.18}
\end{equation}
\hspace*{2cm}\textbullet \emph{Minimal path}

This procedure consists in determining the phase shifts from
the point on the unit circle closest to the value obtained
for the $S$ matrix.

For uncoupled states, we therefore write

\begin{equation}
	\begin{aligned}
		1+i f_{J,1} &= R_J e^{2i\delta_J}\\
		1+i f_{J,2} &= R_{JJ}e^{2i\delta_{JJ}}
	\end{aligned}
	\tag{4.19}
\end{equation}

In the triplet channel, this method may have several variants,
depending on the norm used to measure the distance between
two matrices. One possibility is to note that Eq.~(4.19) is
identical to the parametrization of partial waves above the
production threshold.

\newpage\centerline{\hfill\small -- 34 --}\vskip .2cm
By analogy, we shall write for the coupled states

\begin{equation}
	\begin{aligned}
		S_{J,3} &= R_-\cos 2\epsilon_J\,
		e^{2i\delta_{J-}}
			\\[2mm]
			S_{J,4} &= R_+\cos 2\epsilon_J\,
			e^{2i\delta_{J+}}
			\\[2mm]
			S_{J,5} &= i\sin 2\epsilon_J\,
			e^{i(\delta_{J+}+\delta_{J-}+\alpha)}
		\end{aligned}
		\tag{4.20}
	\end{equation}
	
	This is the way Mac Gregor\supercite{ref47} defines the
	inelasticity coefficients in $NN$ scattering, but here $\alpha$
	and $(R-1)$ measure the deviation of the model from unitarity.
	
	In the particular case where $f_J$ is real, the phase shifts
	are given by
	
	\begin{equation}
		\begin{aligned}
			\tan 2\delta_J &= f_{J,1}\\
			\tan 2\delta_{JJ} &= f_{J,2}\\
			\tan 2\delta_{J\pm} &= f_{J,3,4}\\
			\tan 2\epsilon_J &= f_{J,5}
		\end{aligned}
		\tag{4.21}
	\end{equation}
	but this method can also be applied when $f_J$ has an
	imaginary part. In our calculation, it is straightforward
	to include the imaginary part arising from the fourth-order
	terms, or even from the sixth-order terms, which can be
	calculated very simply using the (perturbative) unitarity
	relation in the direct channel.
	
	For example,
	
	\begin{equation}
		\begin{aligned}
			2\,\operatorname{Im} f_{J,1}^{(4)}
			&=
			\left(f_{J,1}^{(\mathrm{OPE})}\right)^2
			\\[3mm]
			\operatorname{Im} f_{J,1}^{(6)}
			&=
			f_{J,1}^{(\mathrm{OPE})}
			\operatorname{Re} f_{J,1}^{(4)}
		\end{aligned}
		\tag{4.22}
	\end{equation}
	
	For the singlet channel, the various methods described above
	can be illustrated geometrically (see Fig.~16).
	
	\vskip .3cm
	
	\centerline{
		\includegraphics[width=.3\linewidth]{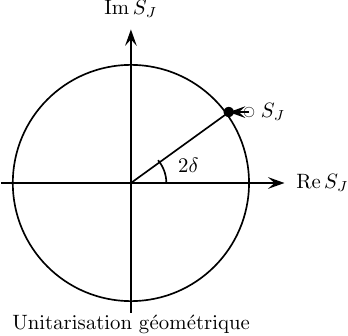}\quad
		\includegraphics[width=.3\linewidth]{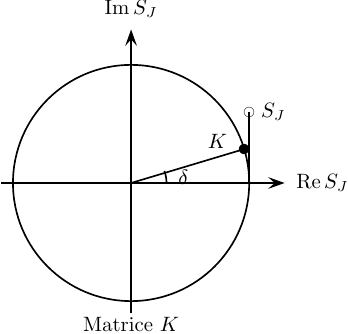}\quad\includegraphics[width=.3\linewidth]{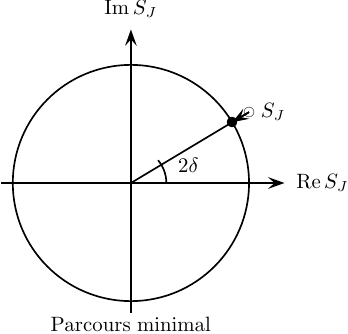}}
	\vskip .2cm
	\centerline{\small \underline{Fig. 16}: Different unitarization procedures. $o$ = calculated $S$ matrix, $\bullet$ unitarized $S$ or $K$ matrix}
	\vskip .2cm
	\newpage\centerline{\hfill\small -- 35 --}\vskip .2cm
	
	Of course, the list above is not exhaustive. For example,
	when only the contributions from the second- and fourth-order
	Feynman diagrams are retained, a Padé approximation can be
	used which, by construction, satisfies unitarity
	\supercite{ref48,ref49}.
	
	In general, the unitarization corrections turned out to be
	small and largely independent of the method used, especially,
	of course, for larger values of $J$ and smaller phase shifts.
	As an example, the values obtained using the various methods
	mentioned above are given below for the ${}^3D_3$ phase shift
	at 330 MeV. This is a rather unfavorable case, since the
	angular momentum is low ($\ell=2$) and the energy is high:
	
	\begin{center}
		\begin{tabular}{lr}
			Geometrical unitarization & 10.7$^\circ$\\
			Minimal path $\operatorname{Im}f_J=0$ & 9.7$^\circ$\\
			$\operatorname{Im}f_J$ = fourth-order terms & 10.4$^\circ$\\
			$K$ matrix & 10.1$^\circ$
		\end{tabular}
	\end{center}
	
	For $\ell=4$, the differences are less than 3\% for
	$T_{\mathrm{lab}}<400$ MeV.
	
	The phase shifts presented in Section II were calculated
	using the $K$-matrix method.
	
	\noindent 2) \underline{Discussion of the results}
	
	\quad a-- \dashuline{Summary of the various contributions}
	
	We shall review, starting with the best-established ones,
	the ingredients used to evaluate the scattering amplitude and,
	from it, the nucleon-nucleon phase shifts.\\
	%
	%
	\hspace*{2cm} \textbullet \emph{Pion exchange and fourth-order contributions}\\
	depend only on the pion-nucleon coupling constant, which
	is well known.
	
	\noindent
	(i) As expected, the pion pole dominates the partial waves
	with high angular momentum, although this is less clearly
	the case for the triplet transitions
	$\ell=J-1\longrightarrow\ell=J-1$, because at low energy
	
	\[
	f_{J,3}^{(\mathrm{O.P.E.})}
	\propto k^{2J+1}
	\]
	whereas the centrifugal barrier requires only
	$
	f_{J,3}\propto k^{2J-1}.
	$
	This anomaly does not occur in the other partial waves
	(except for ${}^1S_0$)\supercite{ref48}. The H waves provide
	a striking illustration of this property of pion exchange
	(see Fig.~18). At 210 MeV, in ${}^1H_5$, ${}^3H_4$ and
	${}^3H_5$, the fourth-order terms account for less than 15\%
	of O.P.E., whereas in ${}^3H_6$ they are of the same order
	of magnitude.
	
	\noindent
	(ii) As for the crossed diagram, which has a double spectral
	representation in $t$ and $\bar{t}$, the ``static'' approximation
	$t=0$, sometimes used\supercite{ref7,ref8,ref9} to calculate
	an equivalent local potential, is inaccurate above 150 MeV.
	In ${}^1G_4$, at 330 MeV, the discrepancy would be 0.5\%
	(30\% of the experimental value) compared with the exact
	calculation of the diagram using the method described in
	Section 1.b.
	
	\noindent
	(iii) The ladder diagram, on the other hand, has only a
	right-hand cut in $t$, and its contribution is simply of the form
	
	\begin{equation*}
		f_J\propto
		\int_{4\mu^2}^{\infty}
		\rho_B(w,t')\,
		Q_J\left(1+\frac{t'}{2k^2}\right)\,\dd t'
	\end{equation*}
	
	\newpage\centerline{\hfill\small -- 36 --}\vskip .2cm
	However, in performing the integration, one must remember
	that the function $\rho_B(w,t')$ has a discontinuity at the
	Mandelstam boundary
	\[
	t'=4\mu^2+\frac{4\mu^4}{w-4m^2}.
	\]
	\hspace*{2cm} $\bullet$ \emph{The double spectral contributions}\\
	which depend on the $\pi N$ phase-shift analyses and on
	their extrapolation into the region $t>4\mu^2$, as discussed
	in Chapter III. We have used the CERN\supercite{ref24}
	(C) and Glasgow\supercite{ref25} (G) analyses, which represent,
	or nearly represent, the largest discrepancy one can find
	in the interpretation of pion-nucleon data.\\
	\hspace*{2cm} $\bullet$ \emph{The subtraction functions in the $(-)$ isospin channel},\\
	which describe two-pion exchange in the P wave.
	The dominance of the $\rho$ meson is quite clear, and all
	the models considered have led to very similar results.
	Those presented here were obtained using the P-wave
	$\pi\pi\longrightarrow N\bar{N}$ amplitude of Höhler,
	Strauss, and Wunder\supercite{ref31}.\\
	\hspace*{2cm} $\bullet$ \emph{The subtraction function $S_1^+(t)$},\\
	which is related to the modulus of the S-wave
	$\pi\pi\longrightarrow N\bar{N}$ amplitude and therefore,
	as discussed in Chapter III, to the pion-pion phase shift
	in the $J=T=0$ channel. We have used two models:
	\begin{itemize}\itemsep -2pt
		\item the first (R) was obtained\supercite{ref32}
		from the BDI pion-pion S-wave solution of Morgan and Shaw.
		\item the second (N.R.) corresponds to a non-resonant
		S wave\supercite{ref35}.
	\end{itemize}
	\noindent
\hspace*{2cm} $\bullet$ \emph{The exchange of the $\omega$ meson}
accounts for part of the three-pion-exchange contribution.
While the ratio
${G_\omega^T}/{G_\omega^V}=-0.42$
is fairly well determined from the isoscalar magnetic moment
of the nucleon, the value of $G_\omega^V$ remains poorly known.
The value
$
{(G_\omega^V)^2}/{4\pi}=4.65
$
used here was derived, through SU(3), from the $\rho$
coupling constant\supercite{ref14}.

\quad {b-- \dashuline{Results}}

The phase shifts calculated with the various models are
shown in Figs. (17) to (20) and compared with the latest
energy-dependent and energy-independent phenomenological
analyses performed by the Livermore group\supercite{ref50}
in the 0--400 MeV range. For comparison, the results of
perturbative calculations at second (O.P.E.) and fourth
(O.P.E. + 4th) order are also shown.

The following general conclusions can be drawn from these
figures\supercite{ref42}:

\noindent
(i) In most cases, the results obtained from the $\pi N$
phase-shift analyses of CERN and Glasgow differ only slightly,
particularly below 200 MeV. In some cases, the difference
is so small that only curve (G) could be plotted. The
discrepancy is more pronounced in the lowest partial waves,
but only above 250 MeV.

\noindent
(ii) The results, on the other hand, are highly sensitive
to the $\pi\pi\longrightarrow N\bar{N}$ S-wave model used,
and clearly favor the non-resonant solution (N.R.). Some
of the excessively attractive phase shifts obtained with
solution (R) would be improved by increasing the coupling
constant $G_\omega^{V\,2}$, but the repulsion due to the
$\omega$ acts over too short a range for this correction
to be satisfactory in all partial waves.

\noindent
(iii) Compared with phenomenology, our results, and in
particular our solution (G + NR), generally provide a
significant improvement over the simple perturbative models
(OPE) and (OPE + 4th).

\newpage\centerline{\hfill\small -- 37 --}\vskip .2cm
\centerline{\includegraphics[width=.45\linewidth,angle=-0.05]{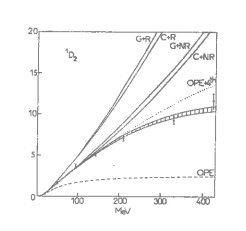}\raisebox{-.1cm}{\includegraphics[width=.45\linewidth,angle=2.5]{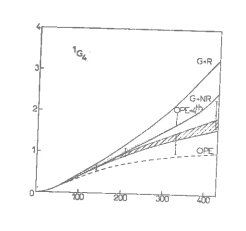}}}

\begin{center}
	{\small	\underline{Fig. 17}: $NN$ phase shifts as a function of the kinetic energy
		(in the laboratory frame) for the
		\underline{even singlet} subspace.}
\end{center}

\begin{description}
	\item[O.P.E. :] one-pion exchange
	
	\item[O.P.E. + 4$^{\mathrm{th}}$ :] one-pion exchange and fourth-order diagrams
	
	\item[C + R :] CERN $\pi N$ phase shifts\supercite{ref24}
	and the ``resonant'' $\pi\pi$ S-wave of Ref. \cite{ref32}
	
	\item[G + R :] Glasgow $\pi N$ phase shifts\supercite{ref25}
	and the ``resonant'' $\pi\pi$ S-wave of Ref. \cite{ref32}
	
	\item[C + R :] CERN $\pi N$ phase shifts and the $\pi\pi$ S-wave
	``non-resonant'' solution of Ref. 35
	
	\item[G + R :] Glasgow $\pi N$ phase shifts and the $\pi\pi$ S-wave
	``non-resonant'' solution of Ref. \cite{ref35}
	
	\item[Shaded area :] M.A.W. energy-dependent analysis
	\supercite{ref50}
	
	\item[Vertical bars :] M.A.W. energy-independent analysis
\end{description}
\newpage\centerline{\hfill\small -- 38 --}\vskip .2cm
\includegraphics[width=.99\linewidth]{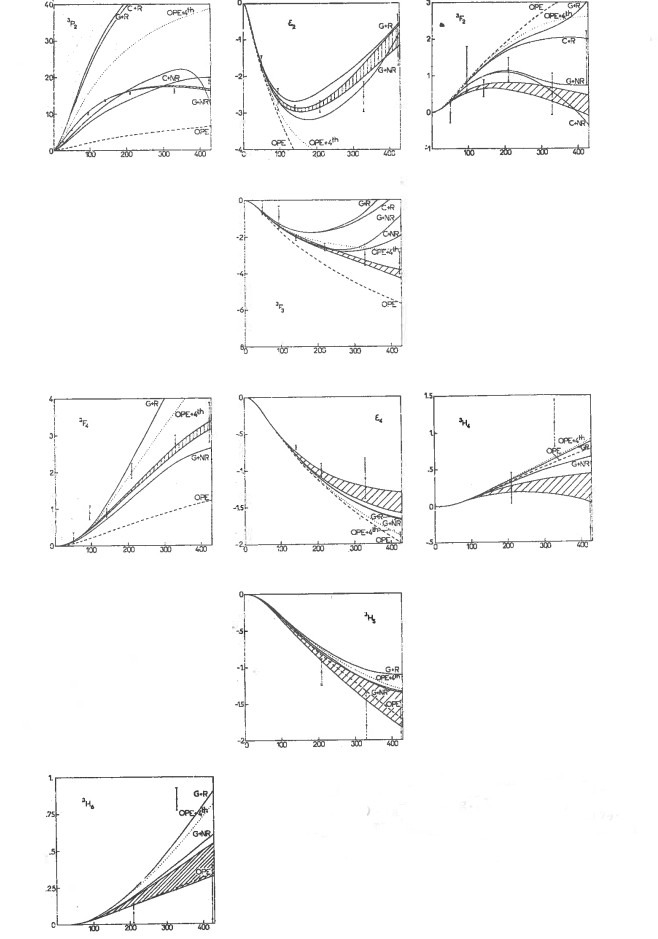}\\[5pt]
\centerline{\small \underline{Fig. 18} : odd triplet subspace (same notation as in Fig. 17)}
\newpage\centerline{\hfill\small -- 39 --}\vskip .2cm
\includegraphics[width=.99\linewidth]{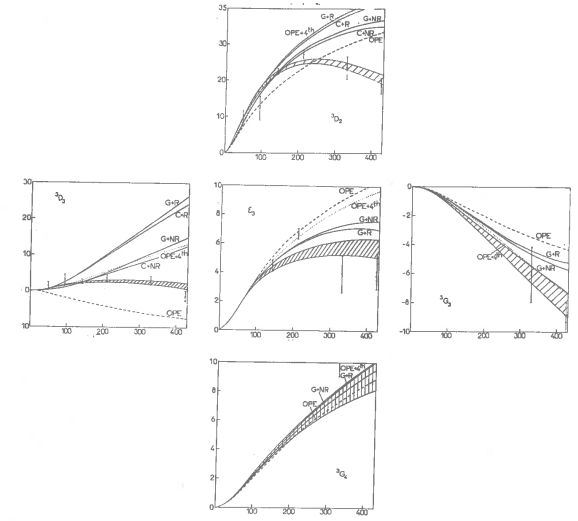}\\[-2pt]
\begin{minipage}{.4\linewidth}
	\includegraphics[width=.99\linewidth]{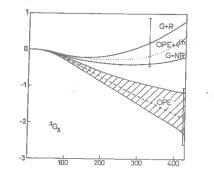}
\end{minipage}
\begin{minipage}{.59\linewidth}
	\centerline{\small \underline{Fig. 19} : even triplet subspace}\centerline{ (same notation as in Fig. 17)}
\end{minipage}	
\vskip -.2cm
\begin{minipage}{.35\linewidth}
	\centerline{\small \underline{Fig. 20} : odd singlet subspace}\centerline{\small (same notation as in Fig. 17)}
\end{minipage}
\begin{minipage}{.64\linewidth}
	\includegraphics[width=.95\linewidth]{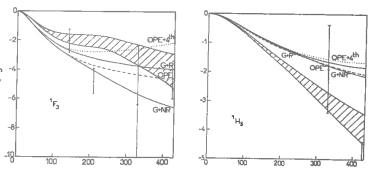}	
\end{minipage}

\newpage\centerline{\hfill\small -- 40 --}\vskip .2cm
Let us now examine the results in greater detail in the
various subspaces.

In the \underline{even singlet} (Fig. 17), ${}^1G_4$ appears
to be dominated by the fourth-order terms (primarily the
crossed diagram among these). In fact, the additional
attraction introduced by the $\pi N$ resonances is almost
exactly compensated by the ``pair suppression'' effect
associated with the non-resonant pion-pion S-wave.

For ${}^1D_2$, the (N.R.) solutions are satisfactory at low
energy; above 100 MeV, the short-range core plays an
important role, and our calculation depends on the
unitarization procedure employed.

In the \underline{odd triplet} (Fig. 18), agreement with
experiment is good, with ${}^3F_3$ providing the best example.
The same holds for ${}^3F_2$, which is particularly interesting
because the difference between models (R) and (N.R.) is
especially pronounced in this channel.

The ${}^3P_2$ and $\epsilon_2$ waves are sensitive to the core,
but the values we obtain, despite the inherent difficulties
associated with unitarization, are rather encouraging.

For the ${}^4H_4$ wave, the two MacGregor analyses
(energy-dependent and energy-independent) differ markedly
at 330 MeV, and no firm conclusion can therefore be drawn
about our results.

In the \underline{isospin-zero states} (Figs. 19 and 20),
the experimental uncertainties are larger, making it difficult
to compare the various models. Our results are compatible
with the phenomenological analyses in channels such as
${}^1F_3$, ${}^3G_3$ and, at low energy,
${}^3D_3$ or ${}^3D_2$.

As for the ${}^3G_5$ and ${}^1H_5$ phase shifts,
our results agree better with an earlier analysis by
MacGregor\supercite{ref51}.

\quad{c- \underline{Comparison with other works}}

There are numerous publications in the literature dealing
with two-pion exchange or, more generally, with the
intermediate-range nuclear forces\supercite{ref43,ref52}.
Here, however, we shall restrict ourselves to works that,
like ours, rely on dispersion-relation theory. In most cases,
the authors have used a formalism that is essentially
equivalent to ours, in that they subtract the S- and P-waves
of the crossed channel from the dispersion integral in order
to account more effectively for pion-pion correlations,
although without emphasizing the need to eliminate certain
asymptotic divergences. It should be noted, moreover, that
no divergences arise in the calculations when the $\pi N$
amplitude is represented by a finite pole expansion in $t$.
The sum is most often restricted to the $N$ and $N_{33}^*$,
but some authors\supercite{ref14,ref41} also include the
$N_{11}^*$ and $N_{31}^*$. In the latter case, the contribution
of the rescattering terms appears to be fairly close to the
one we have estimated by analytically extrapolating the
$\pi N$ amplitude from the physical region. In fact,
a comparison of the nucleon-nucleon phase shifts obtained
by the various groups essentially illustrates their sensitivity
to the $\pi\pi\longrightarrow N\bar{N}$ S-wave model.

Furuichi and Watari\supercite{ref13} showed that the terms
involving the $N_{33}^*$ (Fig. 21) make a strongly attractive
contribution, particularly through their component with
quantum numbers $J=T=0$ in the $t$ channel.

\newpage\centerline{\hfill\small -- 41 --}\vskip .2cm

\begin{center}
	\includegraphics[width=.9\linewidth]{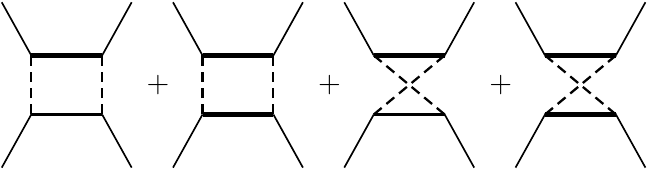}\\[3pt]
	{\small \underline{Fig. 21} : Contribution of the $N^*_{33}$}
\end{center}

According to these authors, this indicates that the correction
due to correlations between the two pions in the S-state
(the right-hand cut of $f_{+}^{0}(t)$, in the terminology
of Chapter III) should be repulsive. This conclusion is
consistent with ours and is confirmed by a study by
Furuichi, Kanada and Watanabe\supercite{ref39}, who used
different values of the phase shift $\delta_0^0(t)$ to construct
the $\pi\pi\longrightarrow N\bar N$ S-wave, thereby ruling out
an overly resonant solution. They nevertheless obtained good
results with an ``up-down'' solution, which does, admittedly,
contain a free parameter associated with the left-hand cut.

Durso\supercite{ref12} obtained good results for peripheral
nuclear phase shifts in 1966 using the Bowcock, Cottingham
and Lurié parametrization\supercite{ref19} of the $\pi N$
amplitude. The latter authors introduced a phenomenological
constant in the function $A^{+}$ to account for pion-nucleon
scattering lengths. The presence of this constant also explains
the agreement obtained in the nucleon-nucleon case (both in
Durso's calculation and in the evaluation of the nuclear
potential by Cottingham and Vinh Mau\supercite{ref7}), and
actually corresponds to a substantial modification of the
$\pi\pi\longrightarrow N\bar N$ S-wave compared with the
earlier descriptions\supercite{ref30} of the $\pi N$ amplitude.
Along the same lines, but within a specific model, Brown
and Durso\supercite{ref34} showed that a proper description
of the pion-nucleon interaction at very low energy helps
reduce the excessive attraction usually found in the
intermediate-range nuclear forces.

In the calculation by Chemtob and Riska\supercite{ref41},
the contribution from two-pion exchanges with $J_t\gg 1$
differs little from the one we obtained using another method.
Moreover, for exchanges with $J_t=0$, they use the
non-resonant S-wave of Ref. \cite{ref35}, which explains
why their results are very similar to ours obtained with the
same $f_{+}^{0}(k)$ (N.R. model).

\newpage\centerline{\hfill\small -- 42 --}\vskip .1cm

\begin{center}
	\underline{{Chapter V}} -- {NUCLEON--NUCLEON INTERACTION
		IN THE CENTRAL REGION}
\end{center}

As we have seen, the calculation of one- or two-pion exchange
allows us to evaluate the long- and intermediate-range nuclear
forces. Part of the three-pion contribution is contained in
the $\omega$-meson pole, but the other three-pion exchanges,
particularly those with isospin $T=1$ in the crossed channel,
as well as the shorter-range forces, cannot, at present,
be evaluated theoretically. We have therefore studied this
part of the interaction phenomenologically. In this chapter,
we describe the methods used and present the results obtained
so far.

\noindent{1) \underline{Phase-shift analysis}}

In nucleon-nucleon phase-shift analyses, the parameters
$\delta_{\ell,s,j}$ and $\epsilon_J$ are adjusted to reproduce
the experimental measurements of cross sections, polarization,
correlations, and so on. This can only be done satisfactorily
by including a large number of partial waves, particularly
at higher energies. In the analyses carried out so far
\supercite{ref50,ref51,ref53}, one assumes that one-pion
exchange (O.P.E.) dominates the highest partial waves;
the coupling constant is then the only parameter to be
adjusted for this contribution, although its value is also
constrained by the study of other reactions. Nevertheless,
enough ``free'' partial waves remain ($\ell_{\max}\sim 6$
at 300--400 MeV) to make their determination difficult,
not least because of the numerical challenges involved.
More importantly, ambiguities may arise, with two or more
solutions reproducing the data with nearly the same accuracy.
In such cases, the choice of the best solution on the basis
of its ``$\chi^2$'' may be called into question by a new
experiment or even by a simple normalization correction.
This can be seen by examining the table below, which
collects some results obtained at different times by the
same group.
\vskip .3cm

\centerline{\begin{minipage}{.58\linewidth}
		\begin{center}
			$I=1\qquad E_{\mathrm{lab}}=425\ \mathrm{MeV}$\\
			\begin{tabular}{c|r|r|r}
				Wave & 1968\supercite{ref51} & 1969\supercite{ref50} & 1974\supercite{ref53}\\[2pt]
				\hline
				${}^{1}S_0$ & $-19.35$ & $-19.00$ & $-17.15\pm1.76$\\
				${}^{1}G_4$ & $1.42$ & $1.96$ & $2.07\pm0.35$\\
				${}^{3}H_6$ & $0.35$ & $0.03$ & $0.57\pm0.41$\\
				$\epsilon_2$ & $-1.43$ & $-0.96$ & $-1.40\pm0.79$\\
				$\epsilon_4$ & $-1.86$ & $-2.19$ & $-2.61\pm0.37$
			\end{tabular}
		\end{center}
	\end{minipage}
	\begin{minipage}{.41\linewidth}
		\begin{center}
			$I=0\qquad E_{\mathrm{lab}}=425\ \mathrm{MeV}$\\[4pt]
			\begin{tabular}{c|r|r}
				Wave & 1968\supercite{ref51} & 1969\supercite{ref50}\\
				\hline
				${}^{1}F_3$ & $-6.01$ & $-4.91\pm1.05$\\
				${}^{3}G_3$ & $1.18$ & $-9.30\pm1.35$\\
				$\epsilon_3$ & $8.98$ & $4.12\pm1.23$\\
				${}^{3}G_4$ & $0.39$ & $11.08\pm1.08$
			\end{tabular}
		\end{center}
	\end{minipage}\hfill}
\vskip .2cm

One way to improve this situation is to constrain the peripheral
phase shifts not by one-pion exchange alone, but by the
combined O.P.E. + T.P.E. (two-pion exchange) contribution
calculated in the preceding chapters. This makes it possible
to reduce $\ell_{\max}$ and to determine phenomenologically
only the lower partial waves.

A recent phase-shift analysis carried out at Saclay\supercite{ref54} has yielded the following preliminary results:
\begin{center}
	\begin{tabular}{c|c|c|c|c}
		long range & intermediate range & short range & number of parameters & $\chi^2$/point\\[2pt]
		\hline
		O.P.E. & free & free & 44 & 1.007\\
		O.P.E. & G + NR & free & 29 & 1.106\\
		O.P.E. & G + R & free & 29 & 1.169\\
		\hline
	\end{tabular}
\end{center}

\newpage\centerline{\hfill\small -- 43 --}\vskip .1cm

The $\chi^2$/point is calculated from proton-proton observables
between 300 and 400 MeV. In the last two solutions, the
phase shifts ${}^{3}H_4$, ${}^{3}H_5$, ${}^{3}H_6$,
${}^{1}G_4$ and $\epsilon_4$ obtained by us (see Chapter IV
or Ref. 42) have been included in the scattering amplitude.

These results are encouraging, since they confirm that our
evaluation of the intermediate-range forces is in satisfactory
agreement with experiment. However, it should be noted that,
in this energy range, the data are not sufficiently restrictive
to discriminate between the (N.R.) and (R) solutions.

A similar study will soon be undertaken, on the one hand,
at very low energies, where the data are abundant and precise,
and hence highly ``selective'', and, on the other hand,
above 400 MeV, by calculating the peripheral inelasticity
coefficients from the double spectral functions
$\gamma_j^I(w,t)$ and treating the others as free parameters.

\noindent
{2) \underline{Search for a semi-phenomenological potential}}

\noindent
\quad {a- \dashuline{Description of the interaction in coordinate space}}

For a phenomenological study of nuclear forces, the interaction can also
be described in terms of a potential in coordinate space, which offers
several obvious advantages.

\noindent
(i) In this form, the physical interpretation of the results
(attraction, repulsion) is straightforward.

\noindent
(ii) The treatment of Coulomb effects, the numerical calculation of
phase shifts, or the quadrupole moment of the Deuteron are relatively
simple in this representation.

\noindent
(iii) The concept of a potential is widely used in Nuclear Physics,
which facilitates comparisons with phenomenological forces and the use
of the interaction constructed here in nuclear structure calculations.

However, this should not obscure a few fundamental difficulties:

\noindent
(i) Taking the non-relativistic limit of a scattering amplitude
constructed from field theory or the $S$-matrix is not as
straightforward as it might appear. It has long been known that certain
``static'', ``adiabatic'', or ``no-recoil nucleon'' approximations
can be problematic\supercite{ref8,ref9}.

\noindent
(ii) The problem of defining a potential in momentum space that
reproduces, in the Born approximation, the theoretical evaluation of
the lightest exchanges is relatively simple. The faithful treatment
of the iteration of these exchanges at higher orders has, however,
been the subject of numerous publications in recent years
\supercite{ref8,ref55}. Their authors have mainly sought to identify
the most suitable wave equation to replace the cumbersome
Bethe--Salpeter equation. They have also shown that, when restricting
oneself, as we do here, to the Lippmann--Schwinger equation,
relativistic corrections can be simulated by modifying the expression
of the potential\supercite{ref21}, although this procedure is not
entirely free of ambiguities.

\noindent
(iii) Calculating the Fourier transform of the potential requires
knowledge of the off-shell matrix elements (which is related to the
previous problem). Moreover, although the momentum transfer $t$ is
naturally conjugate to the relative distance $r$ between the nucleons,
the energy dependence requires the introduction of nonlocal terms,
for which no simple prescription is available.

\newpage\centerline{\hfill\small -- 44 --}\vskip .1cm

\quad{b- \dashuline{Evaluation of the long- and intermediate-range potential}}

The method adopted to evaluate this potential from the peripheral
part of the $S$-matrix, in view of the difficulties discussed above,
was presented in Refs.~\cite{ref21,ref22}. We recall briefly that
the contribution from one-pion exchange leads to the Yukawa potential,
to which is added a term representing the difference between the
two-pion exchanges evaluated using dispersion relations and those
obtained by simply iterating the O.P.E.P. This is done to avoid
double counting in the region $t>4\mu^2$. The long- and
intermediate-range potential constructed in this way appears to
provide a faithful representation of our evaluation of the peripheral
forces, since solving the Schrödinger equation yields, through the
high-angular-momentum phase shifts, the same values as those obtained
by directly projecting the $S$-matrix.

The comparison\supercite{ref21,ref22} with phenomenological potentials
also confirms the conclusions suggested by the examination of the
phase shifts: this approach provides a good description of the nuclear
forces between 1 and 2 fm, particularly for the G + NR and C + NR
solutions (the potential constructed from model R for the S-wave
$\pi\pi\longrightarrow N\bar N$ amplitude appears, in fact, to be
too attractive).

Compared with the preceding chapter and earlier publications
\supercite{ref21,ref56}, we have introduced only minor changes in
the calculation of the intermediate-range forces, retaining in
particular solution (G) for the $\pi N$ phase shifts.

\noindent
(i) For the subtraction functions $S_+^*(t)$ and $S_j^*(t)$, we
have used the S- and P-wave $\pi\pi\longrightarrow N\bar N$
amplitudes recently calculated by Nielsen and Oades
\supercite{ref33}. The corresponding values of $f_+^0(t)$ are given
in Chapter III; they correspond to model $R_3$, which is close to
N.R. in magnitude, but much more satisfactory from a theoretical
standpoint. For the P-wave, the modification compared with the older
model of Hohler, Strauss, and Wunder\supercite{ref31} is not very
significant.

\noindent
(ii) It appeared to us that the coupling constant
$G_\omega^2/4\pi=4.68$, obtained from that of the $\rho$ by means
of SU(3), did not adequately account for the repulsion due to the
$\omega$ meson\supercite{ref56}. In agreement with a recent estimate
\supercite{ref57}, we have adopted the value
$G_\omega^2/4\pi=9.52$, which leads to better results for
nucleon--nucleon interactions.

\quad{c- \dashuline{Nonlocal terms}}

In momentum space, the interaction takes the form
\begin{equation}
	\widetilde V=\sum_\alpha \widetilde V_\alpha(\Delta,E)\,
	\widetilde\Omega_\alpha .
	\tag{5.1}
\end{equation}
where (unlike in Chapter I, where we set $w=4E^2$)
\begin{equation}
	\Delta=\sqrt{-t/4},
	\qquad
	E=\frac{w-4m^2}{4m}.
	\tag{5.2}
\end{equation}
and
\begin{equation}
	\begin{aligned}
		\widetilde\Omega_c&=1,\qquad
		\widetilde\Omega_{SS}=\sigman\cdot\sigmap,\\
		\widetilde\Omega_{SO}&=-\frac{i}{2}
		(\sigman+\sigmap)\cdot(\pp_1\times\pp_2),\\
		\widetilde\Omega_T&=4\left(
		\Delta^2\sigman\cdot\sigmap
		-3(\sigman\cdot\Del)(\sigmap\cdot\Del)\right),\\
		\widetilde\Omega_{SO2}&=
		\sigman\cdot(\pp_1\times\pp_2)\,
		\sigmap\cdot(\pp_1\times\pp_2).
	\end{aligned}
	\tag{5.3}
\end{equation}
are the Okubo--Marshak invariants\supercite{ref58}.

\newpage\centerline{\hfill\small -- 45 --}\vskip .1cm

The energy dependence of the components
$\widetilde V_\alpha(\Delta,E)$ reflects the nonlocality of the
potential. This can also be illustrated by evaluating the partial
Fourier transforms
\begin{equation}
	V_\alpha(E,r)\propto
	\int d^3\Del\,e^{i\Del\cdot\boldsymbol r}\,
	\widetilde V_\alpha(\Delta,E)
	\tag{5.4}
\end{equation}
and examining, at fixed $r$, the variation of $V_\alpha(E,r)$ with
$E$.

The results presented in Ref.~\cite{ref21} show that the energy
dependence is significant for the central and spin-orbit terms
(a 40\% variation for the $C^+$ potential between the $NN$ threshold
and the $NN\pi$ threshold), and considerably smaller for the tensor,
spin-spin, and quadratic spin-orbit terms.

The latter gives rise to another form of nonlocality when one
attempts to express it in coordinate space. Indeed, in addition to
the usual terms
\begin{equation}
	\Omega_{SO2}=
	\frac{1}{2}\left(
	\sigman\cdot\boldsymbol L\,\sigmap\cdot\boldsymbol L+
	\sigmap\cdot\boldsymbol L\,\sigman\cdot\boldsymbol L
	\right)
	\tag{5.5}
\end{equation}
there appear\supercite{ref59,ref22} velocity-dependent tensor
couplings
\begin{equation}
	\Omega_p=
	(\sigman\cdot\pp)(\sigmap\cdot\pp)
	-(\sigman\cdot\sigmap)\,\pp^2.
	\tag{5.6}
\end{equation}

It appears impossible to neglect all these nonlocal contributions
(and this is quickly confirmed by a few numerical tests). We have
therefore solved the Schrödinger equation with a potential of the
form $V(E,r)$, fixing $E$ to the incident kinetic energy at which
the phase shifts are to be calculated.

Several simplifications are nevertheless possible.


\noindent
(i) The quadratic spin-orbit components other than
$\Omega_{SO2}$ are short-ranged, and it appeared that they could
easily be omitted and incorporated into the phenomenological
description of the core.

\noindent
(ii) The energy dependence of the tensor, spin-spin, and especially
quadratic spin-orbit potentials is weak for $r\gtrsim 1$ fm, and
can therefore be neglected.

\noindent
(iii) Finally, the energy dependence is approximately linear between
0 and 300 MeV, i.e.,
\begin{equation*}
	V_\alpha(E,r)\simeq U_\alpha(r)+E\,W_\alpha(r).
\end{equation*}

We adopted this approximation for some of our solutions, since it
appears to greatly facilitate the use of an energy-dependent
potential in a nuclear structure calculation, where $E$ is not
well-defined. Indeed, in the case of a linear dependence, the
interaction can be expressed in an equivalent form as a
velocity-dependent force\supercite{ref60,ref61},
\begin{equation}
	\left[
	\frac{\pp^2}{2m}+U(r)+E W(r)
	\right]\psi=E\psi
	\quad\Longrightarrow\quad
	\left[
	\frac{1}{\sqrt{1-W(r)}}\frac{\pp^2}{2m}
	\frac{1}{\sqrt{1-W(r)}}+
	\frac{U(r)}{1-W(r)}
	\right]\varphi=E\varphi .
	\tag{5.7}
\end{equation}
where
$\varphi=\sqrt{1-W(r)}\,\psi$.

\newpage\centerline{\hfill\small -- 47 --}\vskip .1cm

The table below indicates which nonlocal terms have been retained
in the long- and intermediate-range potential for the various
solutions considered. The last one is by far the simplest, since
the energy dependence is linear, occurs only in the central and
spin-orbit terms, and the quadratic spin-orbit component reduces
to the usual operator $\Omega_{SO2}$.

\begin{small}\begin{center}
		\underline{Table V}\\
		Characteristics of the theoretical potential and number of
		parameters describing the core for each isospin state.\\[3pt]
		D.E.:
		energy dependence of the potential as it appears in the
		theoretical calculations\\
		D.L.E.:
		linear approximation to the above energy dependence\\
		I.E.:
		energy-independent potential (obtained from the previous one
		by setting $E=0$)\\[9pt]
		\renewcommand{\arraystretch}{1.35}
		\setlength{\tabcolsep}{4pt}
		\small
		\begin{tabularx}{\textwidth}{|>{\centering\arraybackslash}m{0.65cm}|
				>{\raggedright\arraybackslash}m{1.65cm}|
				*{4}{>{\centering\arraybackslash}X|}|
				>{\centering\arraybackslash}m{2.15cm}|}
			\hline
			\multicolumn{2}{|c|}{Solution} &
			Central & Spin-orbit & Tensor & Spin-spin &
			\shortstack{Quadratic\\spin-orbit}\\
			\hline
			
			\multirow{2}{*}{\Large 1} &
			theor. pot. &
			D.E. & D.E. & D.E. & D.E. &
			\shortstack{complete\\I.E.}\\
			\cline{2-7}
			& \shortstack[l]{number of\\core\\parameters} &
			2 & 2 & 2 & 2 &
			\shortstack{2\\correlated}\\
			\hline
			
			\multirow{2}{*}{\Large 2} &
			theor. pot. &
			D.L.E. & D.L.E. & I.E. & D.L.E. &
			$\Omega_{SO2}$ alone\\
			\cline{2-7}
			& \shortstack[l]{number of\\core\\parameters} &
			2 & 1 & 1 & 2 & 1\\
			\hline
			
			\multirow{2}{*}{\Large 3} &
			theor. pot. &
			D.L.E. & D.L.E. & I.E. & I.E. &
			$\Omega_{SO2}$ alone\\
			\cline{2-7}
			& \shortstack[l]{number of\\core\\parameters} &
			2 & 1 & 1 & 1 & 1\\
			\hline
		\end{tabularx}
\end{center}\end{small}

\quad d- \dashuline{Parametrization of the core}

The potential $V_{\mathrm{th}}(E,r)$ obtained from our theoretical
study is valid only at distances greater than approximately
$1\,\mathrm{fm}$. We have supplemented it with a constant core by
writing
\begin{equation}
	V(E,r)=(p_1+E p_2)f(r)+[1-f(r)]V_{\mathrm{th}}(E,r).
	\tag{5.8}
\end{equation}

Continuity is ensured by the function
\begin{equation}
	f(r)=\frac{1}{1+(\alpha r)^\beta},
	\qquad
	\alpha=1.25\,\mathrm{fm}^{-1},
	\qquad
	\beta=10.
	\tag{5.9}
\end{equation}
which vanishes rapidly for $r>\alpha^{-1}$, thereby preventing
any modification of the ``theoretical'' region. Conversely,
$V_{\mathrm{th}}(E,r)$ plays no role for $r<\alpha^{-1}$.

\newpage\centerline{\hfill\small -- 48 --}\vskip .1cm

The energy dependence (parameter $p_2$) has been introduced in
the central term; in solutions 2 and 3, this term therefore depends
linearly on $E$ at all values of $r$, namely
\begin{equation}
	\begin{aligned}
		V^{c}(E,r)={}&
		\left[p_1^{c}f(r)+[1-f(r)]U_c(r)\right]
		+E\left[p_2^{c}f(r)+[1-f(r)]W_c(r)\right].
	\end{aligned}
	\tag{5.10}
\end{equation}

For the spin-orbit, tensor, and spin-spin components, we have
either allowed a slight energy dependence or imposed $p_2=0$,
which explains why the number of parameters is not the same for
all our models, as indicated in the table above.

It should be noted that, when performing transformation (5.7),
the factor $\sqrt{1-W(r)}$ that renormalizes the wave function
should not vanish. This places, in particular, a constraint on the
energy dependence of the spin-orbit potential in the core, which
must remain weak. This is why, for solution 3, we imposed
$p_2^{\mathrm{so}}=0$, although the theoretical part contains a
nonlocal contribution $E W_{\mathrm{so}}(r)$.

In solution 1, we sought to include all quadratic spin-orbit terms,
namely
\[
V_{S02}\,\Omega_{S02}
+\bigl(V_{\sigma q}\,\Omega_q+\mathrm{h.c.}\bigr)+\cdots
\qquad \text{(see Ref.~\cite{ref22})}
\]

In this case, the radial equations involve the two functions
$V_{\sigma q}(r)$ and $V_{S02}(r)$, both derived from the
momentum-space potential $\widetilde{V}_{S02}(\Delta)$ and related
by
\[
V_{S02}(r)=-\frac{1}{r}\frac{d}{dr}V_{\sigma q}(r).
\]

The parametrization (5.8) is then not very convenient, since
the derivative $df/dr$ is nearly a delta function. We therefore
wrote
\begin{equation}
	\begin{aligned}
		V_{\sigma q}(r)
		&=q_1 f(r)+[1-f(r)]V_{\mathrm{th}}^{\sigma q}(r),\\
		-rV_{S02}(r)
		&=q_2 f(r)+[1-f(r)]
		\frac{dV_{\mathrm{th}}^{\sigma q}}{dr}.
	\end{aligned}
	\tag{5.11}
\end{equation}
thus abandoning, at short distances, the strict derivative
relation, while retaining it qualitatively (if $q_1>0$,
$V_{\sigma q}$ decreases between 0 and 1 fm, and we impose
$q_2<0$). This is an indirect way of introducing two correlated
parameters into the quadratic spin-orbit potential.

For solutions 2 and 3, we simply wrote
\begin{equation}
	V_{S02}(r)=p_1 f(r)+[1-f(r)]V_{\mathrm{th}}^{SO2}(r).
	\tag{5.12}
\end{equation}

\newpage\centerline{\hfill\small -- 49 --}\vskip .1cm

\quad{e- \dashuline{Fitting the core parameters}}

The parametrized potential is inserted into the Schrödinger
equation, which is projected onto partial waves and solved
numerically; see, for example, Ref.~\cite{ref62}. This allows us
to calculate the phase shifts $\delta_{\ell,s,j}(E)$, the mixing
angles $\epsilon_j(E)$, and the low-energy parameters $a$ and
$r_0$, and to compare them quantitatively with the results of the
most recent phenomenological analyses
\supercite{ref50,ref53,ref62,ref64}, by evaluating
\begin{equation}
	\begin{aligned}
		\chi^2={}&
		\sum_{E,\ell,s,j}
		\left[
		\frac{\delta_{\ell,s,j}(E)-\delta_{\ell,s,j}^{(0)}(E)}
		{\Delta\delta_{\ell,s,j}^{(0)}(E)}
		\right]^2+\sum_{E,j}
		\left[
		\frac{\epsilon_j(E)-\epsilon_j^{(0)}(E)}
		{\Delta\epsilon_j^{(0)}(E)}
		\right]^2
		+\left[
		\frac{a-a^{(0)}}{\Delta a^{(0)}}
		\right]^2
		+\left[
		\frac{r_0-r_0^{(0)}}{\Delta r_0^{(0)}}
		\right]^2 .
	\end{aligned}
	\tag{5.13}
\end{equation}
where $\delta_{\ell,s,j}^{(0)}(E)$ and
$\Delta\delta_{\ell,s,j}^{(0)}(E)$ are the phase shifts from the
energy-independent solution and the corresponding uncertainties,
at the energies
\[
E_{\mathrm{lab}}=2E=25,\ 50,\ 95,\ 142,\ 210,\ 330\ \mathrm{MeV}.
\]

For isospin $I=0$, we can also evaluate the binding energy,
quadrupole moment, and other properties of the Deuteron, and
compare them with the experimental values.

The experimental phase shifts for $I=1$ are obtained from
proton--proton data and are published after subtraction of the
purely Coulomb contribution (the interference term therefore
remains). We account for this by including the electrostatic
potential in $V(E,r)$ and matching the Schrödinger-equation
solutions asymptotically not to spherical Bessel functions, but
to the Coulomb functions $G_\ell(k,r)$ and $F_\ell(k,r$.
Finally, the vacuum-polarization effect must be taken into account
to accurately calculate the scattering length $a$ and the effective
range $r_0$\supercite{ref65}.

These quantities are the first coefficients in the expansion of the
${}^{1}S_0$ phase shift:
\begin{equation}
	k\operatorname{cotg}\delta
	=-\frac{1}{a}
	+\frac{1}{2}r_0 k^2
	-P r_0^3 k^4
	+Q r_0^5 k^6+\cdots .
	\tag{5.14}
\end{equation}

When performing a Lagrange interpolation for $E=1,2,3,4$ MeV,
and then for $E=1,2,4,8$ MeV, the differences obtained for $a$
and $r_0$ are of the same order as the experimental uncertainties.
The $\chi^2$ associated with these quantities would therefore be
meaningless if such sets of energy values were adopted. We have
thus investigated the precise energies at which the measurements
used to determine $a$ and $r_0$ experimentally were performed
\supercite{ref66}. We hope that this will make it meaningful to
attempt to reproduce these values to within 1\% using our potential.

Finally, the function $\chi^2$, which depends on the adjustable
parameters of the core, was minimized using an appropriate
numerical program\supercite{ref67}.

\newpage\centerline{\hfill\small -- 49 --}\vskip .1cm

\quad f-- \dashuline{Results and discussion}

\hspace*{2cm}\textbullet \emph{General considerations}

Overall, the results are good, since we obtain a $\chi^2$ per
point of 2.2, which is comparable, from this standpoint, to the
best phenomenological potentials. Some of these potentials are
local, making them easier to use, but ours has the advantage of
involving fewer free parameters. Indeed, it is worth emphasizing
that, throughout our calculations, we have refrained from
modifying the theoretical part of the potential, even within the
uncertainties associated with its evaluation (for example, the choice
between the CERN and Glasgow A solutions). We adopted this
approach for the sake of clarity and to avoid further complicating
the already demanding numerical part of this work. In contrast,
the intermediate-range forces in conventional phenomenological
potentials are fairly unconstrained, through the freedom in the
coupling constants or even in the exchanged masses, which are
not all taken from the meson spectrum.

A detailed comparison between our potential and phenomenological
forces would be tedious and often difficult because of the
nonlocal terms we have introduced. These forces are often
classified according to the hardness of their core (see, for
example, Ref.~68 for a fairly recent review). In short, our
potential belongs to the category of ``soft-core'' potentials:
it is less attractive at intermediate range and considerably less
repulsive at short distances than, for example, the Hamada--Johnson
potential. Tables at the end of this chapter give the values of the
core parameters for our various solutions. The first two solutions
contain nonlocal terms in the spin-orbit, spin-spin, and/or tensor
components and therefore require additional parameters to describe
the core region, compared with solution No.~3. In seeking the best
agreement with experiment, we therefore allowed the first six
parameters to be readjusted:
\[
(p_1^C,\ p_2^C,\ p_1^{SO},\ p_1^{SS},\ p_1^T,\ p_1^{SO2}),
\]
but the values obtained were fairly close to those of solution
No.~3. The latter can therefore be regarded as a good
approximation, retaining only the essential nonlocal terms.

We now examine, subspace by subspace, the results obtained for
the nucleon--nucleon phase shifts. For solution No.~3, these
are shown in Figs.~22 and 23.

\hspace{2cm}\textbullet\ \emph{Even singlet subspace}
($I=1$)

The tensor and spin-orbit potentials do not contribute in this
subspace, which considerably simplifies the calculations. The
${}^{1}G_4$ wave and the ${}^{1}D_2$ wave at low energies
depend only on the theoretical part of the potential, and we
recover, to a very good approximation, the phase shifts obtained
by directly projecting the $S$-matrix in momentum space.

By adjusting the core of the potential $V_C-3V_{SS}$, it is easy
to reproduce the ${}^{1}S_0$ phase shift, which is attractive at
low energies and repulsive around 300 MeV. However, it is generally
more difficult to reproduce $a$ and $r_0$ accurately, since these
also depend on the intermediate-range forces. The good agreement
obtained for these quantities is therefore partly due to our
calculation of two-pion exchanges.

\newpage\centerline{\hfill\small -- 50 --}\vskip .1cm

Compared with the analysis of M.A.W.\supercite{ref50}, our results
for solution No.~3 are as follows, corresponding to an average
$\chi^2$ of 2.16 per point. For solution No.~1, the $\chi^2$
per point is 1.11, while for solution No.~2 it is 1.95.

\begin{center}
	\renewcommand{\arraystretch}{1.45}
	\setlength{\tabcolsep}{7pt}
	\begin{tabular}{r@{\qquad}c@{\qquad}c@{\qquad}c}
		$E_{\mathrm{lab}}$ & ${}^{1}S_0$ & ${}^{1}D_2$ & ${}^{1}G_4$\\
		25  & $49{,}30\ (49{,}60\pm0{,}26)$
		& $0{,}74\ (0{,}74\pm0{,}03)$
		& $0{,}04\ (\mathrm{O.P.E.})$\\
		50  & $39{,}80\ (39{,}43\pm0{,}44)$
		& $1{,}76\ (1{,}67\pm0{,}10)$
		& $0{,}17\ (\mathrm{O.P.E.})$\\
		95  & $27{,}08\ (25{,}53\pm1{,}09)$
		& $3{,}57\ (3{,}71\pm0{,}24)$
		& $0{,}44\ (\mathrm{O.P.E.})$\\
		142 & $16{,}84\ (16{,}70\pm0{,}54)$
		& $5{,}32\ (5{,}11\pm0{,}16)$
		& $0{,}73\ (0{,}62\pm0{,}06)$\\
		210 & $4{,}84\ (5{,}42\pm0{,}53)$
		& $7{,}44\ (7{,}05\pm0{,}28)$
		& $1{,}15\ (1{,}00\pm0{,}10)$\\
		330 & $-12{,}03\ (-10{,}53\pm1{,}33)$
		& $9{,}69\ (9{,}26\pm0{,}40)$
		& $1{,}94\ (1{,}22\pm0{,}23)$
	\end{tabular}
\end{center}

\begin{equation*}
	a = -7{,}817\ \mathrm{fm}
	\quad (-7{,}819\pm0{,}009),\\
	\qquad r_0 =  2{,}747\ \mathrm{fm}
	\quad ( 2{,}820\pm0{,}044).
\end{equation*}
(O.P.E. indicates that, in the phenomenological analysis, this
phase shift is fixed to the value given by one-pion exchange.
These points have not been included in the calculation of
$\chi^2$.)

A substantial part of the $\chi^2$ comes from the ${}^{1}G_4$
wave at 210 and 330 MeV, where the experimental uncertainties
are very small, although there is a noticeable discrepancy between
the energy-dependent and energy-independent solutions of M.A.W.
We are, moreover, in better agreement with the values recently
proposed by A.H.R.\supercite{ref53}.

\hspace*{2cm}\textbullet\ \emph{Odd triplet subspace}
($I=1$)

This subspace allows us to determine the cores of the spin-orbit
and tensor potentials, as well as that of the combination
$V_C+V_{SS}$. The $V_{SO2}$ potential contributes to the odd
triplet subspace, but also to the even singlet subspace; we have,
of course, imposed the same values for its parameters in both
subspaces.

Compared with M.A.W., the $\chi^2$ values per point obtained are
\vskip -.4cm
\[
\begin{array}{ll}
	\renewcommand{\arraystretch}{0.8}
	\text{solution No.~1} & 1{,}9\\[2pt]
	\text{solution No.~2} & 2{,}4\\[2pt]
	\text{solution No.~3} & 2{,}6
\end{array}
\]

It is traditionally rather difficult for a potential to reproduce
all the P waves accurately
(${}^{3}P_0$, ${}^{3}P_1$, ${}^{3}P_2$). Additional nonlocal
terms help in this respect, which explains why the agreement
with experiment is better for the first solution. On the other
hand, we found, without regret, when testing intermediate models
between Nos.~1 and 2, that the complications associated with the
quadratic spin-orbit terms do not lead to any significant
improvement in the description of proton--proton scattering.

A large part of our $\chi^2$ in the odd triplet subspace is due
to the $J=4$ waves. These discrepancies could, of course, be
attributed to the uncertainties associated with the theoretical
part of the potential. However, it is also worth noting that these
waves are not very well established experimentally, as indicated
by the discrepancy between the energy-dependent and
energy-independent analyses of M.A.W., and by the difference
between the latter and the more recent analysis of A.H.R.

\newpage\centerline{\hfill\small -- 51 --}\vskip .1cm

The phase shifts obtained with solution No.~3 are summarized in
the table below and compared with the energy-independent
analysis of M.A.W.

\noindent
\begin{minipage}{.49\textwidth}
	\centering
	\renewcommand{\arraystretch}{1.55}
	\begin{tabular}{rrl}
		$E_{\mathrm{lab}}$ && ${}^{3}P_0$\\
		25  & $8{,}84$&\quad $(\,8{,}52\pm0{,}31\,)$\\
		50  & $11{,}75$&\quad $(\,10{,}58\pm0{,}66\,)$\\
		95  & $10{,}19$&\quad $(\,10{,}95\pm1{,}94\,)$\\
		142 & $6{,}04$&\quad $(\,6{,}29\pm0{,}49\,)$\\
		210 & $-0{,}60$&\quad $(-0{,}86\pm0{,}56)$\\
		330 & $-11{,}48$&\quad $(-12{,}52\pm1{,}59)$
	\end{tabular}
\end{minipage}
\begin{minipage}{.49\textwidth}
	\centering
	\renewcommand{\arraystretch}{1.55}
	\begin{tabular}{rrl}
		$E_{\mathrm{lab}}$ && ${}^{3}P_1$\\
		25  & $-5{,}13$&\quad $(-5{,}04\pm0{,}15)$\\
		50  & $-8{,}65$&\quad $(-8{,}39\pm0{,}29)$\\
		95  & $-13{,}21$&\quad $(-13{,}68\pm0{,}45)$\\
		142 & $-17{,}05$&\quad $(-17{,}04\pm0{,}16)$\\
		210 & $-21{,}91$&\quad $(-22{,}22\pm0{,}32)$\\
		330 & $-29{,}40$&\quad $(-28{,}79\pm1{,}19)$
	\end{tabular}
\end{minipage}

\begin{small}
	\renewcommand{\arraystretch}{1.55}
	\setlength{\tabcolsep}{5pt}
	
	\begin{center}
		\begin{tabular}{r@{\qquad}c@{\qquad}c@{\qquad}c}
			$E_{\mathrm{lab}}$ & ${}^{3}P_2$ & $\epsilon_2$ & ${}^{3}F_2$\\[2pt]
			25  & $2{,}44\ (\,2{,}45\pm0{,}08\,)$
			& $-0{,}85\ (\mathrm{O.P.E.})$
			& $0{,}11\ (\mathrm{O.P.E.})$\\
			50  & $5{,}73\ (\,5{,}76\pm0{,}14\,)$
			& $-1{,}79\ ( -1{,}63\pm0{,}19)$
			& $0{,}36\ ( -0{,}02\pm0{,}28)$\\
			95  & $10{,}60\ (10{,}08\pm0{,}34)$
			& $-2{,}72\ ( -2{,}63\pm0{,}25)$
			& $0{,}79\ (\,1{,}19\pm0{,}61)$\\
			142 & $13{,}77\ (13{,}68\pm0{,}10)$
			& $-3{,}01\ ( -2{,}86\pm0{,}07)$
			& $1{,}10\ (\,0{,}65\pm0{,}23)$\\
			210 & $15{,}60\ (15{,}67\pm0{,}23)$
			& $-2{,}86\ ( -2{,}82\pm0{,}16)$
			& $1{,}28\ (\,1{,}16\pm0{,}33)$\\
			330 & $13{,}96\ (16{,}25\pm0{,}55)$
			& $-2{,}19\ ( -2{,}56\pm0{,}41)$
			& $0{,}86\ (\,0{,}49\pm0{,}57)$
		\end{tabular}
	\end{center}
	
	\begin{center}
		\begin{tabular}{r@{\qquad\qquad}c}
			$E_{\mathrm{lab}}$ & ${}^{3}F_3$\\[2pt]
			25  & $-0{,}24\ (\mathrm{O.P.E.})$\\
			50  & $-0{,}72\ ( -0{,}38\pm0{,}37)$\\
			95  & $-1{,}55\ ( -0{,}91\pm0{,}49)$\\
			142 & $-2{,}23\ ( -2{,}05\pm0{,}16)$\\
			210 & $-3{,}02\ ( -2{,}58\pm0{,}20)$\\
			330 & $-4{,}11\ ( -3{,}58\pm0{,}58)$
		\end{tabular}
	\end{center}
	
	\begin{center}
		\begin{tabular}{r@{\qquad}c@{\qquad}c@{\qquad}c}
			$E_{\mathrm{lab}}$ & ${}^{3}F_4$ & $\epsilon_4$ & ${}^{3}H_4$\\[2pt]
			25  & $0{,}02\ (\mathrm{O.P.E.})$
			& $-0{,}05\ (\mathrm{O.P.E.})$
			& $0{,}00\ (\mathrm{O.P.E.})$\\
			50  & $0{,}13\ (\,0{,}21\pm0{,}15)$
			& $-0{,}21\ (\mathrm{O.P.E.})$
			& $0{,}03\ (\mathrm{O.P.E.})$\\
			95  & $0{,}46\ (\,0{,}89\pm0{,}19)$
			& $-0{,}54\ (\mathrm{O.P.E.})$
			& $0{,}11\ (\mathrm{O.P.E.})$\\
			142 & $0{,}91\ (\,0{,}90\pm0{,}12)$
			& $-0{,}86\ ( -0{,}68\pm0{,}03)$
			& $0{,}22\ (\mathrm{O.P.E.})$\\
			210 & $1{,}69\ (\,2{,}02\pm0{,}19)$
			& $-1{,}24\ ( -0{,}99\pm0{,}09)$
			& $0{,}38\ (\,0{,}24\pm0{,}21)$\\
			330 & $3{,}25\ (\,2{,}77\pm0{,}23)$
			& $-1{,}69\ ( -1{,}11\pm0{,}28)$
			& $0{,}61\ (\,1{,}12\pm0{,}32)$
		\end{tabular}
	\end{center}
\end{small}

\newpage\centerline{\hfill\small -- 52 --}\vskip .1cm

\begin{small}
	\renewcommand{\arraystretch}{1.55}
	\setlength{\tabcolsep}{6pt}
	
	\begin{center}
		\begin{tabular}{r@{\qquad\qquad}c}
			$E_{\mathrm{lab}}$ & ${}^{3}H_5$\\[2pt]
			25  & $-0{,}02\ (\mathrm{O.P.E.})$\\
			50  & $-0{,}08\ (\mathrm{O.P.E.})$\\
			95  & $-0{,}31\ (\mathrm{O.P.E.})$\\
			142 & $-0{,}55\ (\mathrm{O.P.E.})$\\
			210 & $-0{,}83\ (-1{,}07\pm0{,}18)$\\
			330 & $-1{,}35\ (-1{,}80\pm0{,}46)$
		\end{tabular}
	\end{center}
	\vskip -.5cm
	\begin{center}
		\begin{tabular}{r@{\qquad}c@{\qquad}c@{\qquad}c}
			$E_{\mathrm{lab}}$ & ${}^{3}H_6$ & $\epsilon_6$ & ${}^{3}K_6$\\[2pt]
			25  & $0{,}00\ (\mathrm{O.P.E.})$
			& $-0{,}00\ (\mathrm{O.P.E.})$
			& $0{,}00\ (\mathrm{O.P.E.})$\\
			50  & $0{,}01\ (\mathrm{O.P.E.})$
			& $-0{,}03\ (\mathrm{O.P.E.})$
			& $0{,}00\ (\mathrm{O.P.E.})$\\
			95  & $0{,}05\ (\mathrm{O.P.E.})$
			& $-0{,}12\ (\mathrm{O.P.E.})$
			& $0{,}02\ (\mathrm{O.P.E.})$\\
			142 & $0{,}11\ (\mathrm{O.P.E.})$
			& $-0{,}23\ (\mathrm{O.P.E.})$
			& $0{,}04\ (\mathrm{O.P.E.})$\\
			210 & $0{,}24\ (0{,}15\pm0{,}13)$
			& $-0{,}40\ (\mathrm{O.P.E.})$
			& $0{,}09\ (\mathrm{O.P.E.})$\\
			330 & $0{,}52\ (0{,}77\pm0{,}15)$
			& $-0{,}66\ (\mathrm{O.P.E.})$
			& $0{,}20\ (\mathrm{O.P.E.})$
		\end{tabular}
	\end{center}
\end{small}

\quad. \underline{Even triplet subspace} $I=0$

In the even triplet subspace, fitting the properties of the
Deuteron provides strong constraints on the parameters of the
potential at threshold ($E\simeq0$). We have reproduced the
binding energy exactly and obtained a good value for the
quadrupole moment $Q$, which is interesting since $Q$ is
fairly sensitive to the intermediate-range forces. Calculating
the magnetic moment using a velocity-dependent potential raises
some difficulties\supercite{ref69}, and in our opinion, the
predicted value should not be given too much weight, although
it is satisfactory.

The even triplet phase shifts are not known with great precision,
and the results differ from one analysis to another, particularly
at 330 MeV. We have used the M.A.W. results between 25 and
210 MeV and Signell's solution 1\supercite{ref63} at 330 MeV.

The experimental $J=1$ phase shifts, despite their large error
bars, are rather difficult to reproduce with a local potential,
because they exhibit fairly abrupt variations with energy.
A better agreement is obtained with a velocity-dependent central
potential, but the $\chi^2$ can be reduced further if one allows
nonlocality in the spin-orbit and even tensor components
(solution No.~1).

The $\chi^2$ values per point obtained for the even triplet
phase shifts are
\vskip -.4cm
\[
\begin{array}{ll}
	\text{solution No.~1} & 2{,}5\\[3pt]
	\text{solution No.~2} & 3{,}3\\[3pt]
	\text{solution No.~3} & 3{,}3
\end{array}
\]

\newpage\centerline{\hfill\small -- 53 --}\vskip .1cm

In the latter case, the corresponding results are

\begin{small}
	\begin{center}
		\renewcommand{\arraystretch}{1.35}
		\setlength{\tabcolsep}{8pt}
		\begin{tabular}{r@{\qquad}c@{\qquad}c@{\qquad}c}
			$E_{\mathrm{lab}}$ & ${}^{3}S_1$ & $\epsilon_1$ & ${}^{3}D_1$\\[2pt]
			25  & $80{,}14\ (84{,}64\pm2{,}71)$
			& $1{,}87\ (-0{,}23\pm1{,}14)$
			& $-3{,}06\ (-3{,}12\pm0{,}19)$\\
			50  & $62{,}35\ (57{,}11\pm2{,}69)$
			& $2{,}17\ (3{,}53\pm3{,}27)$
			& $-7{,}09\ (-5{,}28\pm1{,}80)$\\
			95  & $44{,}12\ (44{,}98\pm3{,}10)$
			& $2{,}49\ (-1{,}28\pm3{,}79)$
			& $-13{,}19\ (-10{,}62\pm1{,}34)$\\
			142 & $31{,}43\ (29{,}33\pm0{,}93)$
			& $2{,}96\ (4{,}28\pm0{,}96)$
			& $-18{,}29\ (-14{,}44\pm0{,}78)$\\
			210 & $17{,}02\ (13{,}85\pm1{,}50)$
			& $3{,}87\ (6{,}47\pm0{,}70)$
			& $-24{,}15\ (-17{,}94\pm1{,}49)$\\
			330 & $0{,}14\ (5{,}70\pm4{,}20)$
			& $5{,}96\ (8{,}90\pm2{,}70)$
			& $-31{,}82\ (-29{,}20\pm3{,}00)$
		\end{tabular}
	\end{center}
	
	\renewcommand{\arraystretch}{1.35}
	\setlength{\tabcolsep}{7pt}
	
	\begin{center}
		\begin{tabular}{r@{\qquad\qquad}c}
			$E_{\mathrm{lab}}$ & ${}^{3}D_2$\\[2pt]
			25  & $4{,}02\ (\mathrm{O.P.E.})$\\
			50  & $9{,}73\ (9{,}36\pm2{,}24)$\\
			95  & $18{,}20\ (12{,}32\pm3{,}23)$\\
			142 & $24{,}05\ (22{,}54\pm0{,}79)$\\
			210 & $28{,}59\ (27{,}31\pm1{,}23)$\\
			330 & $30{,}51\ (27{,}60\pm4{,}00)$
		\end{tabular}
	\end{center}
	
	\begin{center}
		\begin{tabular}{r@{\qquad}c@{\qquad}c@{\qquad}c}
			$E_{\mathrm{lab}}$ & ${}^{3}D_3$ & $\epsilon_3$ & ${}^{3}G_3$\\[2pt]
			25  & $0{,}05\ (\mathrm{O.P.E.})$
			& $0{,}60\ (\mathrm{O.P.E.})$
			& $-0{,}06\ (\mathrm{O.P.E.})$\\
			50  & $0{,}30\ (1{,}50\pm0{,}94)$
			& $1{,}74\ (\mathrm{O.P.E.})$
			& $-0{,}28\ (\mathrm{O.P.E.})$\\
			95  & $1{,}24\ (3{,}22\pm1{,}22)$
			& $3{,}61\ (\mathrm{O.P.E.})$
			& $-0{,}96\ (\mathrm{O.P.E.})$\\
			142 & $2{,}51\ (2{,}52\pm0{,}64)$
			& $5{,}10\ (4{,}44\pm0{,}40)$
			& $-1{,}83\ (\mathrm{O.P.E.})$\\
			210 & $4{,}31\ (3{,}67\pm1{,}01)$
			& $6{,}60\ (6{,}70\pm0{,}36)$
			& $-3{,}19\ (\mathrm{O.P.E.})$\\
			330 & $6{,}52\ (3{,}50\pm1{,}70)$
			& $8{,}17\ (8{,}60\pm1{,}80)$
			& $-5{,}48\ (-4{,}70\pm2{,}60)$
		\end{tabular}
	\end{center}
	
\begin{center}
	\begin{tabular}{r@{\qquad}c@{\qquad}c@{\qquad}}
		$E_{\mathrm{lab}}$ & ${}^{3}G_4$ & ${}^3G_5$\\[2pt]
		25  & $0{,}18\ (\mathrm{O.P.E.})$&$-0,01\ (\mathrm{O.P.E.)}$\\
		50  & $0{,}79\ (\mathrm{O.P.E.})$&$-0,05\ (\mathrm{O.P.E.)}$\\
		95  & $2{,}21\ (\mathrm{O.P.E.})$&$-0,16\ (\mathrm{O.P.E.)}$\\
		142 & $3{,}76\ (\mathrm{O.P.E.})$&$-0,25\ (\mathrm{O.P.E.)}$\\
		210 & $5{,}83\ (\mathrm{O.P.E.})$&$-0,30\ (\mathrm{O.P.E.)}$\\
		330 & $8{,}88\ (0{,}8\pm1{,}60)$&$-0,19\ (1,60\pm 0,70)$\\
	\end{tabular}
\end{center}
\end{small}

\newpage\centerline{\hfill\small -- 54 --}\vskip .1cm

\noindent
\underline{Deuteron properties}

\begin{tabular}{@{}l}
	-- binding energy \qquad $E_D=-2{,}2246\ \mathrm{MeV}$
	\qquad $(-2{,}2246\ \mathrm{MeV})$\\
	-- quadrupole moment \qquad $Q_D=0{,}2900\ \mathrm{fm}^2$
	\qquad $(0{,}2875\ \mathrm{fm}^2)$\\
	-- D-state admixture \qquad $P_D=6{,}75\,\%$\\
	-- asymptotic ratio \qquad $D/S=2{,}93\,\%$\\
	-- magnetic moment \qquad $\mu=0{,}8392\,\mu_N$
	\qquad $(0{,}8574\,\mu_N)$
\end{tabular}

\noindent
\underline{Low-energy parameters}

\begin{tabular}{@{}l}
	-- scattering length \qquad $a_{np}=5{,}418\ \mathrm{fm}$
	\qquad $(5{,}413\ \mathrm{fm})$\\
	-- effective range \qquad $r(-E_D,0)=1{,}753\ \mathrm{fm}$
	\qquad $(1{,}748\ \mathrm{fm})$
\end{tabular}

\vskip .3cm

\qquad .\underline{Odd singlet subspace}

This subspace is poorly constrained experimentally. Signell
\supercite{ref63}, for example, gives two solutions for the
${}^{1}P_1$ phase shift at 330 MeV, $-43.3^\circ$ and
$-27.6^\circ$! Moreover, even for the P wave, this subspace is
not very sensitive to the core of the interaction. Indeed, it is
quickly apparent that even substantial changes in the parameters
of the short-range potential have little effect on the calculated
phase shifts. In the even singlet subspace, for example, the
repulsive core is compensated by the attractive long- and
intermediate-range components, so that the resulting phase shift
is highly sensitive to changes in the forces in either region.
In the odd singlet subspace, by contrast, the interaction is
repulsive at all distances, and the cumulative effect is less
sensitive to variations in the height of the core. Needless to
say, in the ${}^{1}F_3$ and ${}^{1}H_5$ waves, the centrifugal
potential dominates at short distances.

Under these conditions, the $\chi^2$ method does not allow us
to determine the short-range odd-singlet potential
unambiguously, unlike what we found in the other subspaces.

With a core of approximately 200 MeV for the combination
$V_C-3V_{SS}$, we obtain the following results (solution No.~3):

\begin{small}
	\begin{center}
		\renewcommand{\arraystretch}{1.6}
		\setlength{\tabcolsep}{7pt}
		\begin{tabular}{r@{\qquad}c@{\qquad}c@{\qquad}c}
			$E_{\mathrm{lab}}$ & ${}^{1}P_1$ & ${}^{1}F_3$ & ${}^{1}H_5$\\[2pt]
			25  & $-7{,}23\ (-4{,}00\pm0{,}75)$
			& $-0{,}47\ (\mathrm{O.P.E.})$
			& $-0{,}03\ (\mathrm{O.P.E.})$\\
			50  & $-11{,}32\ (-9{,}15\pm2{,}30)$
			& $-1{,}23\ (\mathrm{O.P.E.})$
			& $-0{,}19\ (\mathrm{O.P.E.})$\\
			95  & $-16{,}49\ (-10{,}80\pm2{,}90)$
			& $-2{,}25\ (\mathrm{O.P.E.})$
			& $-0{,}54\ (\mathrm{O.P.E.})$\\
			142 & $-20{,}89\ (-18{,}20\pm1{,}30)$
			& $-3{,}25\ (-2{,}02\pm0{,}79)$
			& $-0{,}91\ (\mathrm{O.P.E.})$\\
			210 & $-26{,}45\ (-22{,}20\pm1{,}95)$
			& $-4{,}33\ (-4{,}80\pm0{,}86)$
			& $-1{,}36\ (\mathrm{O.P.E.})$\\
			330 & $-34{,}88\ (-43{,}30\pm2{,}10)$
			& $-6{,}17\ (-2{,}60\pm1{,}60)$
			& $-1{,}98\ (-2{,}50\pm1{,}60)$
		\end{tabular}
	\end{center}
\end{small}

\noindent
$\chi^2/\text{point}=5{,}6$.

Most of the $\chi^2$ comes from the controversial points at
$E_{\mathrm{lab}}=330$ MeV. Here, the values in parentheses
are those of Signell's solution 1\supercite{ref63}, while at the
five lower energies they are taken from M.A.W.\supercite{ref50}.

\newpage\centerline{\hfill\small -- 55 --}\vskip .1cm

\centerline{\includegraphics[width=\linewidth]{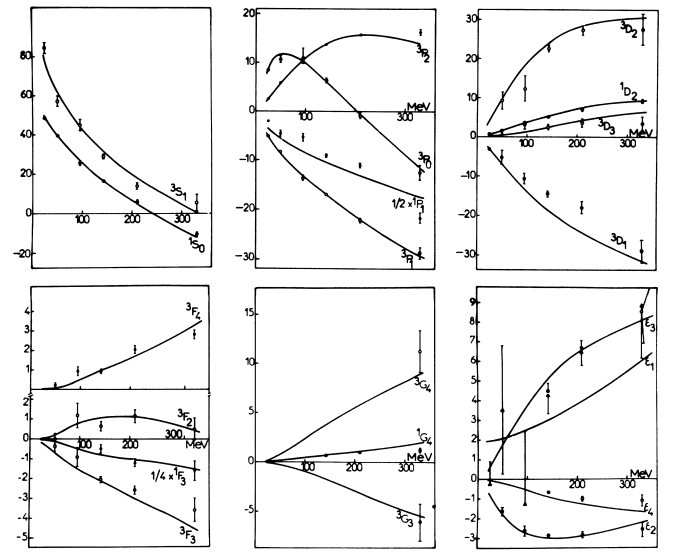}\hfill}

\centerline{\small \underline{Fig. 23}: Phase shifts obtained with solution No.~3}

\newpage\centerline{\hfill\small -- 56 --}\vskip .1cm

\begin{small}
	\centering
	{Core parameters of the potential for isospin $I=1$
		(even-singlet and odd-triplet subspaces).
		The parameters are defined in Eqs.~(5.8), (5.11), and (5.12).}
	\vskip .2cm
	\renewcommand{\arraystretch}{1.6}
	\setlength{\tabcolsep}{10pt}
	
	\begin{tabular}{|>{\centering\arraybackslash}m{3.2cm}|
			>{\centering\arraybackslash}m{3.1cm}|
			>{\centering\arraybackslash}m{3.1cm}|
			>{\centering\arraybackslash}m{3.1cm}|}
		\hline
		Potential & Solution 1 & Solution 2 & Solution 3 \\
		\hline
		
		Central
		\newline Singlet
		&
		$\begin{aligned}
			p_1^{\mathrm{CS}} &= 45{,}6\ \mathrm{MeV}\\
			p_2^{\mathrm{CS}} &= 0{,}915
		\end{aligned}$
		&
		$\begin{aligned}
			p_1^{\mathrm{CS}} &= 43\ \mathrm{MeV}\\
			p_2^{\mathrm{CS}} &= 0{,}858
		\end{aligned}$
		&
		$\begin{aligned}
			p_1^{\mathrm{CS}} &= 43\ \mathrm{MeV}\\
			p_2^{\mathrm{CS}} &= 0{,}952
		\end{aligned}$
		\\
		\hline
		
		Quadratic
		\newline spin-orbit
		&
		$\begin{aligned}
			q_1 &= -2{,}3\ \mathrm{MeV\,fm^2}\\
			q_2 &= 23{,}9\ \mathrm{MeV\,fm}
		\end{aligned}$
		&
		$\displaystyle p_1^{\mathrm{SO2}}=28\ \mathrm{MeV}$
		&
		$\displaystyle p_1^{\mathrm{SO2}}=33\ \mathrm{MeV}$
		\\
		\hline
		
		Spin-orbit
		&
		$\begin{aligned}
			p_1^{\mathrm{SO}} &= -308\ \mathrm{MeV}\\
			p_2^{\mathrm{SO}} &= -0{,}058
		\end{aligned}$
		&
		$\displaystyle p_1^{\mathrm{SO}}=-286\ \mathrm{MeV}$
		&
		$\displaystyle p_1^{\mathrm{SO}}=-286\ \mathrm{MeV}$
		\\
		\hline
		
		Tensor
		&
		$\begin{aligned}
			p_1^{\mathrm{T}} &= 68\ \mathrm{MeV}\\
			p_2^{\mathrm{T}} &= -0{,}17
		\end{aligned}$
		&
		$\displaystyle p_1^{\mathrm{T}}=33\ \mathrm{MeV}$
		&
		$\displaystyle p_1^{\mathrm{T}}=30\ \mathrm{MeV}$
		\\
		\hline
		
		Central
		\newline Triplet
		&
		$\begin{aligned}
			p_1^{\mathrm{CT}} &= 115\ \mathrm{MeV}\\
			p_2^{\mathrm{CT}} &= 0{,}957
		\end{aligned}$
		&
		$\begin{aligned}
			p_1^{\mathrm{CT}} &= 86{,}4\ \mathrm{MeV}\\
			p_2^{\mathrm{CT}} &= 0{,}863
		\end{aligned}$
		&
		$\begin{aligned}
			p_1^{\mathrm{CT}} &= 77{,}4\ \mathrm{MeV}\\
			(p^{\mathrm{CT}} &= p^{\mathrm{CS}})
		\end{aligned}$
		\\
		\hline
	\end{tabular}
\end{small}

\vskip 1.5 cm

\begin{small}
	\centering
	{Core parameters of the potential for isospin $I=0$
		(even-triplet and odd-singlet subspaces).}
	\vskip .2cm
	\renewcommand{\arraystretch}{1.6}
	\setlength{\tabcolsep}{10pt}
	
	\begin{tabular}{|>{\centering\arraybackslash}m{3.2cm}|
			>{\centering\arraybackslash}m{3.1cm}|
			>{\centering\arraybackslash}m{3.1cm}|
			>{\centering\arraybackslash}m{3.1cm}|}
		\hline
		Potential & Solution 1 & Solution 2 & Solution 3 \\
		\hline
		
		Central
		\newline Singlet
		&
		$\begin{aligned}
			p_1^{\mathrm{CS}} &= 250\ \mathrm{MeV}\\
			p_2^{\mathrm{CS}} &= 0{,}940
		\end{aligned}$
		&
		$\begin{aligned}
			p_1^{\mathrm{CS}} &= 237\ \mathrm{MeV}\\
			p_2^{\mathrm{CS}} &= 0{,}940
		\end{aligned}$
		&
		$\begin{aligned}
			p_1^{\mathrm{CS}} &= 200\ \mathrm{MeV}\\
			p_2^{\mathrm{CS}} &= 0{,}998
		\end{aligned}$
		\\
		\hline
		
		Quadratic
		\newline spin-orbit
		&
		$\begin{aligned}
			q_1 &= -3\ \mathrm{MeV\,fm^2}\\
			q_2 &= 16\ \mathrm{MeV\,fm}
		\end{aligned}$
		&
		$\displaystyle p_1^{\mathrm{SO2}}=25\ \mathrm{MeV}$
		&
		$\displaystyle p_1^{\mathrm{SO2}}=25\ \mathrm{MeV}$
		\\
		\hline
		
		Spin-orbit
		&
		$\begin{aligned}
			p_1^{\mathrm{SO}} &= 50\ \mathrm{MeV}\\
			p_2^{\mathrm{SO}} &= -0{,}02
		\end{aligned}$
		&
		$\displaystyle p_1^{\mathrm{SO}}=50\ \mathrm{MeV}$
		&
		$\displaystyle p_1^{\mathrm{SO}}=50\ \mathrm{MeV}$
		\\
		\hline
		
		Tensor
		&
		$\begin{aligned}
			p_1^{\mathrm{T}} &= -30\ \mathrm{MeV}\\
			p_2^{\mathrm{T}} &= 0{,}02
		\end{aligned}$
		&
		$\displaystyle p_1^{\mathrm{T}}=-29{,}7\ \mathrm{MeV}$
		&
		$\displaystyle p_1^{\mathrm{T}}=-29{,}7\ \mathrm{MeV}$
		\\
		\hline
		
		Central
		\newline Triplet
		&
		$\begin{aligned}
			p_1^{\mathrm{CT}} &= 16{,}6\ \mathrm{MeV}\\
			p_2^{\mathrm{CT}} &= 0{,}998
		\end{aligned}$
		&
		$\begin{aligned}
			p_1^{\mathrm{CT}} &= 9{,}92\ \mathrm{MeV}\\
			p_2^{\mathrm{CT}} &= 0{,}998
		\end{aligned}$
		&
		$\begin{aligned}
			p_1^{\mathrm{CT}} &= 10{,}92\ \mathrm{MeV}\\
			(p^{\mathrm{CT}} &= p^{\mathrm{CS}})
		\end{aligned}$
		\\
		\hline
	\end{tabular}
\end{small}

\newpage\vskip -0.05cm\centerline{\hfill\small -- 57 --}\vskip .1cm

\centerline{\underline{\uppercase{Conclusion}}}

We first calculated the exchange of two pions within the framework
of dispersion relations, using the currently known properties of
the $\pi$ mesons. (It is particularly important to note that the
values we used for the modulus $|f_+^0(k)|$ of the S-wave
$\pi\pi\longrightarrow N\bar N$ amplitude are confirmed by the
recent results obtained in Copenhagen~\cite{ref33} and Paris
\cite{ref70}, based on the latest analyses of pion--nucleon and
pion--pion phase shifts.) At this stage, our aim was not to
reproduce the nucleon--nucleon scattering data as accurately as
possible, but rather to establish connections between experimental
information from different branches of hadronic physics. We
believe that we have succeeded in demonstrating the consistency
of the results for the $\pi N$, $\pi\pi$, and $NN$ systems,
thereby once again highlighting the value of dispersion relations
in the study of elementary particles.

The long- and intermediate-range interaction constructed in this
way was then supplemented with a core fitted empirically to the
properties of the Deuteron and to the experimental phase shifts
between 0 and 300 MeV. We have already distinguished two possible
approaches to this semi-phenomenological study.

It can first be viewed as an indirect test of the theoretical
investigation presented in the preceding chapters. In this sense,
we have succeeded, using very few free parameters to describe
the short-range forces, in reproducing the entire set of
low-energy nucleon--nucleon scattering results. This provides
support for our evaluation of two-pion exchanges.

The other viewpoint is more ambitious: if one places a high degree
of confidence, a priori, in the dispersion-relation method, one
may consider the potential to be well determined beyond 1 fm and
regard the analysis of scattering data as a means of identifying
the essential features of the short-range forces. (This is not
unreasonable, since, apart from the odd-singlet subspace, we
found practically no ambiguity in the parameters describing the
core region, at least within the functional form adopted.
In other words, the intermediate-range forces and one-pion
exchange determine the size of the core; this can also be
appreciated by comparing the phenomenological potentials
\cite{ref68}.)

The results then reveal two salient features:

(i) Our potential has a soft core, which is quite different from
the infinite repulsion often assumed and seems intuitively more
consistent with experiments and high-energy models suggesting
that the nucleon is composite.

(ii) This repulsion increases rather rapidly with the relative
velocity of the colliding nucleons. Already in the intermediate
region, our theoretical study had revealed a noticeable energy
dependence of the potential; in the core region, the nonlocal
terms are even more important.

A natural extension of our work is to investigate the implications
of these properties in systems containing several nucleons, or,
more simply, in the case of the off-shell nucleons involved, for
example, in proton--proton bremsstrahlung. Nuclear-structure
calculations are facilitated by the presence of a soft core, which
improves the convergence of the perturbation series, and fairly
good results have already been obtained with phenomenological
S.S.C. (super-soft-core) potentials. The most novel feature is
the strong energy dependence of the forces, and it will be very
interesting to see the results of calculations undertaken with
our potential for nuclear matter and for certain nuclei.

\newpage\vskip -0.05cm\centerline{\hfill\small -- 58 --}\vskip- .1cm	

\end{document}